\documentclass[graybox, vecphys, natbib]{svmult}

\usepackage{mathptmx}       
\usepackage{helvet}         
\usepackage{courier}        
\usepackage{type1cm}        
\usepackage{makeidx}         
\usepackage{graphicx}        
\usepackage{multicol}        
\usepackage[bottom]{footmisc}
\usepackage{hyperref}        
\usepackage{soul}            

\makeindex             
                       
\def \lesssim {\mathrel{\vcenter
     {\offinterlineskip \hbox{$<$}\hbox{$\sim$}}}}
\def \gtrsim {\mathrel{\vcenter
     {\offinterlineskip \hbox{$>$}\hbox{$\sim$}}}}

\newcommand{\apj}{Astrophys. J.}%

\newcommand{\apjl}{Astrophys. J. Lett.}%
\newcommand{\apjs}{Astrophys. J. Supp.}%
\newcommand{\aap}{Astron. Astrophys.}%
\newcommand{\mnras}{Mon. Not. Roy. Astron. Soc.}%
\newcommand{\prl}{Phys. Rev. Lett.}    
\newcommand{\prd}{Phys. Rev. D}    
\newcommand{\prc}{Phys. Rev. C}    
\newcommand{\araa}{Annual Review of Astronomy and Astrophysics}
\newcommand{\nat}{Nature}
\newcommand{\physrep}{Physics Reports}
\newcommand{\pasj}{Publ. Astron. Soc. of Japan}
\newcommand{\jcap}{Journal of Cosmology and Astroparticle Physics}

\begin{document}
\bibliographystyle{spbasic}
\title*{Dynamics and equation of state dependencies of relevance for nucleosynthesis in supernovae and neutron star mergers}
\titlerunning{Equation of state effects in supernovae and compact object mergers}
\author{Hans-Thomas Janka$^1$ \thanks{corresponding author} and Andreas Bauswein$^2$}
\authorrunning{Hans-Thomas Janka and Andreas Bauswein}
\institute{$^1$ Max-Planck-Institute for Astrophysics, Karl-Schwarzschild-Str.~1, 85748~Garching, Germany\\ \email{thj@mpa-garching.mpg.de}\\
$^2$ GSI Helmholtzzentrum f\"ur Schwerionenforschung, Planckstra\ss e~1, 64291~Darmstadt, Germany\\ \email{a.bauswein@gsi.de}}
%
%
\maketitle
\abstract{Neutron stars (NSs) and black holes (BHs) are born when the final collapse of the stellar core terminates
the lives of stars more massive than about 9 solar masses. This can trigger the powerful ejection of a large fraction 
of the star's material in a core-collapse supernova (CCSN), whose extreme luminosity is energized 
mostly by the decay of radioactive isotopes such as $^{56}$Ni and its daughter nucleus $^{56}$Co. When evolving 
in close binary systems, the compact relics of such infernal catastrophes spiral towards each other on orbits 
gradually decaying by gravitational-wave emission. Ultimately, the violent collision of the two components 
forms a more massive, rapidly spinning remnant, again accompanied by the ejection of considerable amounts 
of matter. These merger events can be observed by high-energy bursts of gamma rays 
with afterglows and electromagnetic transients called kilonovae, which radiate the energy released
in radioactive decays of freshly assembled rapid neutron-capture elements. By means of their mass ejection and
the nuclear and neutrino reactions taking place in the ejecta, both CCSNe and compact object mergers (COMs) 
are prominent sites of heavy-element nucleosynthesis and play a central role in the cosmic cycle of matter
and the chemical enrichment history of galaxies. The nuclear equation of state (EoS) of NS matter,
from neutron-rich to proton-dominated conditions and with temperatures ranging from about zero to roughly 100\,MeV,
is a crucial ingredient in these astrophysical phenomena. It determines their dynamical processes, their
remnant properties even at the level of deciding between NS or BH, and the properties of the associated 
emission of neutrinos, whose interactions govern the thermodynamic conditions and the neutron-to-proton ratio 
for nucleosynthesis reactions in the innermost ejecta. This chapter discusses corresponding EoS dependent
effects of relevance in CCSNe as well as COMs.}


\section{Dynamical events of neutron stars and their equation of state}
\label{sec:NSs+EoS}

Astrophysical phenomena involving neutron stars (NSs) and black hole (BHs) generate the most extreme conditions
in the universe after the Big Bang. The temperatures, densities, degrees of neutronization, and magnetic 
field strengths realized in the interior of these relics of stellar evolution and in matter in their immediate  
surroundings cannot be explored in terrestrial experiments. Correspondingly, the description of physical processes 
that involve such objects heavily rests on theoretical models for the dynamics and of the relevant microphysics. 
The same holds true for the prediction of associated 
observable signals and the interpretation of such measurements. Because of their extraordinary conditions,
compact stellar remnants are therefore unique laboratories of nuclear physics, gravitational physics, and particle 
physics, in particular of high-energy neutrino physics.

\begin{figure}
\includegraphics[width=1.0\columnwidth]{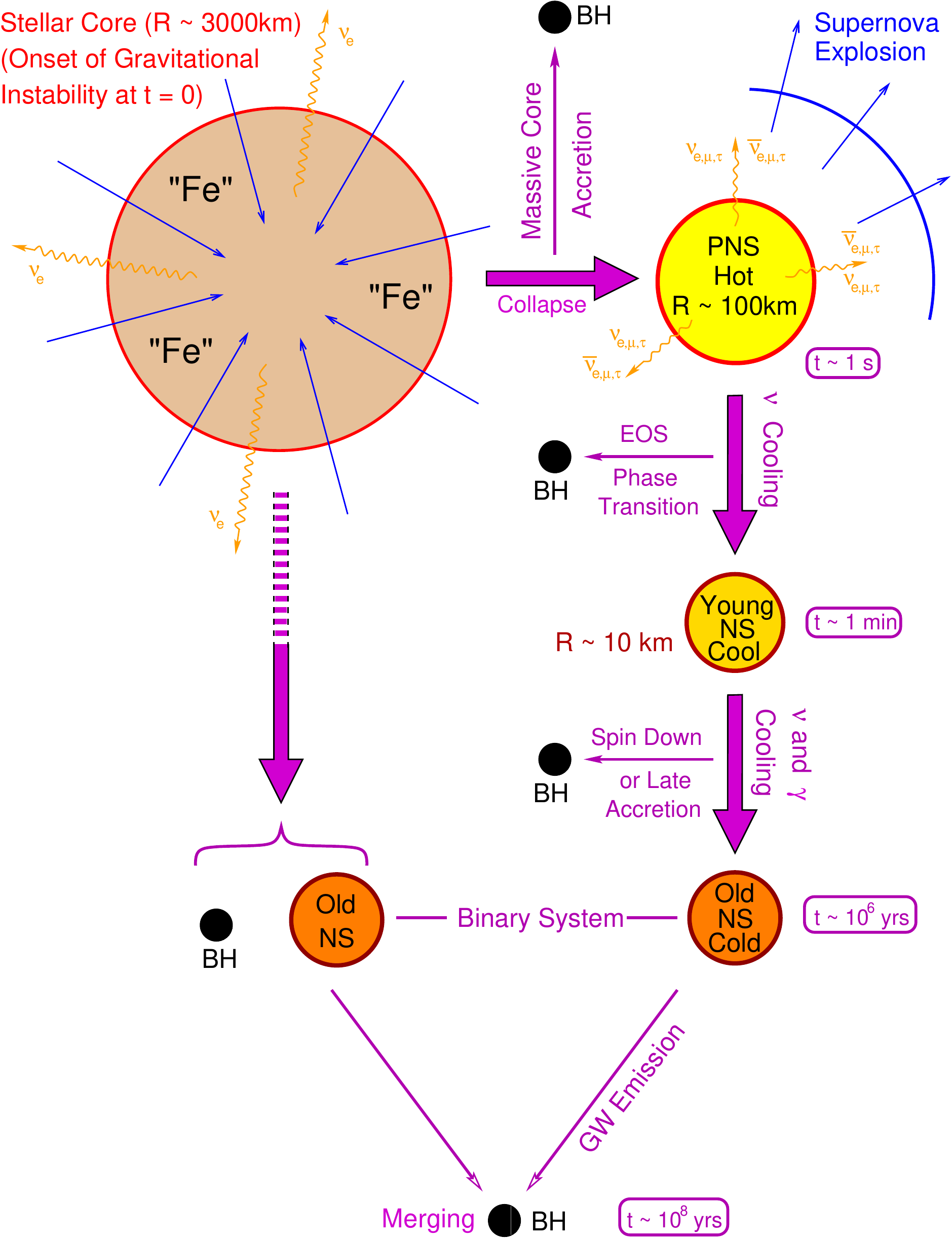}
\caption{Evolution paths from collapsing massive stars to NSs and stellar-mass BHs. The gravitational
instability of the degenerate core (mostly composed of iron-group elements) of a massive star 
can either lead to the ``direct'' formation of
a BH by continuous accretion of matter onto the transiently formed proto-NS (PNS) without any 
concomitant CCSN explosion. If a successful explosion is launched, an initially hot
PNS cools by intense emission of neutrinos and antineutrinos of all flavors. On the way to an old,
cold NS, a phase transition in the high-density EoS, spin-down by angular momentum loss (e.g., through
magnetic fields), or late accretion of matter that does not achieve to get unbound in the CCSN explosion
can lead to the delayed collapse of the PNS or young NS to a BH. In close binary systems the compact 
remnants spiral towards each other by GW emission to merge after tens of millions to billions of years,
giving birth to a BH or, in some cases, a very massive NS. Mass ejection during the CCSN explosion
and COM yields important contributions to the cosmic formation of heavy chemical elements. The
dynamics of these processes and the associated emission of neutrinos are governed by the properties
of the high-density EoS in PNSs and NSs.}
\label{fig:NSBHevol}
\end{figure}

This chapter will provide an introduction to the dynamical scenarios that lead to NS and BH formation by the 
collapse of massive stars and the final collisions of two such bodies after their secular inspiral in binary
systems over tens of millions to billions of years (Figure~\ref{fig:NSBHevol}). These violent cosmic events can 
generate densities up to more than $2\times 10^{15}$\,g\,cm$^{-3}$ (nearly 8 times higher than the nuclear
saturation density of $n_0\approx 0.16$ nucleons\,fm$^{-3}$ or $\rho_0 \approx 2.66\times 10^{14}$\,g\,cm$^{-3}$).
Temperatures up to roughly 100\,MeV (about $10^{12}$K) can be reached inside the new-born NSs (the so-called proto-NSs; 
PNSs) and during the mergers of two NSs, and matter that gets explosively expelled from the close vicinity of the
compact objects starts its expansion with temperatures that can exceed $\sim$1\,MeV (about $10^{10}$\,K), where 
nuclear statistical equilibrium 
(NSE) is established. In such environments neutrino emission and absorption processes play a pivotal role, and the 
neutron-to-proton ratios can therefore range from close to zero up to nearly unity. The discussion in this chapter 
will be focused on the influence of the high-density equation of state (EoS) of hot NS matter on the dynamical 
evolution of core-collapse supernovae (CCSNe) and compact object mergers (COMs) and on how these dependencies 
affect the production of heavy chemical elements in their ejecta. The nucleosynthesis in the expelled material can 
also open new ways to gain more information about the EoS of NS matter, which is still incompletely known, in particular 
also at non-zero temperatures.

Traditionally, the mass-radius relation of
cold NSs, which is a characteristic property of the NS structure and depends on the pressure-density relation 
of the dense-matter EoS, is constrained by
measuring lower limits of the maximum NS mass as well as determinations or estimates of NS radii. 
These can be obtained through astronomical observations of close binary-star systems, mainly double
NS binaries, white dwarf-NS binaries, and X-ray or optical binaries where a NS or BH accretes mass from a 
nondegenerate, gas-donating companion. The currently tightest lower bounds on the maximum NS 
mass come from massive pulsars in relativistic or very compact binaries: $1.91\pm 0.02\,M_\odot$ for PSR~J1614-2230
\citep{Arzoumanian2018}, $2.01\pm 0.04\,M_\odot$ for PSR~J0348+0432 \citep{Antoniadis+2013}, $2.14^{+0.10}_{-0.09}\,M_\odot$ for PSR~J0740+6620 \citep{Cromartie+2020}, $2.13\pm 0.04\,M_\odot$ for PSR~J1810+1744
\citep{Romani+2021}, and $2.35\pm 0.17\,M_\odot$ for the ``black widow'' pulsar PSR~J0952-0607 \citep{Romani+2022}
($1\sigma$ confidence intervals). Combining information from the last case with reanalysis of other pulsars,
a minimum value of 2.09\,$M_\odot$ for the maximum NS mass was concluded at $3\sigma$ confidence \citep{Romani+2022}.

A new multimessenger window to the universe opened up in recent years with the now-possible
measurements of gravitational wave (GW) signals from COM events \citep{GWTC-3:2021arXiv211103606T}, 
in particular from the binary NS mergers of GW170817
\citep{Abbott+2017PhRvL.119p1101A,Abbott+2019PhRvX...9a1001A} and GW190425 \citep{Abbott+2020ApJ...892L...3A}, 
the former being associated with detections of a short gamma-ray-burst, GRB~170817A 
\citep{Troja+2017,Abbott+2017ApJ...848L..12A,Abbott+2017ApJ...848L..13A}, and an ultraviolet, optical,
and near-infrared electromagnetic transient, AT~2017gfo, a so-called kilonova 
\citep[e.g.,][]{Pian+2017,Tanvir+2017,Smartt+2017,Arcavi+2017,Cowperthwaite+2017}. The GW signal from the
NS-NS inspiral phase provided information on the tidal deformability of the NSs and thus new, important 
clues on the cold high-density EoS. The LIGO Scientific Collaboration together with the Virgo Collaboration 
deduced a total mass of the binary system of $2.74^{+0.04}_{-0.01}\,M_\odot$ and individual masses of 
(1.36--1.60)\,$M_\odot$ for the primary component and (1.17--1.36)\,$M_\odot$ for the secondary component
(90\% credible intervals),
if the two NS spins are restricted to low values in the range of observed Galactic binary NSs \citep{Abbott+2017PhRvL.119p1101A}. 
In addition, they could determine the radii of both NSs to be
in the range from about $10.8^{+2.0}_{-1.7}$\,km to $11.9^{+1.4}_{-1.4}$\,km at the 90\% credible level,
depending on the description of macroscopic NS properties or a pressure-density parametrization of the EoS,
respectively \citep{Abbott+2018PhRvL.121p1101A}.

State-of-the-art radius determinations were recently also obtained through  
observations and modelling of X-rays emitted from hot regions on the surface of rotating NSs, making use of
data from the {\em Neutron Star Interior Composition ExploreR} (NICER), an X-ray telescope installed on the
International Space Station. For the isolated millisecond pulsar PSR~J0030+0451 \citet{Riley+2019} inferred a gravitational mass of $1.34^{+0.15}_{-0.16}\,M_\odot$
and an equatorial radius of $12.71^{+1.14}_{-1.19}$\,km (approximately 16\% and 84\% quantiles), whereas \citet{Miller+2019} estimated a gravitational mass of $1.44^{+0.15}_{-0.14}\,M_\odot$ and an equatorial circumferential radius of $13.02^{+1.24}_{-1.06}$\,km (68\% confidence). And for PSR~J0740+6620, which is among 
the NSs with the highest well determined mass values, \citet{Riley+2021} constrained the mass to be
$2.072^{+0.067}_{-0.066}\,M_\odot$ and the equatorial radius to be $12.39^{+1.30}_{-0.98}$\,km (posterior 
credible intervals bounded by the 16\% and 84\% quantiles). Combining these NICER determinations and the NS
mass-radius constraints from GW170817, \citet{Raaijmakers+2021} inferred 95\% credible radius ranges of $12.33^{+0.76}_{-0.81}$\,km 
or $12.18^{+0.56}_{-0.79}$\,km for a 1.4\,$M_\odot$ NS, considering two different parameterizations of the high-density EoS and 
confirming previous bounds derived on grounds of a Bayesian analysis of a heterogeneous data set from X-ray observations (Type-I X-ray bursters and transient low-mass X-ray binaries) by \citet{Steiner+2010}.

Explosive astrophysical events that lead to high-temperature conditions such as CCSNe and binary NS mergers
offer promising perspectives to obtain novel constraints on the intrinsic physics also of hot NS matter. This 
concerns the properties of the nuclear EoS as well as the neutrino reactions in such matter. GW signals and 
neutrino signals are messengers originating most directly from the high-density regions, reflecting effects 
associated with the bulk motions of the plasma (in GWs and neutrinos) as well as the thermodynamical state 
of the emitting regions (in neutrinos). Both of these signals have already demonstrated their high potential, for
GWs in connection with the NS merger event of GW170817, which occurred in the S0 galaxy NGC~4993 at a 
distance of $40\pm 4$\,Mpc (redshift of $z = 0.00968$) \citep{Pian+2017,Smartt+2017}, and for neutrinos in 
association with the once-in-a-hundred-years event of Supernova (SN)~1987A, which was the explosive death of 
the blue supergiant star Sanduleak~$-$69\,202 in the Tarantula Nebula of the Large Magellanic Cloud. 

Two dozen electron
antineutrinos from SN~1987A captured in the large underground water and scintillator experiments of 
Kamiokande~II \citep{Hirata+1987}, IMB \citep{Bionta+1987}, and Baksan \citep{Alekseev+1987} proved the ---at 
least transient--- existence of a NS formed in the collapse of the stellar core. The integrated energy of the 
neutrino burst of approximately $\sim$10\,s duration provided 
rough constraints on the NS mass and its EoS, although with huge uncertainties because of the poor statistics of
the measured neutrinos \citep[e.g.,][]{Burrows1990ARNPS,Burrows1990book}. 
A future CCSN in the Milky Way, however, which is expected to happen 1--3 times 
per century on average \citep[e.g.,][]{Diehl+2006}, will yield splendid information on the time, energy, and
potentially also flavor evolution of the emitted neutrinos \citep[for a recent review, see][]{Mirizzi+2016}.

In the case of GW170817 the distance of the source permitted to measure the GW signal only from the inspiral
phase before the merger of the two NSs, because the final plunge and subsequent ring-down with their 
higher-frequency emission were too weak to be registered by the LIGO and Virgo detectors active at that 
time. Nevertheless, extremely useful limits on the tidal deformability parameter linked to the cold EoS
of the pre-merger NSs could be deduced from the signal \citep{Abbott+2019PhRvX...9a1001A} and provided
valuable input for state-of-the-art constraints on the dense matter EoS and NS properties 
\citep[e.g.,][]{Fattoyev+2018,Raaijmakers+2021}.
With the successively enhanced sensitivity of the GW interferometers, the number of detected NS-NS mergers is 
determined to grow and, with a sufficiently long period of measurements also close events will be
discovered. These will ultimately permit to extract the ring-down GW signal with its characteristic peak frequency from
quadrupole oscillations of the massive merger remnant 
\citep[e.g.,][]{Oechslin+2007,Bauswein+2010,Stergioulas+2011,Bauswein+2012PRL,Bauswein+2012PhRvD,Hotokezaka2013a,Takami+2014,Takami+2015,Bernuzzi2015, Rezzolla+2016}
and with further EoS-dependent features probing the high-temperature, high-density physics and potential phase transitions in the
super-nuclear regime.

Also the fact that the kilonova AT~2017gfo was observed in connection to GW170817 can be used to infer constraints on
the NS EoS. The electromagnetic emission of this astronomical transient over more than 10 days was powered by the 
radioactive decay heating of rapid neutron-capture (``r-process'') nuclei in approximately 0.05\,$M_\odot$ of ejected 
material \citep[e.g.,][]{Kasen+2017,Smartt+2017,Perego+2017}. This implies a NS-NS merger remnant that has not
collapsed to a BH immediately after the collision but, if the remnant finally collapsed, its gravitational instability 
is likely to have happened only considerably (tens of milliseconds
or more) after the instant of the merging. Only if the merger product survived for some time, 
a sufficient amount of mass could be added to the dynamical ejecta from the collision phase itself 
\citep[which can account for up to roughly 0.02\,$M_\odot$;][]{Bauswein+2013,Hotokezaka2013} by subsequent secular mass loss 
of the remnant through magnetohydrodynamic and viscous effects as well as neutrino-driven mass ejection \citep[e.g.][]{Fernandez2013,Perego2014,Just+2015,Wu+2016,Fernandez+2016,Shibata+2017,Radice+2018}. Making use of this argument, \citet{Bauswein+2017}
deduced lower limits on the radii of (cold) 1.6\,$M_\odot$ and maximum-mass NSs of $10.68^{+0.15}_{-0.04}$\,km
and $9.60^{+0.14}_{-0.03}$\,km, respectively. Referring to the observation of the kilonova and 
its properties, in particular its large ejecta mass, the preceding GRB~170817A, 
and the general theoretical understanding of the merger dynamics, upper mass limits for 
non-rotating (spherical) NSs were inferred by 
\citet{Margalit+2017} ($\sim$\,2.17\,$M_\odot$ at 90\% confidence), \citet{Shibata+2017} (2.15--2.25\,$M_\odot$), \citet{Rezzolla+2018} (between $2.01\pm 0.04\,M_\odot$ and $2.16^{+0.17}_{-0.15}\,M_\odot$), and \citet{Ruiz+2018} (2.16--2.28\,$M_\odot$).  

This chapter aims at explicating these and other links between the nucleosynthesis occurring
in CCSNe and COMs on the one hand and characteristic features of the high-density EoS in NSs
on the other hand.

\begin{figure}
\includegraphics[width=1.0\columnwidth]{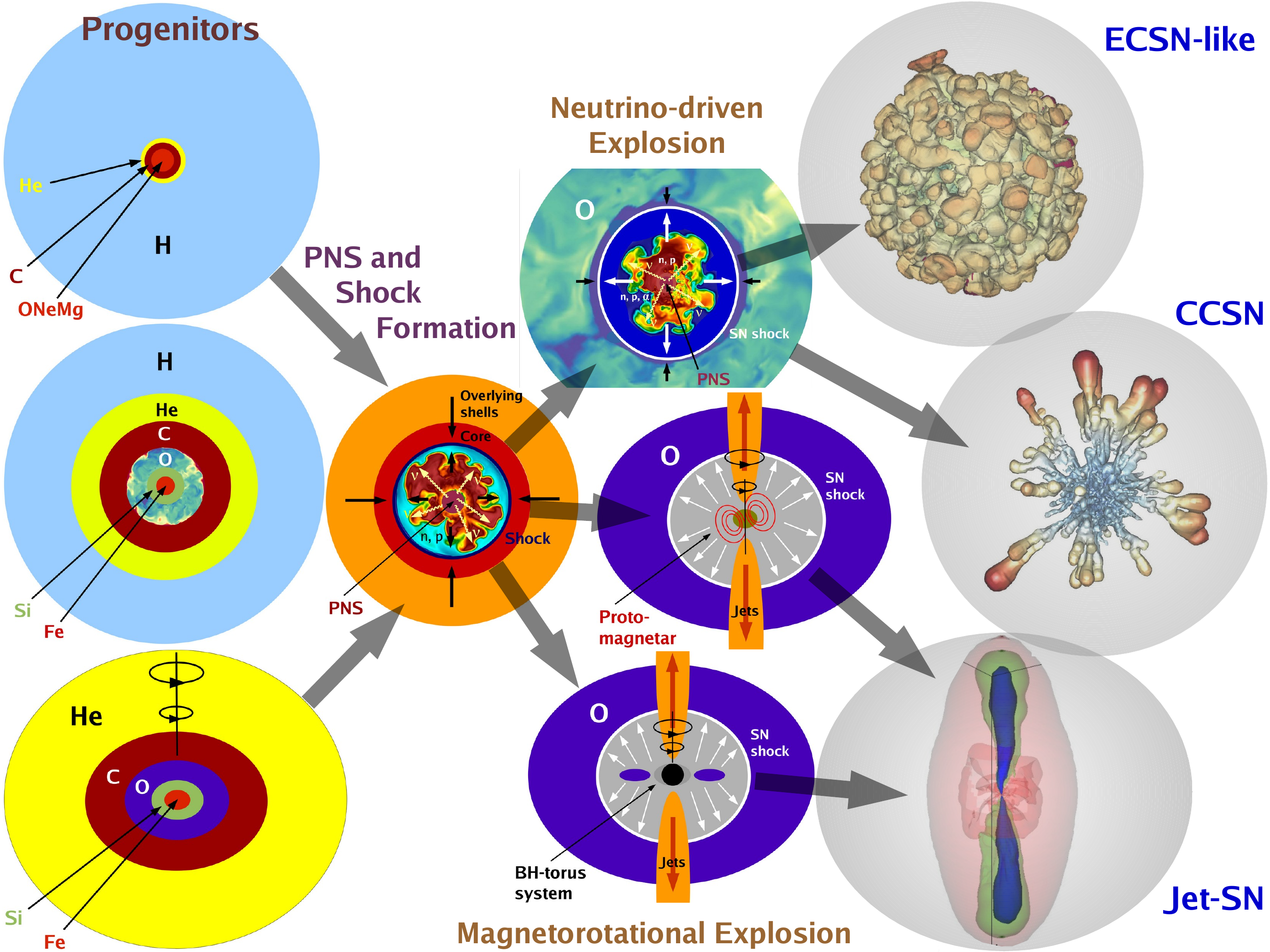}
\caption{Theoretical typology of SNe from massive stars. Different progenitors with their
onion-shell structure indicated schematically by colored layers (not drawn to scale; {\em left column}) 
lead to different kinds of massive-star explosions ({\em right}). All types have in common that the
degenerate stellar core, either composed of oxygen, neon, and magnesium in stars near
the low-mass end of core-collapse progenitors (around 9\,$M_\odot$ for solar metallicity)
or containing iron in more massive progenitors (up to the pair-instability regime), collapses 
to a neutrino-emitting PNS, which begins to assemble at the moment of core bounce, simultaneously 
with the formation of a strong shock wave ({\em second column}). Successful explosions can take place
through the neutrino-driven mechanism or, in rapidly spinning progenitors, by a magnetorotational
mechanism, which requires the efficient amplification of magnetic fields either in a proto-magnetar
or in an accretion disk around a newly formed BH; this scenario is accompanied by the launch of two 
polar jets ({\em third column}). Low-mass progenitors with deep, tenuous, and very extended 
hydrogen envelopes thus lead to events similar to electron-capture-SNe of ONeMg-core progenitors 
(denoted by ECSN-like), more massive iron-core progenitors
produce CCSNe, and rapidly rotating progenitors, which have often stripped their hydrogen envelopes, 
can end their lives by jet-driven or jet-associated SNe ({\em right column}). BHs are expected to form 
in rare cases of CCSNe as well as in jet-SNe besides those stellar core-collapse events that do not end 
in successful explosions. The asymmetric structures interior to the semi-transparent envelopes in the right
column indicate anisotropic distributions of metals mixed outward from the SN core, which happens
nearly spherically with small-scale variations for ECSN-like events and highly asymmetrically for CCSNe.
In jet-SNe these metal-rich structures are dwarfed by the global, prolate deformation 
associated with the much faster polar jets}
\label{fig:SNtypes}
\end{figure}

\section{Core-collapse supernovae}\label{sec:ccsn}

\subsection{Dynamical evolution and types of explosions}

This section will provide a brief summary of the evolution of massive stars and their final ends 
in SN explosions. Theoretical and numerical research, most recently culminating in direct 3D 
simulations, and a wide spectrum of partly game-changing observational discoveries 
have led to increasingly deeper insights over more than 50 years, in particular 
also with respect to the enormous diversity of stellar death scenarios and the plethora of 
phenomena that can play a role during the post-main-sequence evolution (e.g., interactions of 
the components in binary stars). The wealth of literature in this wide field of research cannot be 
adequately accounted for here. Therefore the references provided will be constrained
to a selection of reviews reflecting the development of the knowledge over the decades, earlier
papers defining novel directions of the field, and some recent publications that provide information
of the current status of the rapidly growing understanding. The selection will unavoidably be biased 
by personal views of the authors and it will be constrained by the goal to lay a very basic foundations 
for the subsequent discussion of SN nucleosynthesis and its dependences on the nuclear EoS of hot 
PNS matter.

Massive stars evolve over millions of years in hydrostatic equilibrium, assembling increasingly 
heavier chemical elements through a sequence of nuclear burning stages 
\citep[for reviews, see, e.g.,][]{Woosley+2002,Heger+2003,Woosley+2005}. Thus these stars develop
an onion-shell structure, in which the ashes of these burning phases accumulate in stratified layers
that surround a degenerate core, whose stability against gravitational collapse is sustained by 
the quantum mechanical fermion pressure of degenerate electrons. In stars with birth 
(zero-age-main-sequence; ZAMS) masses of more than roughly 7--9\,$M_\odot$ (depending on 
the initial metal content or ``metallicity'' of the stellar plasma) the degenerate core undergoes a 
catastrophic implosion at the end of the star's stable life to form a proto-NS (PNS) and to 
potentially initiate a SN explosion. The collapse sets in when the gas of electrons becomes 
highly relativistic and the mass of the degenerate stellar core approaches the Chandrasekhar limit.
In the lowest-mass progenitors of SNe ($M_\mathrm{ZAMS}\lesssim 9\,M_\odot$), whose cores are 
composed of oxygen, neon, and magnesium, the gravitational instability
of the stellar core is triggered by electron captures predominantly on $^{20}$Ne 
\citep[e.g.,][]{Kirsebom+2019,Zha+2019,Suzuki+2019}. In the iron cores of more massive
progenitors, which possess higher entropies and are hotter, photo-disintegration reactions of 
iron nuclei through thermal gamma rays initiate a fatal contraction before electron captures 
(first on nuclei, later also on an increasing number of free protons) cause a runaway process 
that ultimately leads to the dynamical collapse \citep[e.g.,][]{Langanke+2003,Janka+2007}. 

In both cases the infall stops only when super-nuclear densities are reached at the center and the 
EoS stiffens by the phase transition to homogeneous nuclear matter, generating dominant abundances of 
free neutrons and protons instead of heavy nuclei. The subsonically collapsing 
central part of the stellar core (the ``inner core'' with a mass of 0.4--0.5\,$M_\odot$) overshoots
its equilibrium density, the repulsive component of the strong force between the nucleons drives its 
re-expansion, and this so-called core bounce sends a strong shock wave into the supersonically
infalling outer layers of the stellar core. This shock wave propagates outward in mass and
radius, but on its way it loses enormous amounts of energy by photo-dissociation of heavy nuclei 
into free nucleons (consuming more than 8\,MeV per nucleon). More energy loss happens 
by the emission of a bright flash 
(``shock break-out burst'') of electron neutrinos, which are released through rapid electron captures 
by the abundant free protons behind the shock when the latter reaches neutrino-transparent
low-density conditions. In combination, both effects weaken the shock and lead to shock stagnation
within milliseconds after core bounce, implying that the bounce shock is unable to initiate a 
successful explosion because it transforms to a standing accretion shock. This means that not only 
the preshock velocities but also the postshock velocities of the stellar medium become negative. 
Therefore the newly forming NS continues to grow in mass but no ejection of matter develops at
this stage.

Different mechanisms have been suggested to revive the stalled SN shock. The most widely 
accepted ones are the delayed neutrino-heating mechanism and magneto-rotationally driven 
explosions (Figure~\ref{fig:SNtypes}). A mechanism based on quark deconfinement is a more recent
non-standard idea for achieving explosions, which, however, requires specific assumptions 
about the behavior of the super-nuclear EoS in hot, quite massive PNSs.

\subsubsection{Neutrino-driven explosions}
\label{sec:neutrinoexplosions}

The neutrino-driven mechanism taps the enormous gravitational binding energy ---several
$10^{53}$\,erg and thus several hundred times the explosion energy of a canonical CCSN--- 
that is set free when the degenerate stellar core collapses to a NS \citep{Colgate+1966}. 
Neutrinos and antineutrinos of all flavors, which are produced by particle reactions in the 
hot PNS interior, escape in huge numbers (about $10^{58}$ in total) from the PNS surface 
(``neutrinosphere''), carry away this binding energy, and govern the gradual conversion of the 
initially proton-rich matter of the PNS to the final, neutron-dominated medium of the cold, 
deleptonized NS. On their way of leaving the collapsing stellar 
core, these neutrinos can transfer a small fraction (on the order of typically one percent) of 
their total energy to the layer behind the stagnant shock wave. This is accomplished mainly 
through electron neutrino and antineutrino captures on free neutrons and protons, respectively, 
\begin{eqnarray}
\nu_e + n &\longrightarrow& e^- + p\,, \label{eq:nuab1}\\
\bar\nu_e + p &\longrightarrow& e^+ + n\,, \label{eq:nuab2}
\end{eqnarray}
which dominate over the inverse processes
\begin{eqnarray}
e^- + p &\longrightarrow& \nu_e + n\,, \label{eq:nuem1}\\
e^+ + n &\longrightarrow& \bar\nu_e + p\,, \label{eq:nuem2}
\end{eqnarray}
in the so-called gain layer, which is the region between the gain radius and the SN shock, 
where neutrino heating exceeds neutrino cooling \citep{Bethe+1985,Bethe1990}.
In order to power the SN explosion, neutrinos thus tap the thermal and degeneracy energy 
that is built up in the PNS by the conversion of gravitational energy during stellar core collapse.
The efficiency of the energy transfer is typically around 5--10\%, measuring the 
fraction of the luminosity radiated in $\nu_e$ and $\bar\nu_e$ that is absorbed by the matter 
in the gain layer. The neutrino energy deposition continues for several seconds before the 
explosion energy approaches its saturation level for the CCSNe of massive progenitors 
\citep[e.g.,][]{Bollig+2021}.

These neutrino-powered explosions develop pronounced large-scale asymmetries in their 
innermost ejecta, connected to non-radial hydrodynamic instabilities that occur in the 
postshock layer due to convective overturn of neutrino-heated gas 
\citep{Herant+1994,Burrows+1995,Janka+1996} and due to a generic, 
global, oscillatorily growing accretion instability of the shock, the standing accretion 
shock instability \citep[SASI;][]{Blondin+2003,Blondin+2007,Foglizzo+2015}. 
Both assist the neutrino-heating mechanism by a corresponding buoyancy-driven
expansion of the postshock matter or by shock expansion due to SASI sloshing and spiral
modes, which have been interpreted in terms of effects associated with turbulent mass 
motions on the largest scales. The violent flows drive the shock outward and therefore
increase the volume and mass in the gain layer, thus enabling more neutrino-energy 
deposition and the runaway expansion of the shock that is necessary for a successful 
SN explosion. The large-scale asymmetries imprinted on the postshock matter,
which are often dominated by dipolar and quadrupolar modes 
\citep[see, e.g.,][]{Burrows+2020,Mueller2020}, are preserved in terms of later 
asymmetries of the heavy-element distributions (in particular of the intermediate-mass 
elements between neon and the iron group) on the largest scales, i.e., in opposite 
hemispheres and different octants \citep{Wongwathanarat+2013,Katsuda+2018}. 
CCSNe are thus expected to be globally asymmetric in their
interior, in particular w.r.t. the metal core (Figure~\ref{fig:SNtypes}). They also
exhibit extensive radial mixing of the metals into the helium shell and hydrogen envelope
of the exploding stars because the initial ejecta asymmetries trigger entrainment and
fragmentation at the 
composition-shell interfaces of the progenitors due to the growth of secondary 
(Rayleigh–Taylor, Kelvin-Helmholtz, and Richtmyer-Meshhov) instabilities near the 
shell boundaries after the passage of the outgoing SN shock \citep[for more details on 
different aspects of the CCSN problem, see recent reviews and comprehensive studies
on this topic and references 
therein, e.g.,][]{Janka2012,Burrows2013,Janka+2016,Janka2017,Mueller2020,Stockinger+2020,
Sandoval+2021,Burrows+2021}. 

Sometimes a distinction is made between the SN explosions of the lowest-mass 
progenitors with either ONeMg cores or iron cores (and representatives in the ZAMS
mass range below $\sim$10\,$M_\odot$) and the CCSNe of more massive stars
\citep[e.g.,][]{Mueller2016}. The former cases are termed electron-capture SNe (ECSNe) 
or ECSN-like events, respectively, because they possess characteristic differences in 
their progenitor and explosion properties compared to CCSNe, although in ECSN-like 
events as well as CCSNe 
the explosions are thought to be powered by neutrino energy transfer. The 
degenerate cores of the mentioned low-mass stars have lower entropies and temperatures
than the iron cores of more massive progenitors, for which reason their electron chemical 
potentials are particularly high. This explains why electron captures on $^{20}$Ne yield 
the first initiation of the gravitational instability in the case of ONeMg cores, whereas
photo-disintegration and the corresponding NSE shift are more relevant for the more
massive iron cores in more massive progenitors. In addition to this difference in the
physics that triggers the core collapse, the particularly steep density gradient at the 
edge of the degenerate core, i.e., at the interface between ONeMg or Fe core and the
overlying shells, enables different explosion dynamics, despite the explosion mechanism
being based on neutrino-heating in both ECSN-like events and CCSNe. Since the SN shock
accelerates outward as it travels down the steep density gradient in ECSN-like cases, 
the explosion develops shortly after bounce, i.e., the accelerated outward expansion 
of the SN shock starts within only some 10\,ms up to about 100\,ms instead of being 
delayed for several 100\,ms in more massive stars. Correspondingly, convective overturn
sets in concomitantly to the explosion but it is not crucial for the success of the
SN blast. The short phase of postshock convection leads to only modest, small-scale 
anisotropies in the ejecta without the presence of dominant dipolar or quadrupolar 
asymmetry modes. After only a short phase of convective overturn, a neutrino-driven 
wind sets in, in which neutrino-heated gas is blown off from the PNS surface by neutrino 
heating essentially in a spherically symmetrical way and provides additional push to 
the explosion. Because the wind contains only little mass (typically several 
$10^{-3}\,M_\odot$) and its power is low, the final explosion energy is usually also low
---around $10^{50}$\,erg instead of the canonical value of about $10^{51}$\,erg of CCSNe---
and reaches its terminal value within less than one second,
the ejecta velocities are small, the SN is globally spherical, and also the 
radial mixing of metals in ECSN-like events is weak (Figure~\ref{fig:SNtypes}). 
Consistent with that, the NS kicks associated with the weakly asymmetric, low-energy
explosions are only of the order of 10\,km\,s$^{-1}$ and thus much below the typical 
natal kicks of NSs of several 
100\,km\,s$^{-1}$ \citep[e.g.,][]{Stockinger+2020,Kozyreva+2021,Kozyreva+2022}.

\subsubsection{Magnetorotational explosions}\label{sec:magex}

In contrast to the neutrino-heating mechanism, the magneto-rotational mechanism makes use 
of the energy of rotation in a rapidly and differentially spinning PNS that is formed during 
the collapse of a fast-rotating progenitor with an average iron-core rotation period shorter
than roughly 10\,s, corresponding to iron-core angular momenta close to $10^{49}$\,erg\,s 
or more. Under such circumstances magnetic fields are amplified not only by compression and 
in turbulent flow vortices but also by winding in differentially rotating layers, convective 
dynamo action, and the magnetorotational instability (MRI) to reach strengths of up to
$10^{15}$--$10^{16}$\,G also in the large-scale field of the PNS. Moreover and equally 
relevant, under such circumstances the kinetic energy of rotation amounts to several
$10^{51}$\,erg at a time when the PNS still has a radius around 30\,km, which is typical 
of the phase during which the explosion is expected to develop. The free energy of rotation 
that is released when the PNS evolves from a differentially rotating to a rigidly rotating 
object can be a fair fraction of the total rotation energy, i.e., of the order of several 
$10^{50}$\,erg. Therefore it can make a significant contribution 
to the canonical explosion energy of CCSNe. If the total reservoir of rotational energy 
in the final, maximally spinning NS with $\sim$\,(10--12)\,km radius could be tapped via 
magnetic fields, explosion energies up to several $10^{52}$\,erg could be produced.
This is sufficient to explain the most energetic observed stellar explosions.

The transfer of this energy through magnetic forces can drive fast mass ejection
in jet-like magnetic towers along the poles, initiating an asymmetric, highly prolate 
explosion (Figure~\ref{fig:SNtypes}). Magnetically induced mass ejection is also possible 
in the equator-near directions, but the success or failure of the explosion and the
exact dynamics and properties of the blast and its associated outflows in jets are 
extremely sensitive to the progenitor conditions, in particular to the core rotation and 
the strength and geometry of the magnetic fields prior to collapse. These initial conditions
determine the field amplification in the new-born NS and the field-triggered mass ejection.
For recent progress in the numerical modelling of such events, see, e.g., 
\citet{Bugli+2021} and \citet{Obergaulinger+2022} and 
references therein. Such magnetorotational explosions are 
expected to leave behind highly magnetized PNSs, so-called proto-magnetars, with 
millisecond periods \citep{Aloy+2021}. 

Also BHs can be formed in the collapse of
rapidly spinning, magnetized progenitors, either after a successful explosion or before.
These BHs then continue to accrete matter from the collapsing star through a centrifugally 
stabilized, thick accretion torus. In this so-called collapsar scenario, energy release 
from the torus by neutrinos, magnetohydrodynamic flows, and magnetically or turbulence-mediated 
viscous dissipation of rotational energy may also lead to jet production and mass ejection 
in a stellar explosion \citep{Woosley1993,MacFadyen+1999,MacFadyen+2001,Zhang+2003,Gottlieb+2022a,Gottlieb+2022b}. 
Alternatively or in addition, energy may be extracted from the spin-energy of the 
Kerr BH through the Blandford-Znajek process \citep{Blandford+1977}. 

These highly asymmetric explosions are associated with the terminal collapse of rapidly
spinning progenitors and receive their energy either from a proto-magnetar or accreting 
Kerr BH. They are considered as explanations of rare kinds of stellar death events with 
rates well below about 1\% of the total CCSN rate, namely long-duration gamma-ray bursts 
(GRBs), hypernovae (HNe), and the roughly 10 times less frequent superluminous SNe (SLSNe). 
Such events are characterized
by different distinct observational features. For GRBs there is the detection of a flash
of gamma rays with energy spectra peaking at hundreds of keV to about 
one MeV, which points to the presence of ultarelativistic jets. For HNe 
explosion energies are diagnosed that are up to 50 times higher than those of 
canonical CCSNe and seem to be out of reach for neutrino-powered explosions. And for SLSNe
observations reveal normal explosion energies but extremely high optical luminosities, 
which exceed those of all other SN types by at least a factor of 10 during the peak 
phase of the emission. Long GRBs have been detected with
and without hypernovae and hypernovae also without GRBs. Also cases of CCSN explosions with 
choked jets that cannot make their way to the stellar surface may be possible. These could 
occur in intermediate explosion cases where the SN blast is mainly driven by neutrino 
heating but magnetorotational effects do not achieve to produce powerful jets, perhaps 
because the stellar core does not rotate sufficiently fast and/or the magnetic field 
configuration is not supportive. 

Magnetohydrodynamically driven mass ejection, in particular extremely fast-expanding
jets in magnetorotational CCSNe 
\citep[e.g.][and references therein]{Winteler+2012,Nishimura+2015,Cowan+2021} 
and outflows from collapsar disks \citep[][]{Siegel+2019,Siegel+2022}, are
discussed as potentially neutron-rich environments that could provide favorable conditions 
for the production of r-process elements. However, it is still unclear whether and to 
which extent such phenomena can play a role as rare sources of neutron-capture elements 
primarily in the low-metallicity universe. Since the production of considerable 
amounts of iron in such events is hardly avoidable, their ability to inseminate metal-poor
stars is at least questionable. Moreover, the neutron excess and hydrodynamic stability of 
jets in 3D depend extremely sensitively on the conditions in the progenitor star and may
require fine-tuned assumptions \citep{Moesta+2018,Halevi+2018,Reichert+2022}. Also, collapsar 
tori had originally been discussed as origins of large amounts of radioactive $^{56}$Ni as 
required by hypernova observations \citep{MacFadyen+1999}, a conjecture that receives 
support by more recent models including a detailed treatment of neutrino transport
\citep{Just+2022}.

Currently it is unclear which fraction of CCSNe are caused by the magnetorotational
mechanism and which progenitors are involved. There may be a continuum of behaviors,
depending on the progenitor's core rotation and magnetic field, namely
mainly (or purely) neutrino-driven explosions, explosions caused by a combination of
neutrino and magnetorotational effects, and extreme SNe exploding only because the
magnetorotational mechanism provides sufficient energy. Rapid rotation in the stellar
iron core with periods of less than $\sim$10\,s is not expected to be very common,
because mass loss in stellar winds leads to the loss of large amounts of angular 
momentum during advanced stages of the stars' evolution, slowing down the core rotation
when magnetic fields are taken into account to establish an efficient coupling of core
and envelope \citep[see, e.g.,][]{Heger+2005}. This is considered as an argument why
white dwarfs are observed to rotate slowly and why pulsars are estimated to have 
relatively long rotation periods (typically tens to hundreds of milliseconds) at the 
time of their birth \citep[for a synopsis of such arguments with relevant references, 
see][]{Janka+2022}. Moreover, asteroseismological measurements suggest more spin-down
and slower interior rotation of red supergiants than expected from current stellar
evolution models including magnetic fields. Together with the finding that CCSNe with 
energies too high to be explained by the neutrino-driven mechanism (i.e., more than 
roughly $2\times 10^{51}$\,erg) are extremely rare, all these facts lead to the 
conclusion that rapid core rotation does not seem to be widespread among pre-collapse 
stars. However, binary interaction is thought to affect the progenitors of a major 
fraction of CCSNe \citep{Sana+2012} and might have an influence on the angular momentum 
of massive stars prior to core collapse in many ways, potentially spinning their 
degenerate cores up or down.

\begin{figure}
\includegraphics[width=1.0\columnwidth]{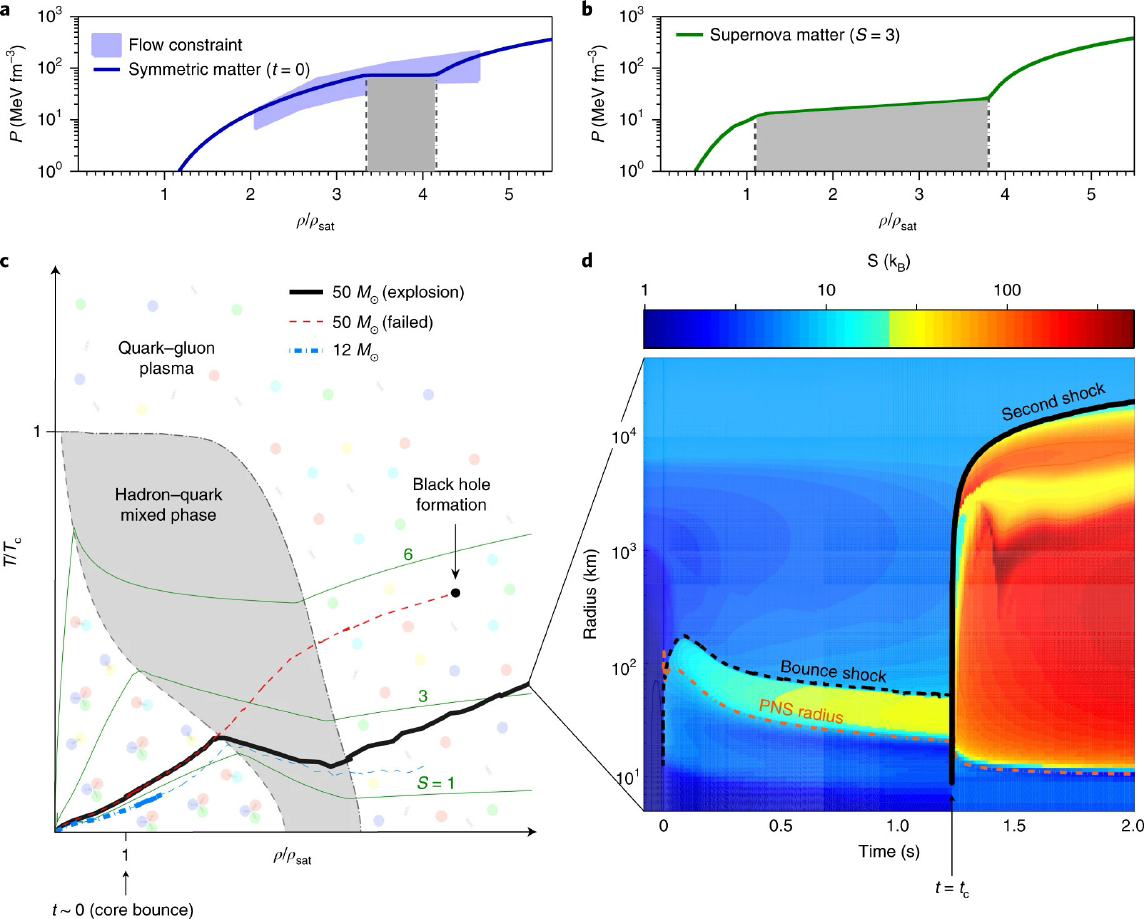}
\caption{Hadron-quark phase transition in a PNS and induced SN explosion. The two upper panels
display the pressure of the hadron-quark EoS versus density (the latter in units of the 
nuclear saturation density) for isospin-symmetric matter at zero temperature ({\em panel~a}) 
and for SN matter of fixed electron fraction ($Y_e = 0.3$) and an entropy per particle of 
3\,$k_\mathrm{B}$ ({\em panel~b}). Gray shading marks the density interval of the 
hadron-quark mixed phase. {\em Panel~c} provides the temperature-density phase diagram of the 
hybrid EoS with curves of constant entropy indicated by thin solid lines. Also the trajectories
of the time-evolving central conditions of three SN models are displayed, with a 50\,$M_\odot$
model developing an explosion due to the energy release during the collapse 
from the hadronic PNS to a hybrid star with a compact core of quark-gluon plasma.
{\em Panel~d} shows the corresponding space-time diagram, which visualizes the evolution
of the PNS radius and core-bounce shock (dashed lines) and the second shock caused by the 
later (at $t=t_\mathrm{c}$) collapse and bounce due to the hadron-quark phase transition 
(solid black line) (Figure from \citealt{Fischer+2018}; for more details consult this reference;
\copyright Nature Springer. Reproduced with permission)}
\label{fig:QCDexplosion}
\end{figure}

\subsubsection{Stellar explosions triggered by quark deconfinement}
\label{sec:QCDSNe}

Nuclear phase transitions in dense NS matter are an interesting phenomenon that might
occur in Nature and that are relevant for our understanding of the properties of  
compact stellar remnants and the signals (neutrinos, gravitational waves, nucleosynthesis,
explosions) associated with their formation in stellar core-collapse events and their
mergers in compact binaries. 

For example, phase transitions in PNSs are considered as one possible path to form a 
stellar-mass BH after a transiently existing NS has released the energy that enables the 
success of a SN explosion. Alternatively, a NS can also be pushed beyond the mass limit 
for BH formation by the fallback of larger amounts of matter that is initially
accelerated outward but does not achieve to become gravitationally unbound during the SN. 
The destiny of gravitational instability might also happen to a 
rapidly rotating NS with a mass that is too large to be stabilized without centrifugal 
forces. If such a supramassive NS ---or hypermassive NS if differential rotation increases 
its mass even beyond the limit for rigid rotation--- loses angular momentum by mass 
shedding or magnetic field effects, it will become unstable and collapse to a BH. 
These different scenarios provide routes to BH formation in association with SN explosions
and in addition to the ``direct'' BH formation that occurs when the explosion fails, in 
which case the temporarily existing PNS continues to accrete mass from the infalling star
until it implodes to become a BH (see Figure~\ref{fig:NSBHevol}).

Interestingly, it has been shown that a first-order phase transition from nuclear matter
to a quark-gluon plasma can also trigger a SN explosion (even in spherically symmetric
models that do not account for the supportive effects of hydrodynamic instabilities),
if other mechanisms do not succeed \citep{Sagert+2009,Fischer+2018}. The energy needed
for the explosion is released at the expense of gravitational binding energy when the 
phase transition sets in and instigates a dynamical collapse of the PNS to a more compact 
hybrid star with a core of pure quark-gluon plasma. 
When the quark core settles into a new hydrostatic equilibrium, a bounce shock is formed. 
This second shock, if propagating out of the compact star, can revive the first shock 
that is linked to the instant of core bounce when nuclear
saturation density was exceeded in the infalling stellar iron core and a PNS began to 
assemble. Thus a SN explosion can ultimately be launched by the help of the second
shock, provided this shock is powerful enough (Figure~\ref{fig:QCDexplosion}).

However, the exact density regime of the hadron-quark mixed phase, its dependence on
temperature and electron fraction in the PNS medium, and the physical properties of
the phase transition and of the quark phase are not known.  
Considerable fine tuning seems to be required in order to create a second shock wave, to 
obtain sufficient energy release for a viable explosion, and to satisfy the observational
lower bounds for the maximum NS mass, too \citep{Fischer+2018,Zha+2021,Jakobus+2022}.
The first models assumed the hadron-quark phase transition to happen around nuclear
saturation density and yielded moderate explosion energies as well as BH formation for 
compact object masses much below those of the heaviest well-measured NSs. These early
models were also in conflict
with our understanding of the explosion of SN~1987A and the likely existence of a NS 
left behind by this SN. \citet{Fischer+2018} moved the phase transition to higher 
densities and their hadron-quark EoS is able to account for the NS mass limit. They 
could obtain fairly high explosion energies when the phase transition
occurred in massive PNSs that did not collapse to BHs before 
(see Figure~\ref{fig:QCDexplosion}). In contrast, large model 
sets with different progenitors and different EoSs with hadron-quark phase transitions 
showed none \citep{Zha+2021} or very few (two out of 97) and very weak 
explosions \citep{Jakobus+2022}. 

In cases where the second shock is formed, the strong bounce connected to the quark-core 
formation produces a high-amplitude GW signal that lasts for several milliseconds and 
dominates other contributions to the GW emission of the collapsing star by far 
\citep{Zha+2020,Kuroda+2022}. Moreover, the phase-transition induced collapse and bounce 
lead to a second, high-luminosity, short-time (of order one millisecond) neutrino burst, 
followed by pulsational ring-down variations of the 
luminosities of neutrinos and antineutrinos of all flavors. The emission is dominated by
electron antineutrinos produced through positron captures in the previously neutronized 
PNS matter. Both signals are characteristic of the strong hadron-quark phase transition
and are predicted to be measurable for a future Galactic CCSN to provide 
observational evidence of such an interesting phenomenon in the deep interior of the
newly formed compact object. If an explosion is caused by the quark deconfinement,
also the production of r-process elements seems possible in rare SN events 
\citep{Fischer+2020}.

\subsubsection{Closing remarks}

After a successful SN explosion a NS or, more generally, a compact star, is left 
behind and cools by intense neutrino emission until neutrino-transparent, cool 
($T \lesssim 1$\,MeV) conditions are reached 
\citep[e.g.,][]{Burrows+1986,Roberts+2017,Li+2021}. During this Kelvin-Helmholtz 
cooling the gravitational binding energy $E_\mathrm{gb}$ released in the formation 
of the compact star is carried away via neutrinos. It is given as a fraction of the 
mass of the remnant by
\begin{equation}
E_{\nu}\, \approx\, E_{\mathrm{gb}}\, =\, {0.6\beta \over 1 - 0.5\beta}
\,Mc^2 
\label{eq_egrav}
\end{equation}
\citep[fit for non-rotating stars according to][]{Lattimer+2001},
where $M = M_0 - E_{\mathrm{gb}}/c^2$ is the gravitational
mass of the object with rest mass $M_0$, and
$\beta = GM/(Rc^2) = R_\mathrm{s}/(2R)$ is the compactness parameter
for stellar radius $R$ and Schwarz\-schild radius $R_\mathrm{s}$.
Typically, $R\sim (2.5....3)R_\mathrm{s}$ and therefore about 10\%
of the star's rest mass or several $10^{53}\,$ergs are lost in neutrinos.

As mentioned above and graphically represented in Figure~\ref{fig:NSBHevol}, during 
the Kelvin-Helmholtz cooling or during the subsequent evolution, the compact star may 
collapse to a BH, either due to a 
phase transition, loss of thermal or angular-momentum support, or fallback accretion,
all of which can destabilize the remnant, if the maximum mass determined by the EoS 
is exceeded. Fallback can not only cause the gravitational implosion of the compact 
star but it can also bring back the products of SN nucleosynthesis 
contained in the innermost initial ejecta.
Depending on the mass, rotation rate, and magnetic fields of the core-collapse 
progenitor as well as on the properties of the high-density EoS, a wide spectrum of
evolutionary scenarios with or without SN explosions is therefore possible, summarized 
visually by Figures~\ref{fig:NSBHevol}, \ref{fig:SNtypes}, and \ref{fig:QCDexplosion}.

The discussion of SN scenarios and their EoS-dependent nucleosynthesis in this chapter
will ignore pulsational pair-instability SNe (PPISNe) and pair-instability SNe (PISNe)
\citep[see, e.g.,][]{Heger+2003} of the most massive stars with ZAMS masses of roughly 
more than 70\,$M_\odot$ and 140\,$M_\odot$, respectively
(the exact mass boundaries are sensitive to metallicity, stellar mass-loss evolution, 
reaction rates for 3$\alpha$ and $^{12}$C($\alpha,\gamma$)$^{16}$O, and on the stellar 
rotation). The collapse of such very massive stars 
is triggered by the onset of electron-positron pair
production when the central temperature approaches $\sim$\,$10^9$\,K, and their powerful
SN eruptions are caused by the energy release from explosive burning of oxygen and
silicon. Therefore the high-density super-nuclear EoS does not play a determining role
for the nucleosynthesis yields in their ejecta \citep[for the more complex events
following the final iron-core collapse in PPISNe, see][]{Powell+2021,Rahman+2022}.

\begin{figure}
\includegraphics[width=1.0\columnwidth]{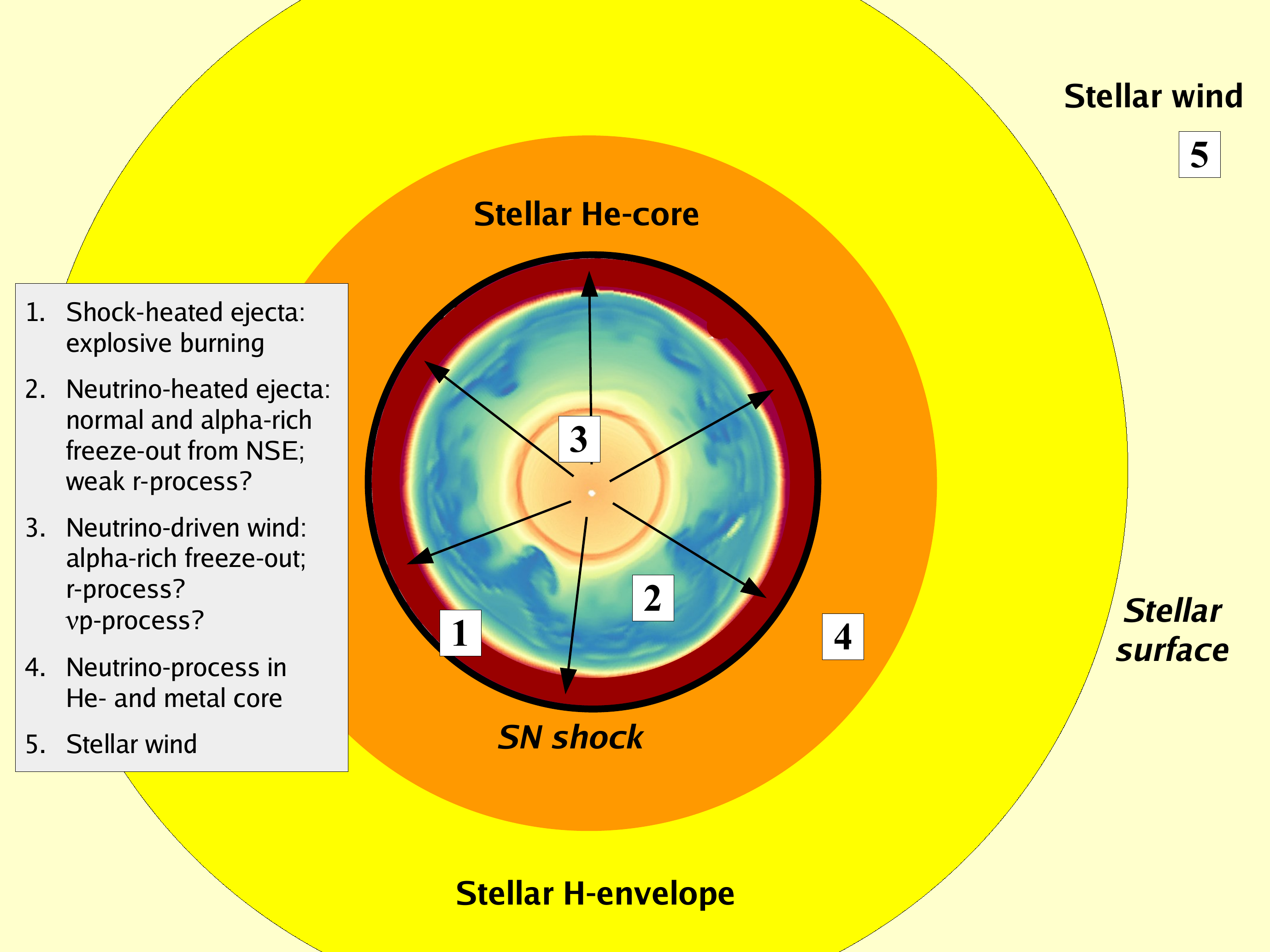}\\
\includegraphics[width=1.0\columnwidth]{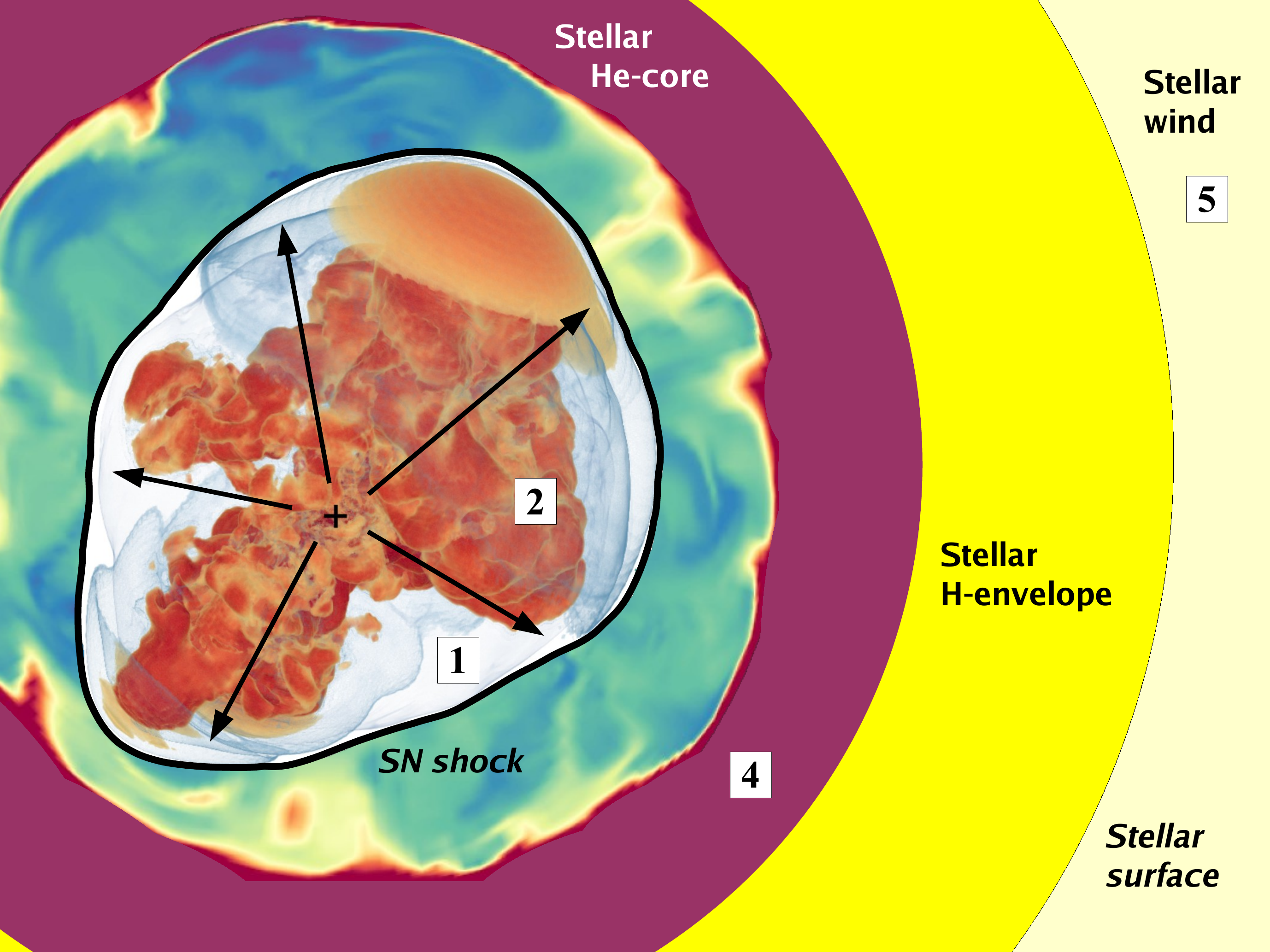}
\caption{Regions of different nucleosynthesis processes in ECSN-like explosions 
of low-mass progenitors ($M_\mathrm{ZAMS}\lesssim 10\,M_\odot$; {\em top}) and iron-core 
CCSNe from more massive progenitors ({\em bottom}). 
Nuclear burning in shock-heated matter (region~1) creates 
intermediate-mass elements and contributes to the production of iron-group species;
normal and alpha-rich freeze-out from NSE in neutrino-heated postshock matter (region~2) 
assembles iron-group and trans-iron nuclei; alpha-rich freeze-out in the neutrino-driven
mass outflow (``neutrino wind''; region~3) from the new-born NS is discussed as potential 
site of an r-process and/or neutrino-proton process; neutrino-nucleus reactions in the 
helium layer and carbon-oxygen core (region~4) can induce the production of various isotopes 
by the so-called neutrino-process; and the matter lost by the star in a stellar wind (region~5) carries
nucleosynthesis products of the sequence of hydrostatic nuclear burning stages prior to core collapse. 
Note that a neutrino-driven wind (region~3) does not seem to be generic to explosions of massive 
progenitors. The stellar shells are not drawn to scale}
\label{fig:SNnucleosynthesis}
\end{figure}

\subsection{Ejecta components and nucleosynthesis}

SN explosions of massive stars contribute to the nucleosynthesis of chemical elements
in the Universe through different parts of their ejecta and by various processes. The 
regions of nucleosynthesis are visualized in Figure~\ref{fig:SNnucleosynthesis} for 
ECSN-like explosions of low-mass progenitors and for CCSNe of more massive iron-core 
progenitors (see Section~\ref{sec:neutrinoexplosions}). Mass loss in the stellar wind 
of the progenitor star ({\bf region~5}) distributes material that the star contained from 
the time of its formation but also nuclei produced by shell burning and the slow neutron
capture process (s-process) and mixed into the outer stellar layers by hydrodynamic
processes like convection, dredge-up, and rotation-induced flows \citep[for reviews of 
the evolution and nucleosynthesis of massive stars and their explosions, see, 
e.g.,][and a wealth of literature referenced therein]{Woosley+2002}.

Nucleosynthesis happening during the SN explosion itself and depending on the conditions
in the ejecta as well as in the newly formed NS takes places in regions~1 to 4. 

\subsubsection{Region~1: shock-heated ejecta}

Region~1 contains the products of explosive nuclear burning of 
silicon, oxygen, neon, and carbon in ejecta that are heated by the outgoing shock.
The material in this region retains its pre-collapse electron fraction $Y_e$, because
the collapse and explosion proceed too quickly for weak reactions (electron captures)
to have an impact.
Assuming that the heat capacity of the plasma behind the shock is dominated by the
radiation field and that nearly isothermal conditions hold in the postshock volume,
the peak temperature achieved as the shock passes a radius $r$ can be estimated to 
good accuracy (except for the innermost region close to the PNS) by
\begin{equation}
    T_\mathrm{s}(r) \approx 1.33\times 10^{10} 
    \left(\frac{E_\mathrm{k,\infty}}{10^{51}\,\mathrm{erg}}\right)^{\! 1/4}
    \left(\frac{r}{10^8\,\mathrm{cm}}\right)^{\! -3/4}\ \mathrm{K}\,,
    \label{eq:Tshock}
\end{equation}
where $E_\mathrm{k,\infty}$ is the final kinetic energy of the explosion at infinity
\citep{Woosley+2002}. If the temperature reaches high enough for the nuclear burning
time scale to be shorter than the hydrodynamical expansion time scale, explosive 
nucleosynthesis will modify the pre-explosion composition. Matter heated to 
$5\times 10^9$\,K and higher (typically for $r \lesssim 3000$--4000\,km) will be 
processed into NSE and become iron-group nuclei during expansion and cooling; silicon
burns on a hydrodynamic time scale between $4\times 10^9$\,K and $5\times 10^9$\,K
($r \lesssim 5000$\,km),
oxygen between $3\times 10^9$\,K and $4\times 10^9$\,K, neon between $2.5\times 10^9$\,K
and $3\times 10^9$\,K, and carbon between $1.8\times 10^9$\,K and $2.5\times 10^9$\,K.
At $r \gtrsim 10000$--15000\,km explosive nuclear burning ceases for all elements
heavier than helium and explosive processing becomes irrelevant. Therefore these
layers are ejected without any appreciable changes of their initial composition;
this concerns most elements lighter than oxygen.
Also isotopes created by the p-process are expected to be produced in the shock-heated
region~1 due to the partial meltdown of s-process nuclei from the star's helium and 
carbon burning stages and from the star's original material at birth.

\begin{figure}
\includegraphics[width=1.0\columnwidth]{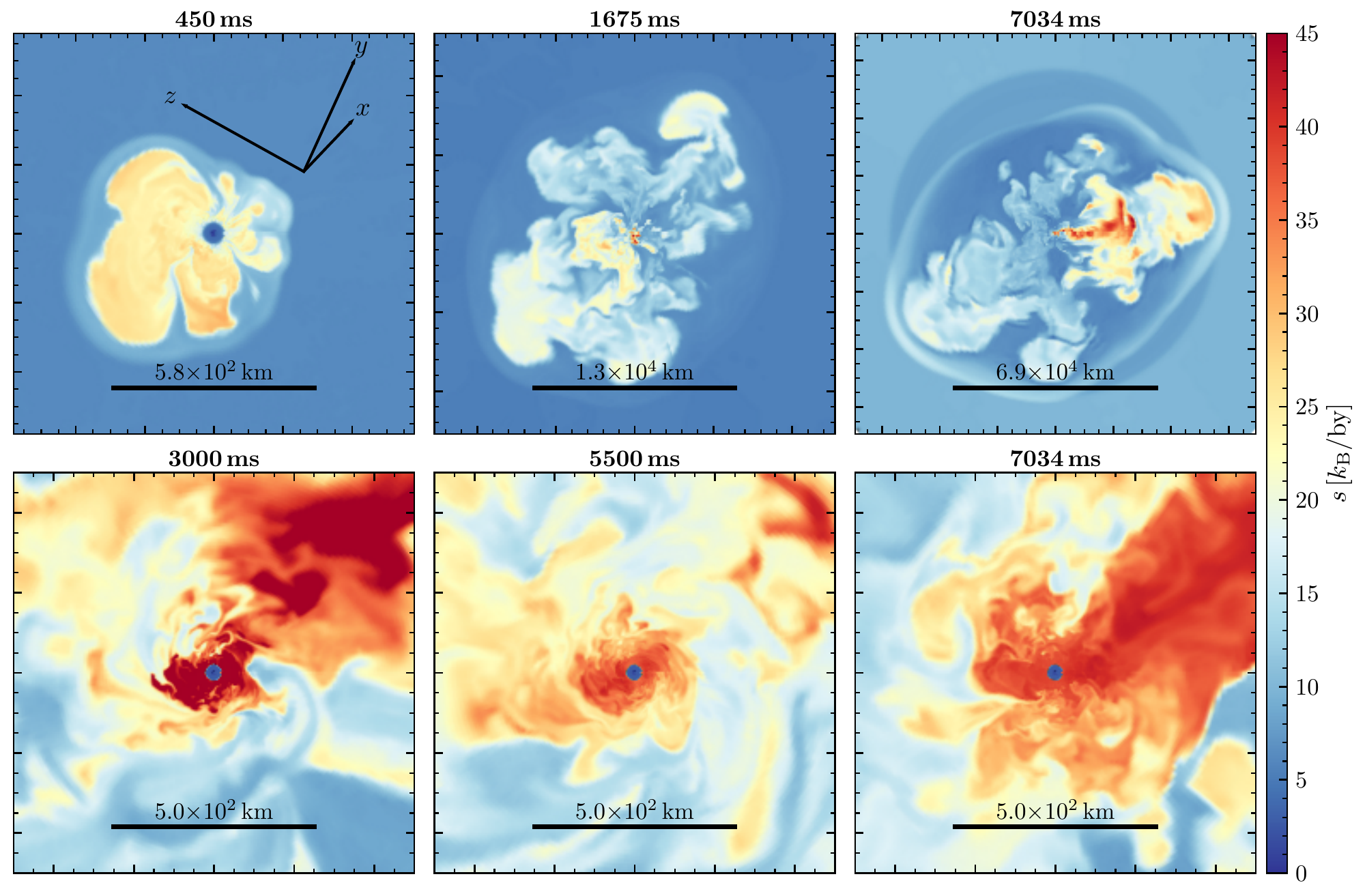}\\
\includegraphics[width=1.0\columnwidth]{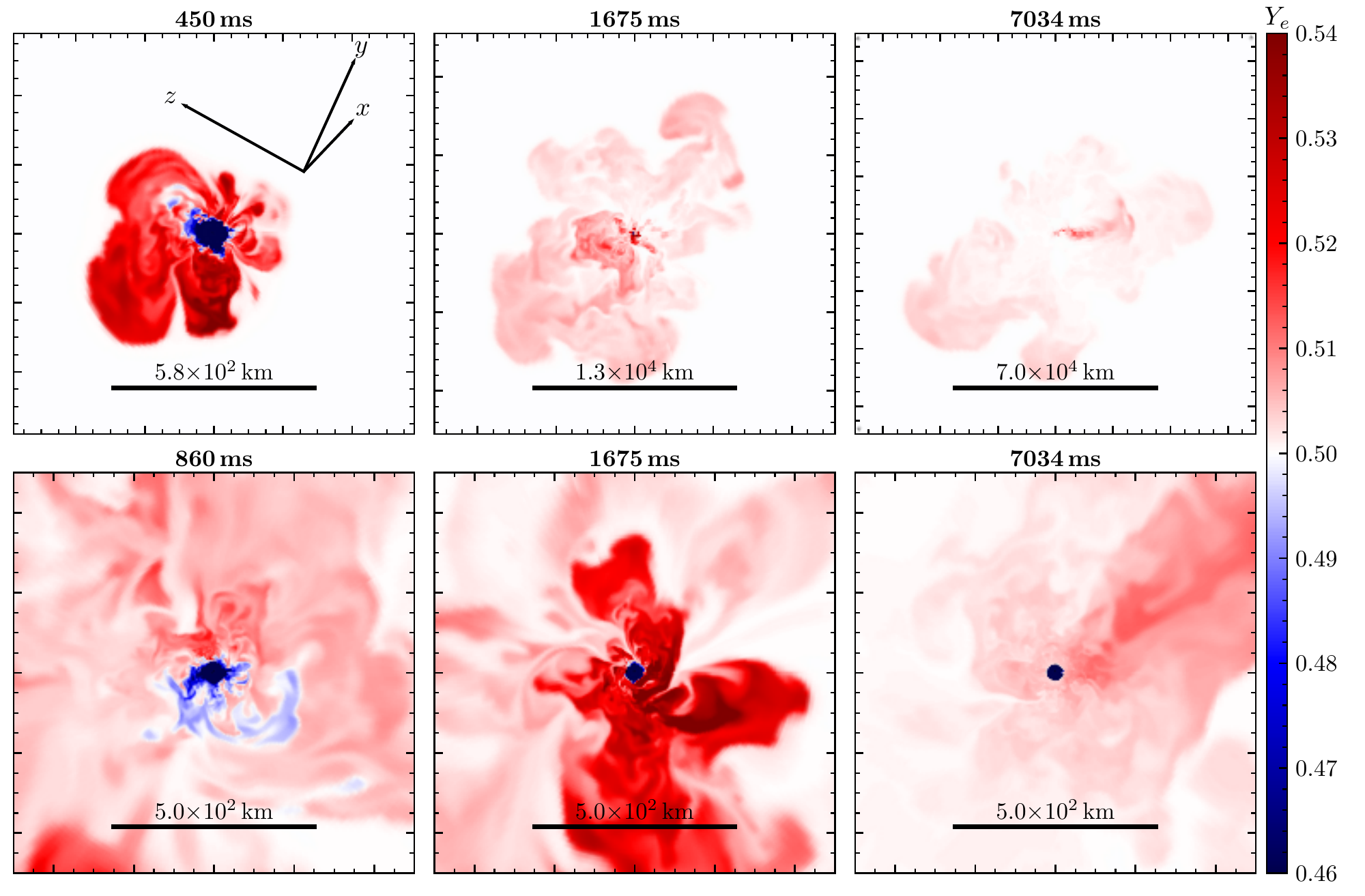}
\caption{Entropy ({\em top}) and electron fraction $Y_e$ ({\em bottom}) distributions 
in cross-sectional cuts at six post-bounce times of a 3D CCSN simulation of a 
19\,$M_\odot$ progenitor, whose explosion sets in at about 0.4\,s after bounce.
Note the different length scales indicated by yard sticks and the different times
specified above the panels. The lower panels show
close-ups of the turbulent vicinity of the PNS. They display lower-entropy downflows 
and high-entropy outflows of neutrino-heated matter until more than 7\,s after bounce.
A spherical neutrino-driven wind of neutrino-heated matter blown off the PNS surface
does not develop until that time (Figure from \citealt{Bollig+2021}; 
\copyright American Astronomical Society. Reproduced with permission)}
\label{fig:s19ejecta}
\end{figure}

\begin{figure}
\begin{center}
\includegraphics[width=0.8\columnwidth]{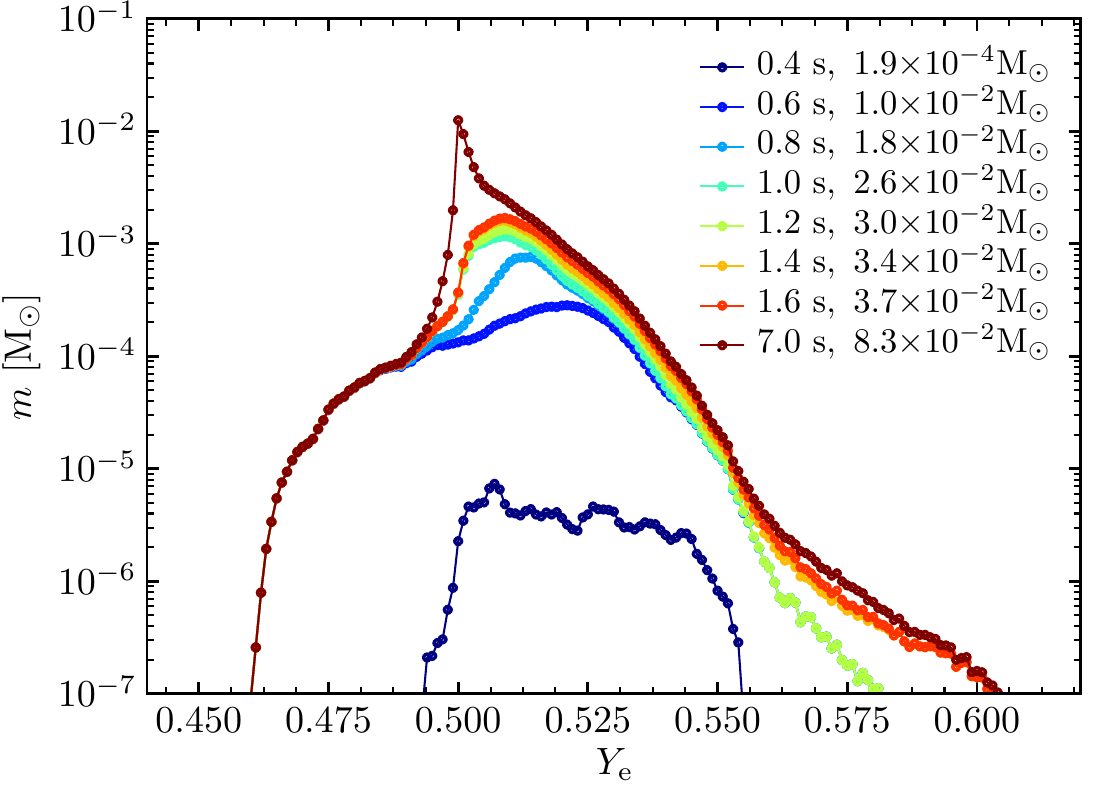}
\end{center}
\caption{Cumulative mass distributions of the ejecta as functions of $Y_e$ until 
different post-bounce times (the integrated ejecta mass is given by the labels) for 
a 3D CCSN simulation of a 19\,$M_\odot$ progenitor (Figure from \citealt{Bollig+2021};
\copyright American Astronomical Society. Reproduced with permission)}
\label{fig:s19-ejectahistogram}
\end{figure}

\subsubsection{Region~2: neutrino-heated ejecta}
\label{sec:nuejecta}

Region~2 contains the neutrino-heated ejecta that have absorbed the 
energy for the explosion via capture reactions of $\nu_e$ and $\bar\nu_e$ 
on neutrons and protons, respectively (Eqs.~\ref{eq:nuab1} and \ref{eq:nuab2}). 
Different from the layers of region~1, the gas in region~2 falls inward to the close 
vicinity of the PNS and resides near the gain radius for a while. Thus it can receive
the energy needed to expand outward again through the irradiation by the intense 
neutrino fluxes from the PNS. At the same time $\nu_e$ and $\bar\nu_e$ absorption, 
together with their inverse processes (Eqs.~\ref{eq:nuem1} and \ref{eq:nuem2}), 
also set the neutron-to-proton ratio that defines the nucleosynthesis conditions
when the gas cools during expansion. Neutrino heating raises the entropy of the
plasma to high values, for which reason buoyancy triggers convective overturn but also
prevents the gas entropy in multi-dimensional simulations to rise as high as it does in
spherically symmetric (1D) explosion models (see Figure~\ref{fig:s19ejecta}). All 
matter in this region reaches NSE with
the initial nuclei being broken down into free nucleons and $\alpha$-particles. The
nucleosynthesis during the explosive ejection proceeds via alpha-rich freeze-out
or normal freeze-out, in which free neutrons and protons recombine to $\alpha$-particles
and heavier nuclei. Because state-of-the-art 2D and 3D CCSN simulations with detailed
neutrino transport show that a dominant fraction of these ejecta obtains a proton excess 
(see Figures~\ref{fig:s19ejecta} and \ref{fig:s19-ejectahistogram}), proton-rich isotopes 
of light trans-iron elements, in particular $^{92}$Mo, can be produced besides iron-group 
nuclei including an appreciable amount of $^{56}$Ni, which adds to the $^{56}$Ni assembled
in region~1 \citep[see][for results from 2D models]{Wanajo+2018}.

\begin{figure}
\includegraphics[width=1.0\columnwidth]{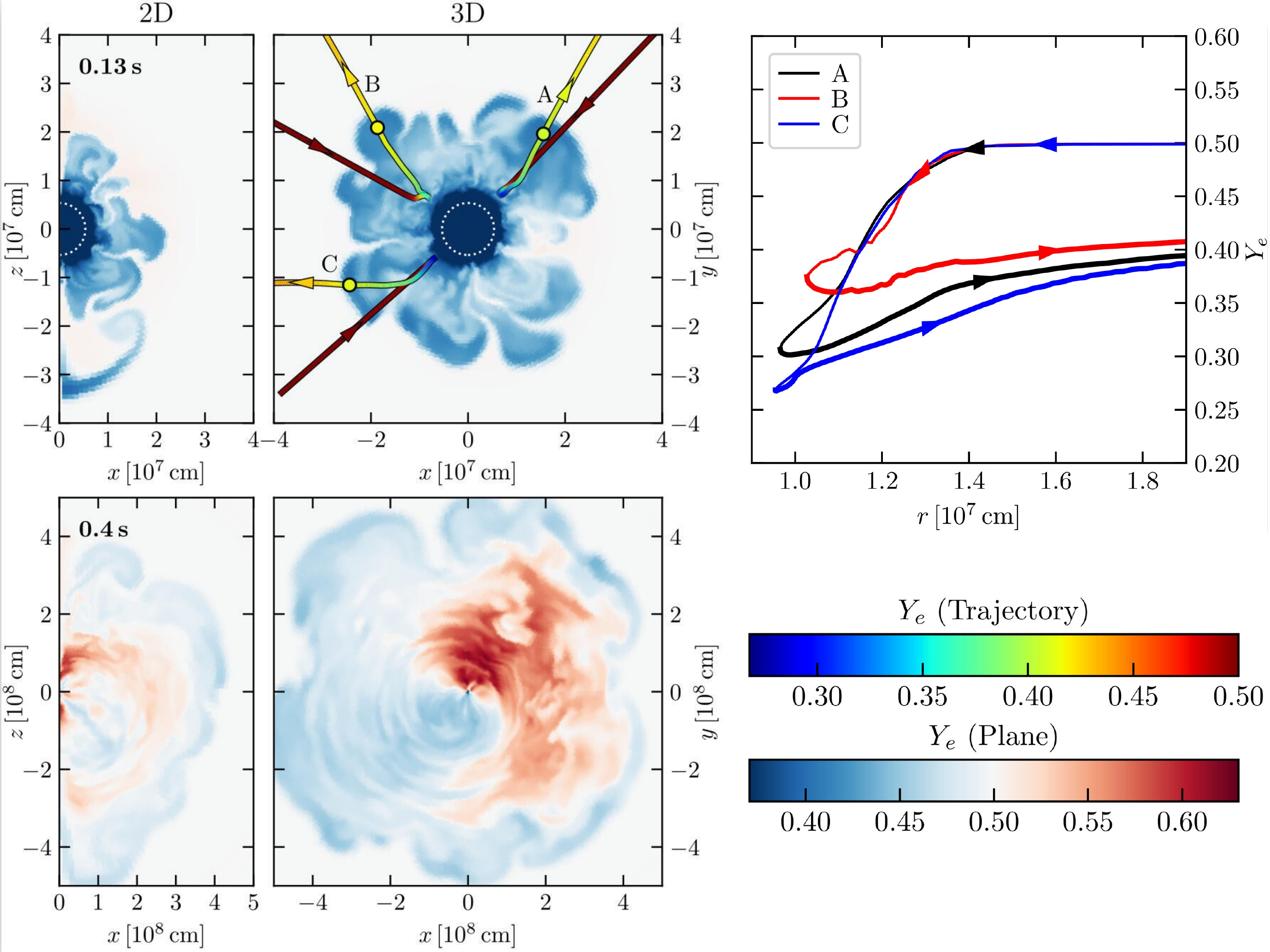}
\caption{Neutron-rich, fast, early ejecta ({\em upper panels}) and LESA-induced hemispheric
asymmetry of the electron fraction in the neutrino-driven wind ejecta ({\em bottom left})
of an ECSN-like explosion of a 9.6\,$M_\odot$ progenitor.
The {\em left panels} display cross-sectional cuts of 2D and 3D explosion models (color coding according
to the color bar on the {\em bottom right}). The {\em upper right panel} shows the evolution of $Y_e$ along 
three representative inflow-outflow trajectories for the earliest, neutron-rich ejecta in the 3D model 
as indicated in the {\em upper left panel} 
(Figure from \citealt{Stockinger+2020}; \copyright Oxford University Press 
on behalf of the Royal Astronomical Society (RAS). Reproduced with permission)}
\label{fig:z9.6ejectaasymmetry}
\end{figure}

\begin{figure}
\includegraphics[width=1.0\columnwidth]{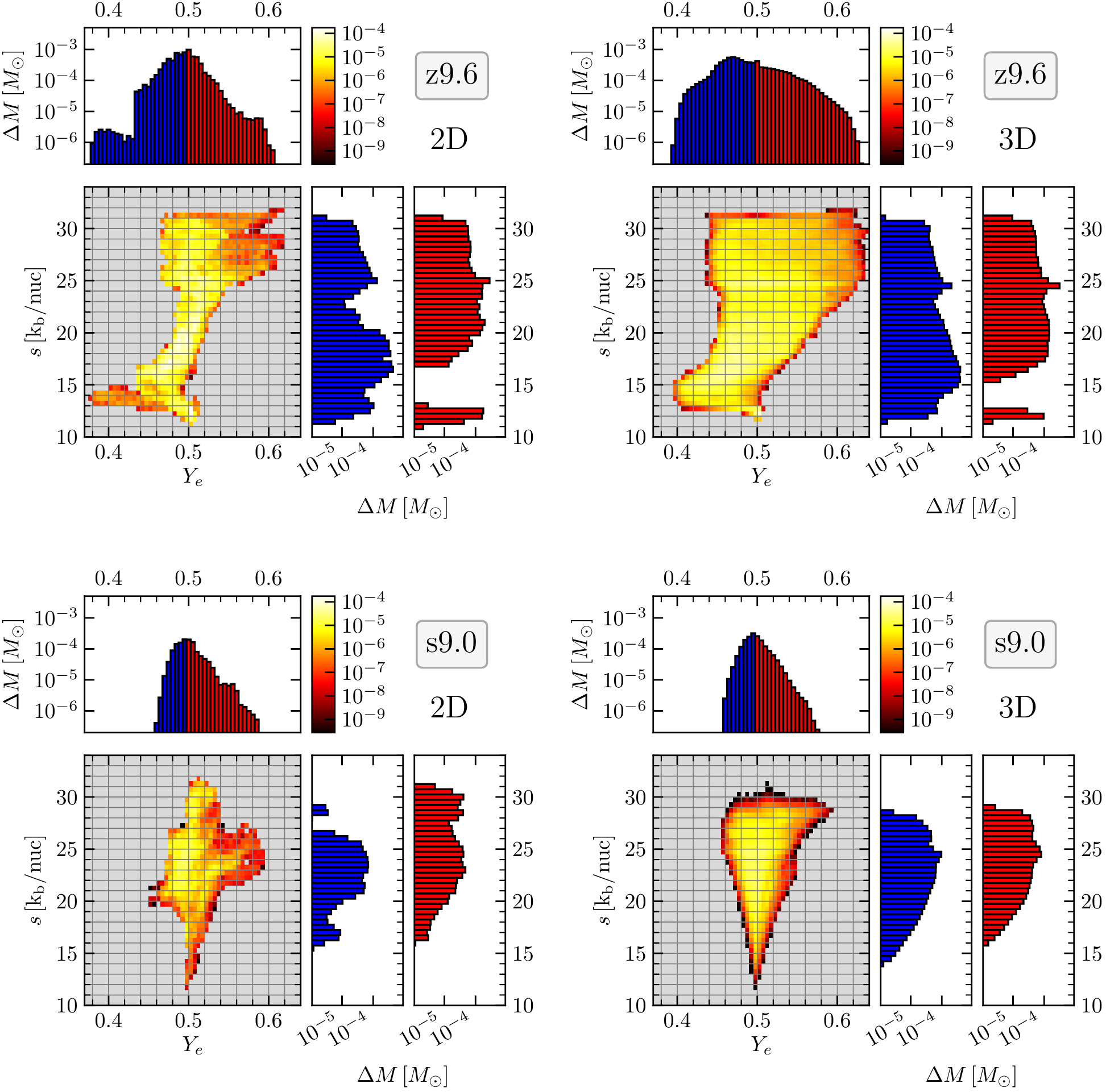}
\caption{Mass distributions of ejecta as functions of $Y_e$ and entropy per nucleon, $s$ 
(in units of Boltzmann's constant), for 2D ({\em left}) and 3D ({\em right}) simulations of 
an ECSN-like explosion of a zero-metallicity 9.6\,$M_\odot$ progenitor ({\em top}) and a 
CCSN explosion of a solar-metallicity 9\,$M_\odot$ progenitor ({\em bottom}). Both 
progenitors have developed iron cores prior to collapse. The central panels in each case 
display the mass distribution in the $Y_e$-$s$-plane (colour coding in units of $M_\odot$ 
according to the colour bar). The corresponding panels above and to the right show the marginal 
distributions with blue and red bars indicating $Y_e < 0.5$ and $Y_e > 0.5$, respectively.
Considerably more neutron-rich matter is ejected due to the fast expansion of the ECSN-like
explosion (Figure from \citealt{Stockinger+2020}; \copyright Oxford University Press on 
behalf of the RAS. Reproduced with permission)}
\label{fig:z9.6+s9.0ejectahistograms}
\end{figure}

An interesting difference exists between ECSN-like explosions and CCSNe. In the
former ones the earliest ejecta expand very quickly behind the SN shock, which 
accelerates and propagates much more rapidly than in the CCSNe of more massive progenitors.
Therefore the exposure of the ejected material 
to the $\nu_e$ and $\bar\nu_e$ fluxes from the PNS is much
shorter and they are able to retain a higher neutron excess, closer to the
neutron-rich state that the gas had near the minimum radius that it reached during its 
infall (see Figure~\ref{fig:z9.6ejectaasymmetry}, upper panels). Compared to CCSNe,
ECSN-like events therefore eject significantly more neutron-rich matter with lower 
minimum values of $Y_e$ (Figure~\ref{fig:z9.6+s9.0ejectahistograms}). Therefore they
can contribute considerably to the production of trans-iron species from Zn to Zr
and could be relevant sources of neutron-rich radioactive isotopes such as $^{48}$Ca
and $^{60}$Fe and possibly also of elements beyond $N = 50$ made by a weak r-process
reaching up to Pd, Ag, and Cd \citep{Wanajo+2011,Wanajo+2013a,Wanajo+2013b}.

\subsubsection{Region~3: neutrino-driven wind}

Region~3 is the neutrino-driven wind (or ``neutrino wind''), which is
a low-density outflow of baryonic matter from the surface layers of the hot PNS.
The neutrino emission during the Kelvin-Helmholtz cooling of the PNS does not 
permit a hydrostatic atmosphere, but neutrino heating exterior to the neutrinosphere,
mainly by $\nu_e$ and $\bar\nu_e$ absorption of Equations~(\ref{eq:nuab1}) and
(\ref{eq:nuab2}) and on a lower level also by neutrino-electron scattering and
neutrino-antineutrino pair annihilation,
leads to mass loss through such a wind \citep[see][and references therein]{Duncan+1986,Woosley+1992,Witti+1994,Qian+1996,Otsuki+2000,Thompson+2001}.
This outflow of matter is quasi-stationary, i.e., slowly evolving with time, and
quasi-spherical for non-rotating or slowly rotating PNSs.
\citet{Qian+1996} provided analytic expressions for the important wind properties,
assuming steady-state conditions in the wind. The mass loss rate is approximately 
given by
\begin{equation}
\dot M\ \approx\ 6\times 10^{-4}\,L_{{\mathrm{tot}},53}^{5/3}
\epsilon_{\nu_e,15}^{10/3} R_6^{5/3} M_{1.4}^{-2}\ \ 
M_\odot\,{\mathrm s}^{-1}\,,
\label{eq:windmdot}
\end{equation}
the asymptotic wind entropy for radiation-dominated conditions is
\begin{equation}
s\ \approx\ 15 + 50\,L_{{\mathrm{tot}},53}^{-1/6}
\epsilon_{\nu_e,15}^{-1/3} R_6^{-2/3} M_{1.4}\ k_{\mathrm B}
\ {\mathrm{per\ nucleon}}\,,
\label{eq:windentropy}
\end{equation}
and the expansion timescale of the wind with velocity $v$ at radius
$r$ where the temperature has dropped to 0.5\,MeV is
\begin{equation}
t_{\mathrm{dyn}}\ \equiv\ \left.{r\over v}\right
|_{T\approx 0.5\,\mathrm{MeV}}\ \approx\
6\times 10^{-3}\,L_{{\mathrm{tot}},53}^{-1}
\epsilon_{\nu_e,15}^{-2} R_6 M_{1.4}\ \ {\mathrm s}\, .
\label{eq:windtimescale}
\end{equation}
Here $L_{{\mathrm{tot}},53}$ denotes the total neutrino luminosity (i.e.,
the sum of the contributions of neutrinos and antineutrinos of all flavors) in
units of $10^{53}\,$erg$\,$s$^{-1}$, $R_6$ the PNS radius
in $10^6\,$cm, $M_{1.4}$ the PNS mass in 1.4$\,M_\odot$,
and $\epsilon_{\nu_e,15}$ is the ratio of neutrino
spectral moments, defined as
$\epsilon_\nu \equiv \left\langle E_\nu^2 \right\rangle/
\left\langle E_\nu \right\rangle$, for electron neutrinos in
units of 15$\,$MeV (the angle brackets denote spectral averages
of powers of the neutrino energies $E_\nu$).

The neutrino-driven wind is a generic feature of spherically symmetric (1D) simulations, 
where explosions have to be triggered artificially in most cases. However, it is witnessed 
in self-consistent, state-of-the-art 3D simulations only for the low-energy ECSN-like 
explosions (upper panel of Figure~\ref{fig:SNnucleosynthesis}). 
The low-mass progenitors of ECSN-like events possess extremely steep density gradients 
exterior to their degenerate cores and correspondingly very low mass-infall rates 
\citep{Stockinger+2020}. Therefore the quickly expanding shock immediately reverses 
their collapse to outflow and creates a low-density bubble around the PNS, into which 
the wind can expand freely. In contrast, in
CCSNe of more massive stars with higher mass-infall rates a neutrino-driven wind does 
not develop until many seconds after the onset of the explosion \citep{Bollig+2021}. 
Instead, in those cases long-lasting accretion downflows 
reach down to the close vicinity of the PNS during all of this long period of 
post-bounce evolution (Figure~\ref{fig:s19ejecta}). The downdrafts are supplied with 
matter swept up by the outgoing SN shock. 
Because of their heavy mass loading they remain intact when they dive inward towards
the PNS and return outward only after having absorbed enough energy from the intense 
neutrino fluxes to become buoyant. Thus they increase the blast-wave energy to values 
of order $10^{51}$\,erg as needed to explain observed energies of canonical CCSNe 
\citep[see][and Section~\ref{sec:neutrinoexplosions}]{Mueller+2017,Bollig+2021}.

As in region~2, the charged-current $\beta$-processes of 
Eqs.~(\ref{eq:nuab1})--(\ref{eq:nuem2}) set the $Y_e$ in the neutrino-driven wind
of region~3. However, because of the low densities (high entropies) and rapid expansion
cooling of the wind, the absorption reactions of $\nu_e$ and $\bar\nu_e$ dominate
quickly over the inverse processes. If the $\nu_e$ and $\bar\nu_e$
capture rates on nucleons (Eqs.~\ref{eq:nuab1} and \ref{eq:nuab2}) are in equilibrium, 
the asymptotic electron fraction $Y_e$ in the wind can be estimated as 
\citep{Qian+1996}
\begin{equation}
Y_e\ \approx\ \left [ 1 +
{L_{\bar\nu_e}\epsilon_{\bar\nu_e}\over
L_{\nu_e}\epsilon_{\nu_e}}\,
Q(\epsilon_{\nu_e},\epsilon_{\bar\nu_e})\,
C(\epsilon_{\nu_e},\epsilon_{\bar\nu_e})\right
]^{-1} \, ,
\label{eq:windye}
\end{equation}
where $L_{\nu_i}$ is the luminosity of neutrino species $\nu_i$
and $\epsilon_{\nu_i}$ is again the ratio of neutrino spectral moments,
$\epsilon_{\nu_i} \equiv \left\langle E_{\nu_i}^2 \right\rangle/
\left\langle E_{\nu_i} \right\rangle$. The factors
$Q(\epsilon_{\nu_e},\epsilon_{\bar\nu_e})$ and $C(\epsilon_{\nu_e},\epsilon_{\bar\nu_e})$
in the denominator of this expression depend on the mean neutrino energies;
$Q$ is a correction factor that accounts for the neutron-proton mass
difference, $\Delta = 1.29\,$MeV, in the $\beta$-reactions, and $C$ contains corrections
due to nucleon recoil and weak magnetism effects \citep{Horowitz+1999,Horowitz2002}.
These factors can be conveniently approximated by
\begin{eqnarray}
Q(\epsilon_{\nu_e},\epsilon_{\bar\nu_e})
&\approx& {1 - 2\Delta/\epsilon_{\bar\nu_e} + 1.2
(\Delta/\epsilon_{\bar\nu_e})^2 \over
1 + 2\Delta/\epsilon_{\nu_e} + 1.2
(\Delta/\epsilon_{\nu_e})^2}\,, \\
C(\epsilon_{\nu_e},\epsilon_{\bar\nu_e})
&\approx& {1 - 8.66\,\epsilon_{\bar\nu_e}/(mc^2)
\over 1+ 1.22\,\epsilon_{\nu_e}/(mc^2)} \, ,
\end{eqnarray}
with $m$ being the nucleon mass. The absorption of $\nu_e$ and $\bar\nu_e$ in the
ejecta drives $Y_e$ to the neutron-rich side, i.e., $Y_e < 0.5$, if
${L_{\bar\nu_e}\epsilon_{\bar\nu_e}\cdot C\cdot Q > L_{\nu_e}\epsilon_{\nu_e}}$.
If one ignores the factor $C$, considers only the linear dependence of $Q$ on
$\Delta/\epsilon_\nu$, and assumes $L_{\bar\nu_e} = L_{\nu_e}$, for example, one
finds the condition $\epsilon_{\bar\nu_e} > \epsilon_{\nu_e} + 4\Delta$. This
means that electron antineutrinos need to be emitted by the PNS with significantly 
harder spectra.

In early numerical models the neutrino-driven wind was found to reach entropies of several 
100\,$k_\mathrm{B}$ per nucleon and $Y_e$ values well below 0.4 after several seconds,
both with trends to more extreme conditions (higher entropies and higher neutron excess)
at later times \citep{Woosley+1994}. 
Therefore the wind environment has been discussed as possible site of r-process
nucleosynthesis for more than a decade 
\citep[e.g.,][and references therein]{Takahashi+1994,Hoffman+1997,Otsuki+2000,Arnould+2007}. 
However, modern, more elaborate SN and PNS cooling simulations with better neutrino transport 
showed that PNS winds do not reach sufficiently high entropies and also remain 
proton-rich instead of developing a large neutron excess \citep[e.g.,][and references therein]{Huedepohl+2010,Fischer+2010,Roberts+2010,Mirizzi+2016}, at least when
strong magnetic fields and rapid rotation do not play a relevant role in the PNS.
Interestingly, in proton-rich ejecta, in particular in those of the neutrino-driven wind 
but potentially also in the neutrino-processed early ejecta of region~2, a $\nu$p-process
might take place when the proton-rich matter is exposed to intense neutrino fluxes. Such a 
possibility requires that the entropy is sufficiently high and the expansion time scale 
is long enough to permit an efficient conversion of protons to neutrons by frequent 
$\bar\nu_e$ absorption reactions
\citep{Froehlich+2006,Pruet+2006,Wanajo2006,Arcones+2012,Wanajo+2018}.

\subsubsection{Region~4: preshock helium shell and metal core}

Region~4 denotes the helium shell and the different layers (Si, O/Ne, C) of the metal 
core prior to the passage of the outgoing SN shock. Well before the shock reaches these
shells, which can take many seconds to minutes for the outermost of the mentioned 
layers, the huge fluxes of neutrinos and 
antineutrinos of all flavors emitted by the PNS irradiate the nuclei in these regions.
This bombardment can cause interesting transmutations via neutrino-induced reactions
\citep{Domogatsky1977,Woosley1977,Domogatskii+1978,Domogatskii+1980,Woosley+1988,Woosley+2002}, 
mainly through charged-current (of $\nu_e$ and $\bar\nu_e$) and neutral-current 
excitation of nuclei that subsequently decay by
the emission of light particles (neutrons, protons, $\alpha$-particles). Thus 
rare isotopes, e.g., $^{11}$B, $^{19}$F, $^{138}$La, and $^{180}$Ta, can be 
produced directly by the neutrino interactions with abundant target nuclei,
and, in addition, the increased abundance of light particles can lead to the
nucleosynthesis of some species such as $^7$Li and $^{26}$Al through a network
of charged-particle reactions. Of course, when the shock passes through the
outer layers, the neutrino-processed abundances will experience some reprocessing
by the shock, which has to be taken into account in detailed calculations
\citep{Woosley+1990}. 

For recent results, accounting for the latest understanding of neutrino-nucleus
cross sections and the crucial neutrino properties, i.e., the time-dependent 
luminosities and spectra of all emitted neutrino species, deduced from 
state-of-the-art CCSN models, see \citet{Sieverding+2018} and, in particular, 
\citet{Sieverding+2019}. Since the production of key species by 
the neutrino process requires neutrino energies that are high enough to 
facilitate neutrino-nucleus interactions and nucleus excitation, the efficiency
of this process is very sensitive to the spectral temperature of the neutrinos.
The synthesis of isotopes through the $\nu$-process is therefore a potential 
thermometer for the neutrinos emitted by CCSNe.

\subsubsection{Closing remarks}

Naturally, all the nucleosynthesis and the dissemination of its products in
regions~1--4 hinge on the success of the CCSN explosion. This implies that the 
nuclear EoS should not lead to fast BH formation but instead it must allow for
ample energy release by neutrinos and/or magnetorotational effects involving a 
sufficiently long-lived PNS. Alternatively, it could mean that a hadron-quark 
phase transition releases enough energy to trigger the explosion, implying
special EoS properties.

More specifically, the EoS has a significant influence on the characteristic 
features of the neutrino emission by the PNS. These, in turn, are crucial for the 
neutrino heating and thus for the viability of the neutrino-driven mechanism.
Also the blast-wave energy of the SN explosion, which determines the nucleosynthesis 
in region~1, depends on the time evolution of the neutrino luminosities and 
specta. The same holds true for the efficiency of the neutrino process in region~4; 
and it also holds true for the neutron-to-proton ratio in the neutrino-heated ejecta 
(region~2) and for the conditions ($Y_e$, entropy, expansion velocity) in the 
neutrino-driven wind (region~3). Therefore, since the nuclear EoS 
controls the cooling and deleptonization of the PNS from its hot
and proton-rich initial state to the final, cold compact remnant, it is also
of great importance for the nucleosynthesis in CCSNe. 

The next section will elaborate on these aspects.

\begin{figure}
\includegraphics[width=1.0\columnwidth]{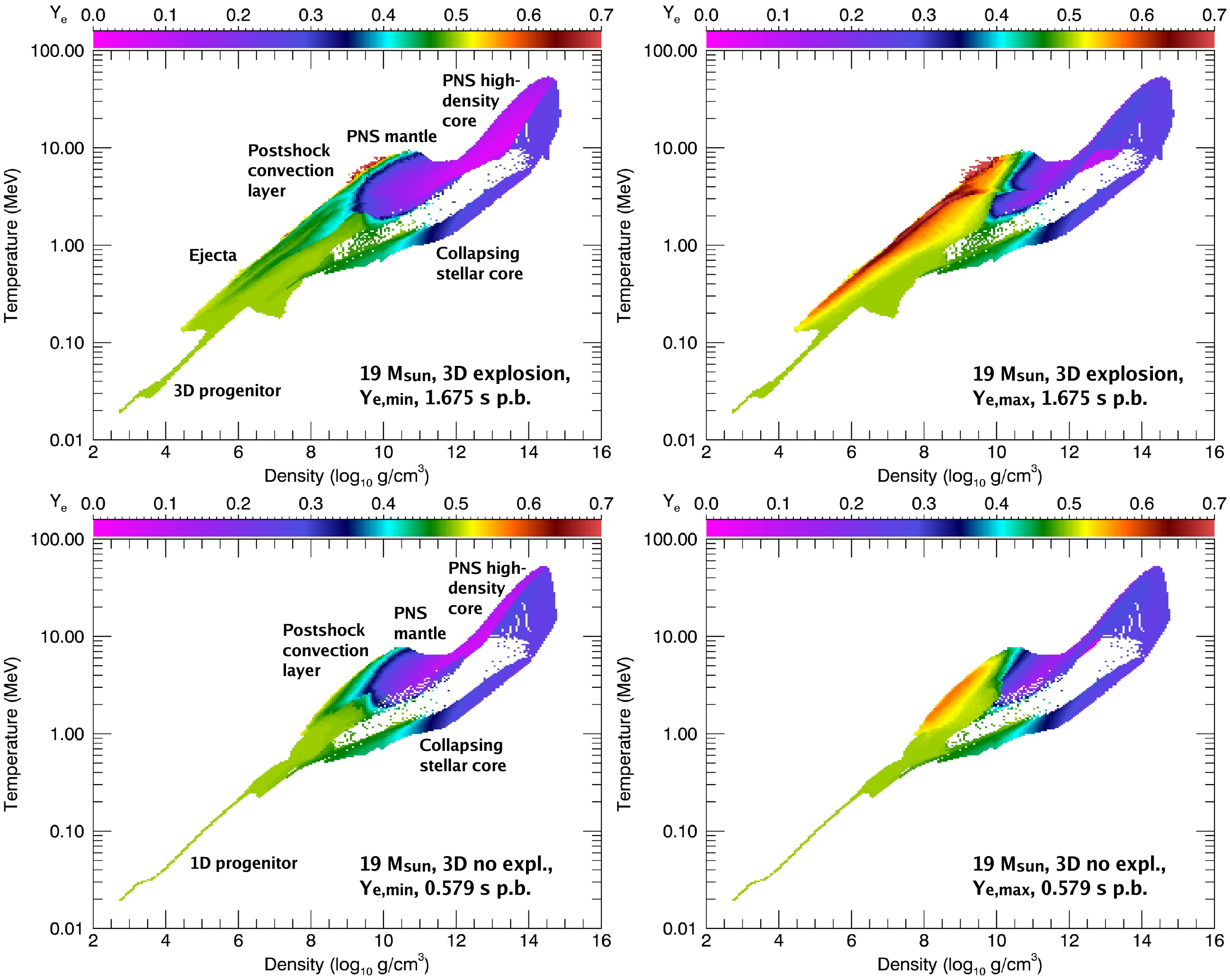}
\caption{Overview of thermodynamically relevant conditions in CCSNe by means of the
minimum ({\em left}) and maximum ({\em right}) values of the electron fraction ($Y_e$) as functions 
of density and temperature in successfully exploding ({\em top}) and non-exploding ({\em bottom})
3D simulations of a 19\,$M_\odot$ progenitor \citep{Bollig+2021} with the LS220 EoS 
\citep{Lattimer+1991}. The two 3D CCSN calculations were evaluated for the entire evolution 
periods until the post-bounce times mentioned in the panels. Note that the empty region 
between the conditions in the interior of the PNS and the conditions in the collapsing
stellar core as well as the high-density, low-temperature region below the stellar core-collapse
branch will be filled during the neutrino-cooling evolution of the PNS
(Figure courtesy of Robert Bollig)}
\label{fig:EOSregions}
\end{figure}

\subsection{Role of the equation of state}

Over decades, a large number of works has investigated the influence of the 
high-temperature nuclear and super-nuclear EoS on the mass limit when PNSs collapse to 
BHs \citep[e.g.,][]{Keil+1995,OConnor+2011,Steiner+2013,Schneider+2020}, 
on the neutrino emission prior to and after the CCSN explosion
\citep[e.g.,][]{Pons+1999,Pons+2001,Marek+2009,OConnor+2013,Schneider+2019,Sumiyoshi+2019,Nakazato+2019,Nakazato+2020,Reed+2020},
and on the GW signals from the death events of massive stars 
\citep[e.g.,][]{Marek+2009,Zha+2020,EggenbergerAndersen+2021,Kuroda+2022}. The relevant 
literature cannot be comprehensively listed here. Many of the mentioned studies ---except those
for GW predictions--- were done in 1D, because corresponding codes are publicly available and
because such calculations are relatively inexpensive and can be performed for large sets of models,
thus systematically exploring parameter dependencies, for example on the nuclear saturation density, 
isoscalar incompressibility modulus, symmetry energy, and effective mass of nucleons. However,
although basic sensitivities of the considered phenomena to variations of the nuclear inputs 
may be carved out, 1D results can also be highly misleading, quantitatively as well as qualitatively. 
Therefore they can provide only limited information on the behavior of more realistic, 
multi-dimensional systems and the real phenomena in Nature (examples will be discussed below). 
Conclusive results that are valid for astrophysical interpretation are not abundantly available 
as yet. In particular, such results are still missing w.r.t.\ the question how CCSN nucleosynthesis
depends on the EoS of PNS matter. Little research has been done on this problem so far.

Figure~\ref{fig:EOSregions} provides an overview of the EoS constraining parameter values of 
density, temperature, and electron fraction $Y_e$ (minimum and maximum values) that were obtained 
during different phases and in different regions of a collapsing $\sim$19\,$M_\odot$ progenitor 
star, its newly formed NS, and the ejecta of its CCSN explosion. These 3D simulations led to 
successful neutrino-driven explosions when they were started from the 3D pre-collapse structure 
of the progenitor, thus taking into account the density and velocity perturbations 
in the convective oxygen-burning shell, but they ended in failed explosions when the initial 
conditions were taken from 1D progenitor profiles \citep{Bollig+2021}. 
The stellar core is characterized by low-entropy, low-temperature conditions in the infall 
phase, the core of the PNS (above a density of roughly $10^{13}$\,g\,cm$^{-3}$) and the mantle
region of the PNS (above a density of several $10^{10}$\,g\,cm$^{-3}$) develop low electron 
fractions, the convective postshock layer (above about $10^8$\,g\,cm$^{-3}$) exhibits a mix of 
low-$Y_e$ and high-$Y_e$ matter in low-entropy accretion downflows and high-entropy buoyant plumes,
respectively. The neutrino-heated CCSN ejecta stick out by their high temperatures and entropies 
and by $Y_e$ values ranging from close to 0.5 to considerably higher numbers 
(see also Figure~\ref{fig:s19-ejectahistogram}). 

The main topic of this section is the question how the properties of these ejecta and the
associated nucleosynthesis are connected to the properties of the high-density EoS inside 
the PNS. This interesting problem is only superficially and fragmentarily explored, 
because only very few self-consistent, first-principle SN 
simulations, in particular in 3D, have so far been carried out with different high-density
EoS models. Moreover, in order to study the impact on the chemical element production, the
explosion simulations need to be followed to sufficiently late times to determine all 
ejecta components and to follow them through the density-temperature regime that is relevant 
for the nucleosynthesis. The discussion of this section can therefore only highlight some
essential connections between the EoS-dependent PNS properties, the corresponding neutrino 
emission, and the explosion and ejecta conditions that hinge on the physics playing a role in 
the center of the SN blast.

\begin{figure}
\includegraphics[width=1.0\columnwidth]{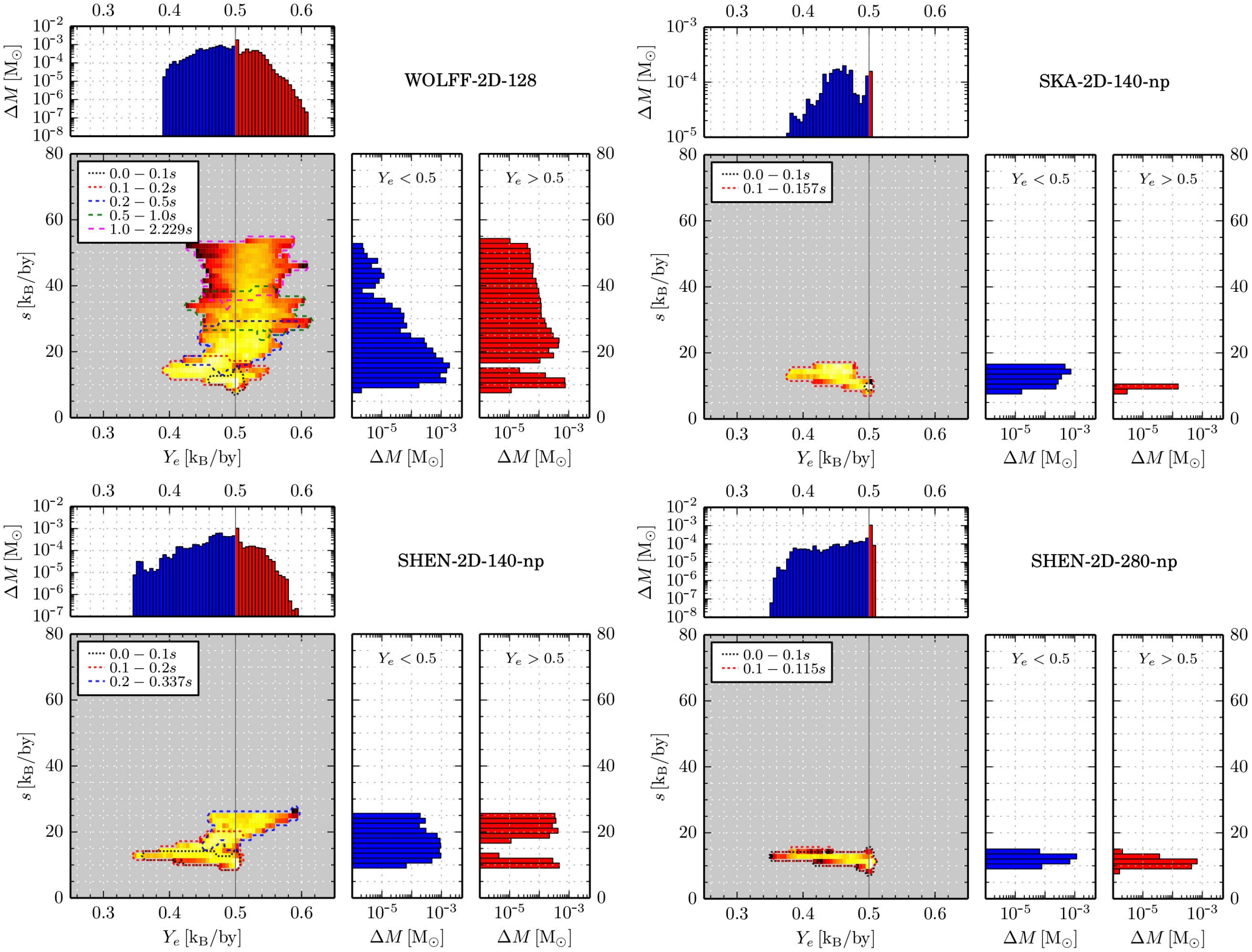}
\caption{Mass distributions of ejecta as functions of $Y_e$ and entropy per nucleon $s$,
plotted similarly to Figure~\ref{fig:z9.6+s9.0ejectahistograms}, for 2D simulations
of ECSN explosions of an 8.8\,$M_\odot$ ONeMg-core progenitor \citep{Groote2014} with 
three different nuclear EoSs: STOS (labelled SHEN) of \citet{Shen+1998,Shen+2011}, WOLFF of
\citet{Hillebrandt+1984} and \citet{Hillebrandt+1985}, and LS-SKa 
\citep[labelled SKA;][J.~Lattimer, 2009, private communication; EoS tables available at 
http://www.astro.sunysb.edu/lattimer/EOS/main.html]{Lattimer+1991}. The time evolution 
is visualized by dashed and dotted lines of different colors in the $Y_e$-$s$-plane, 
which indicate matter that is ejected in different time intervals. This shows that the 
most neutron-rich material is expelled earliest. Except for the simulation with the WOLFF EoS, 
the models were carried only to the onset of the explosion to identify the most neutron-rich
ejecta. Fast-moving, early ejecta with $0.35\lesssim Y_e \lesssim 0.40$, which are expelled 
in rising convective plumes during the first 100--150\,ms after core bounce, exhibit a 
clear sensitivity to the nuclear EoS, whereas the angular resolution
(1.286$^\circ$ versus 0.643$^\circ$ lateral zone size with 140 and 280 equidistant 
angular zones in the range $[0^\circ,180^\circ]$, respectively) has no crucial effect 
on the minimum value of $Y_e$ ({\em lower panels}) (Figure courtesy of Janina von Groote)}
\label{fig:ONeMg-ejectahistograms}
\end{figure}

\subsubsection{Neutron excess in early ejecta of ECSN-like explosions}
\label{sec:ECSNrprocess}

In Section~\ref{sec:nuejecta} it was mentioned that the neutron-to-proton ratio in 
neutrino-heated ejecta differs considerably between ECSN-like explosions and CCSNe of
more massive progenitors. The ejecta expand more slowly in the latter events and 
therefore frequent absorptions of $\nu_e$ and $\bar\nu_e$ lead to an increase of $Y_e$
from pre-ejection conditions of $Y_e \ll 0.5$ to values around 0.5 and even higher. 
Most of the CCSN ejecta therefore possess $Y_e \gtrsim 0.5$ 
\citep[Figures~\ref{fig:s19-ejectahistogram} and \ref{fig:EOSregions} and][]{Wanajo+2018}. 
In contrast, the extremely fast expansion of the SN shock and of the earliest ejecta in
ECSN-like explosions permits this matter to retain a considerable neutron excess with
$Y_e$ possibly as low as 0.35, which could enable the formation of light 
r-process elements up to the palladium and silver region;
see Section~\ref{sec:nuejecta}, Figures~\ref{fig:z9.6ejectaasymmetry} and 
\ref{fig:z9.6+s9.0ejectahistograms}, and references \citet{Wanajo+2011}, 
\citet{Wanajo+2018}, and \citet{Stockinger+2020}. 

The mass distribution of $Y_e$ in the ejecta of ECSN-like explosions depends on 
the dimension of the simulations: expelled material with $Y_e < 0.45$ is not present
in 1D models \citep{Wanajo+2011,Zha+2022}. Also the details of the core structure of 
the progenitor (e.g., whether it was computed in 1D or multi-D) has some noticeable
influence, in particular on the question whether a small amount of matter is 
expelled with $0.35\lesssim Y_e \lesssim 0.40$ \citep{Zha+2022}. This last aspect
is also sensitive to the nuclear EoS applied in the core-collapse and explosion 
modeling, because the ejection of the fastest material hinges extremely sensitively 
on the subtleties of the early postbounce dynamics of the SN shock and on the
rapid development of convective overturn in the 
postshock layer. Also the details of the neutrino emission in the first 100--150 
milliseconds after core bounce have an influence on the $Y_e$ distribution in the
quickly expanding, buoyant plumes (Figure~\ref{fig:ONeMg-ejectahistograms}). 
This was witnessed by \citet{Groote2014}
in general relativistic 2D simulations with state-of-the-art neutrino transport 
when considering a small sample of different EoSs for the hot PNS medium.
More investigations of this aspect with larger sets of EoS models are desirable.

\subsubsection{Relevance of the proto-neutron star radius}

In order to have astrophysical consequences, microscopic EoS properties must lead to macroscopic 
effects in CCSNe. The radius evolution of the hot PNS after core bounce, which depends on the
nonzero-temperature EoS, is one example of such a macroscopic influence of relevance for the
SN evolution. Early 2D simulations of stellar core collapse and the neutrino-driven mechanism 
using a state-of-the-art treatment of neutrino transport and different prescriptions for the 
nuclear EoS suggested that a faster contraction 
of the PNS facilitates shock revival by neutrino heating and the onset of an explosion
\citep{Janka2012}. This result was confirmed with a more approximate description of 
the neutrino physics by \citet{Suwa+2013}. The basic reason is that stronger PNS contraction
releases compression work, which heats the PNS mantle layer and leads to higher virial
temperatures at the neutrinospheres of all neutrino species. Moreover, because of the steepening 
density gradient near the PNS surface, the neutrinospheres move to smaller radii. These effects
in combination permit higher luminosities and harder spectra of the emitted neutrinos, both of 
which enhance the energy deposition by neutrinos in the gain layer behind the stalled shock. 
Therefore, if the PNS contracts faster, an explosion can set in more easily and earlier. 
This behavior is a generically multi-dimensional phenomenon, because in
1D the faster contraction of the PNS leads to stronger recession of the stagnating shock,
despite more neutrino energy deposition by the higher luminosities and mean energies of the 
radiated neutrinos \citep{Janka2012}. In contrast, in the multi-dimensional case, the  
enhanced neutrino heating stirs more violent postshock convection \citep{Murphy+2013}. 
The associated buoyancy (``turbulent'') pressure \citep{Mueller+2015} pushes the shock 
farther out. This expansion of the shock increases the gain layer and thus facilitates a 
positive feedback for even more neutrino heating, enabling outward shock acceleration in 
a runaway process.

\begin{figure}
\begin{center}
\includegraphics[width=0.9\columnwidth]{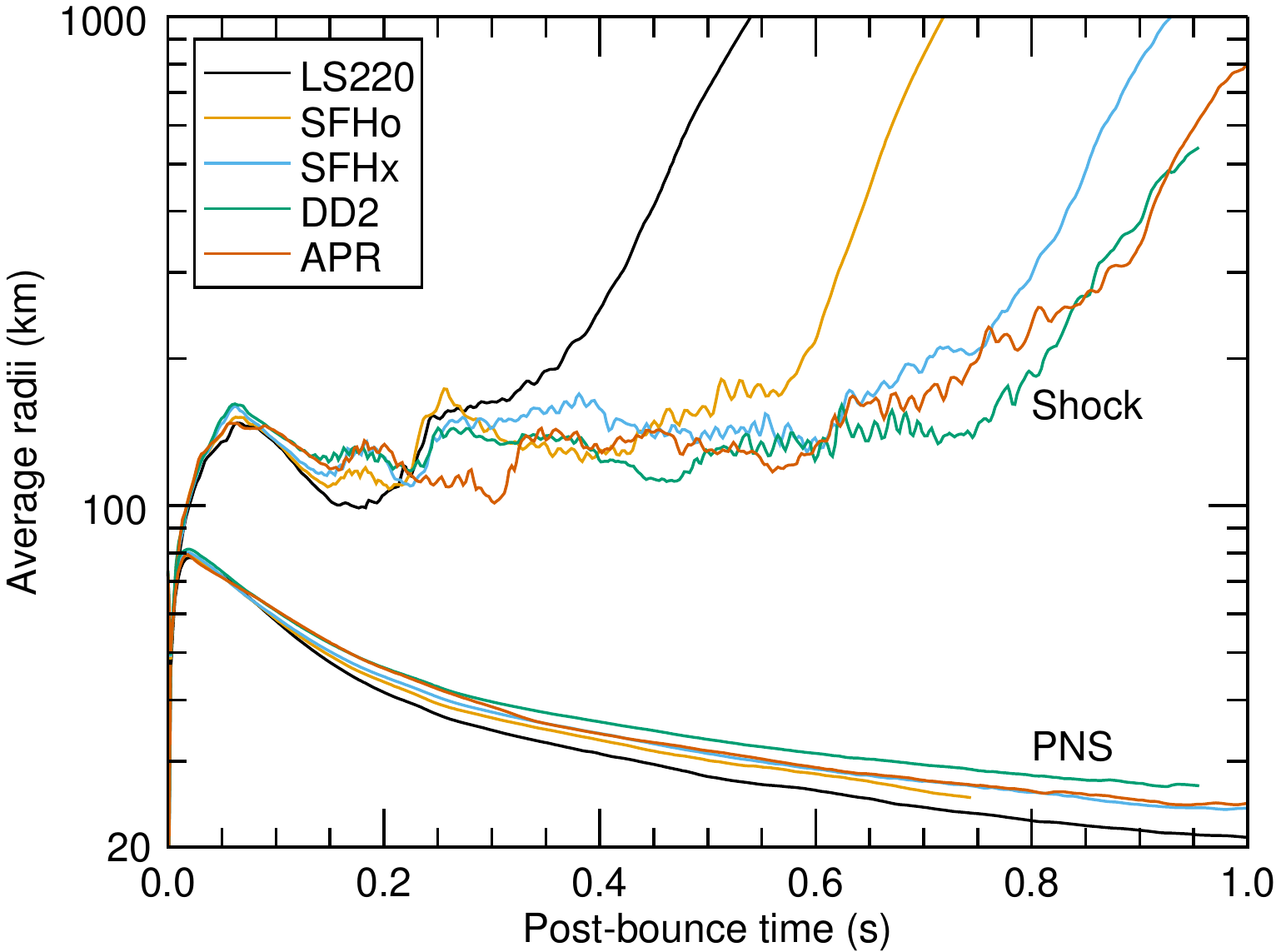}
\end{center}
\caption{Shock radii and PNS radii (spherically averaged) in 3D simulations of CCSN
explosions of a 19\,$M_\odot$ progenitor for different nuclear EoSs widely used in SN
simulations: LS220 of \citet{Lattimer+1991},
SFHo, SFHx of \citet{Steiner+2013} and \citet{Hempel+2010}, DD2 of \citet{Typel+2010,Hempel+2012},
and APR of \citet{Schneider+2019b}. In all cases successful explosions were obtained
in 3D core-collapse simulations started from 3D progenitor conditions \citep{Yadav+2020,Bollig+2021},
but there is a clear correlation between
the onset of the explosion and the contraction of the PNS. The faster the PNS contracts,
the earlier the explosion sets in. This signals the dominant relevance of the PNS radius
evolution over other EoS dependent effects (Figure courtesy of Robert Bollig)}
\label{fig:s19-shockradii}
\end{figure}

Latest 3D CCSN simulations with modern NS EoSs that are fully compatible with the observational
constraints for NS radii and maximum NS mass reviewed in Section~\ref{sec:NSs+EoS}, support 
this tight correlation between the PNS radius evolution and the onset of the explosion 
\citep[for the case of a $\sim$19\,$M_\odot$ star, see Figure~\ref{fig:s19-shockradii}, 
and also see][for a subset of published models]{Bollig+2021}. Remarkably, successful 
neutrino-driven explosions were obtained with all nuclear EoSs tested in the CCSN models, 
provided these simulations were started with 3D progenitor data, i.e., the initial 
conditions were adopted from 3D calculations of convective oxygen-shell burning during 
the last minutes before the core of the progenitor undergoes gravitational collapse 
\citep{Yadav+2020,Bollig+2021}. Corresponding 3D CCSN simulations using 1D progenitor
data did not yield explosions in any of the considered cases. 

Remarkably, 1D SN simulations 
with a parametric treatment for the time evolution of the turbulent kinetic energy  
associated with neutrino-driven convection \citep{Boccioli+2022}
reveal major qualitative and quantitative differences compared to 
the 3D results of Figure~\ref{fig:s19-shockradii}. The 1D results of \citet{Boccioli+2022} 
exhibit a different ordering of the ``explodability'' with varied EoS, for example they 
show only slow explosions for the SFHo EoS and none for the DD2 EoS, in contrast to the
3D models of Figure~\ref{fig:s19-shockradii}. These discrepancies are a warning that 
such 1D calculations should be taken with great caution. They do not only ignore the 
3D nature of pre-collapse convection in the progenitors, but they also attempt to 
describe the effects of non-stationary, anisotropic, non-radial, large-scale and
large-amplitude variations in convective
down- and outflows in the postshock layer by time-dependent equations for averaged
turbulent quantities in a linear approximation, using a mixing-length-theory (MLT) 
approach for the energy flux. 
Such approximations in spherically symmetric models oversimplify the physics 
of postshock turbulence in the non-perturbative regime in an uncontrolled manner. 
Moreover, the employed 1D treatment does not even conserve energy globally 
\citep{Mueller2019,Boccioli+2021}, and its results depend highly sensitively on the values 
of the parameters used in the evolution equation for the turbulent energy. There is no 
guarantee, for example, that MLT parameter values adjusted by comparison to a specific 
3D CCSN simulation are valid also for models with different progenitor masses and different 
nuclear EoSs.  

The pivotal importance of the PNS radius evolution for the onset of explosion is mediated by 
a tight connection between the neutrino emission properties and the PNS radius as described 
above. This physical link was also witnessed when muons were included in the EoS of the hot 
PNS medium and in the neutrino transport, which caused a softening of the high-density EoS 
due to the production of additional particles with considerable rest mass
\citep{Bollig+2017}. And it was again found when strange-quark 
contributions to the nucleon spin were included in the neutral-current neutrino 
scattering rates with neutrons and protons \citep{Melson+2015b}. The overall effect of the 
latter modification was only on the order of 10\% in a single reaction of neutrinos with the
nucleons in the stellar plasma, leading to easier escape of muon and tau neutrinos and 
antineutrinos from the PNS. 
This causes a faster contraction of the PNS, a corresponding temperature rise in the 
neutrinospheric region also of electron neutrinos and antineutrinos, and thus again to more 
postshock heating by $\nu_e$ and $\bar\nu_e$ that leave the neutrinosphere and undergo 
a final charged-current interaction behind the SN shock through the $\beta$-processes of
Eqs.~(\ref{eq:nuab1}) and (\ref{eq:nuab2}). Although the influence of strange-quark corrections
in the neutrino-nucleon scattering rates was assumed to be on the extreme side compared 
to the latest experimental and theoretical limits for such an effect, the 3D CCSN simulations
by \citet{Melson+2015b} nevertheless demonstrated that such a moderate change in the neutrino
opacities was sufficient to convert a failed 20\,$M_\odot$ CCSN explosion into a successful one. 

In fact, the effective mass of nucleons at densities above roughly 10\% of the nuclear 
saturation density was shown to have a strong influence on the radius of hot PNSs, 
although it plays a subdominant role for the properties of cold NSs \citep{Schneider+2019,Yasin+2020}. 
Bigger values of the effective nucleon mass lead to reduced thermal pressure and
thus more compact outer layers of the PNSs, smaller neutrinosphere and PNS radii, and
therefore higher neutrinospheric temperatures. Correspondingly and consistently with the 
3D models of Figure~\ref{fig:s19-shockradii} as well as the works mentioned in the previous 
paragraph, \citet{Schneider+2019}, \citet{Yasin+2020}, and \citet{EggenbergerAndersen+2021} 
therefore also obtained easier and earlier explosions when the PNS contracts faster for 
greater values of the effective nucleon mass. Again, this is in conflict with the findings by
\citet{Boccioli+2022}, who, applying their 1D treatment of turbulence, did not obtain
any clear correlation between PNS (and neutrinosphere) radius, effective mass of nucleons, 
and explosion behavior, but instead reported a better correlation between the viability of
the explosion and the value of the central entropy per nucleon 5\,ms after core bounce.
Once more we warn the reader that approximate descriptions of turbulent effects in 
spherically symmetric CCSN models may cause artifacts in disagreement with direct 3D 
simulations.

Summarizing the situation one can say that CCSN models with different input physics
---except those of \citet{Boccioli+2022}--- agree in their result that a faster PNS 
contraction tightly correlates with a greater likelihood of explosion by the 
neutrino-driven mechanism. 
This outcome was obtained in a similar manner for a variety of physical effects
that can trigger the accelerated shrinking of the PNS. In the specific case of varied 
effective nucleon masses in the super-nuclear EoS, it was also found that larger effective 
masses of the nucleons do not only lead to higher luminosities and mean energies of the
radiated neutrinos \citep{Schneider+2019} but also to bigger amplitudes and 
frequencies of the GW emission \citep{EggenbergerAndersen+2021} and to an earlier collapse
of the hot PNS to a BH at a lower maximum mass, if the SN explosion fails
\citep{Schneider+2020}.

\begin{figure}
\includegraphics[width=1.0\columnwidth]{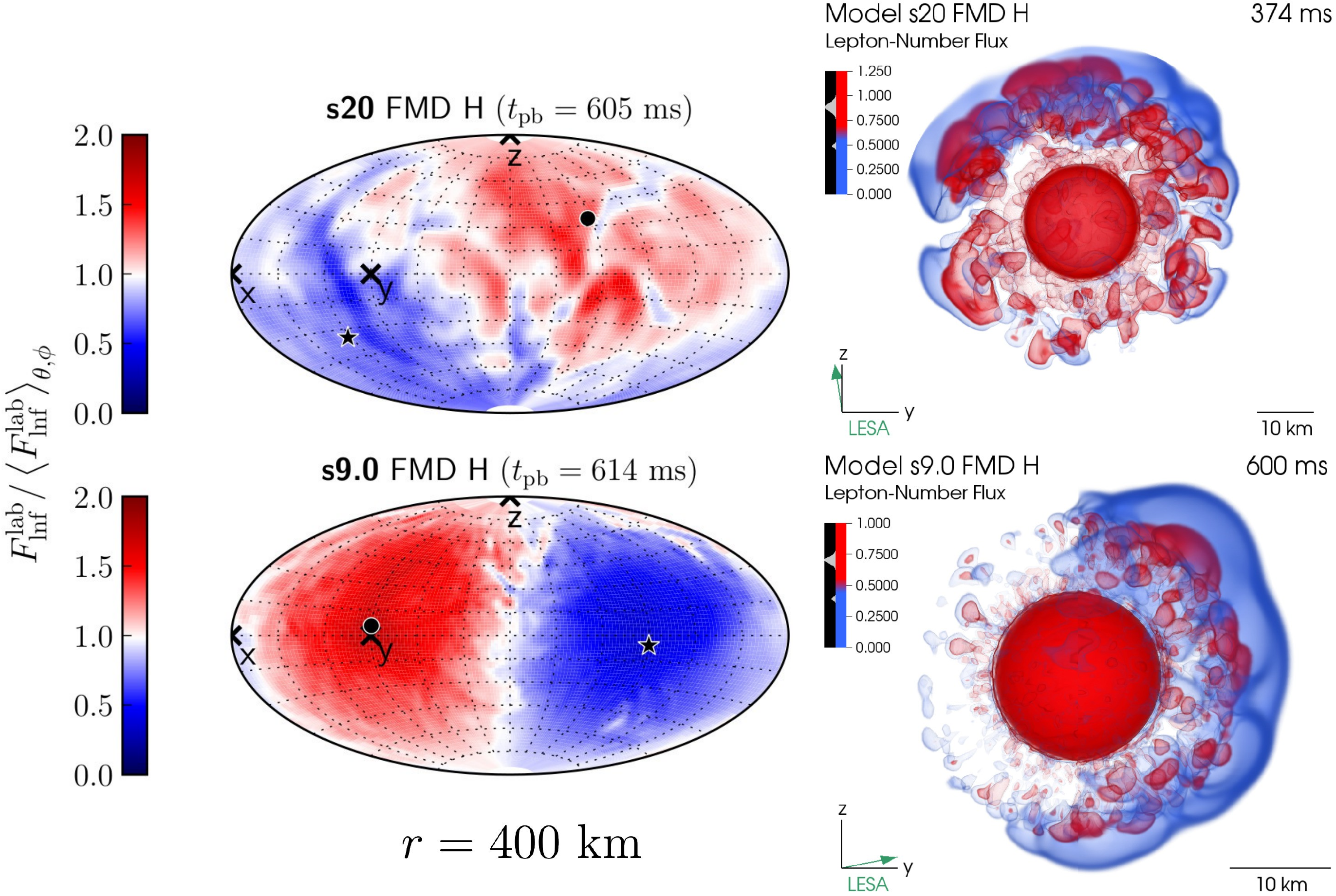}
\caption{Dipolar lepton-number emission asymmetry (LESA) of the neutrino lepton-number
flux (electron neutrinos minus antineutrinos) in 3D CCSN simulations of 20\,$M_\odot$ ({\em top})
and 9.0\,$M_\odot$ ({\em bottom}) progenitors. The Aitoff projections on the {\em left} visualize
the flux density in the lab frame normalized by the angular average at a radius of 400\,km and 
about 0.6\,s after core bounce with the bullets indicating the direction of the LESA dipole 
moments and the stars the opposite directions. The volume renderings on the {\rm right} show the 
electron lepton-number flux density in the lab frame normalized by $10^{43}$\,cm$^{-2}$\,s$^{-1}$ 
in the interior of the PNS at the times labelled in the plots. The LESA dipole direction is
indicated by the green arrow in the tripod in the lower left corner of each panel. The
bubble-like structures are connected to PNS convection in a layer of roughly 10\,km width
and display a clear hemispheric asymmetry with stronger convection on the side of the LESA
dipole direction (Figure from \citealt{Glas+2019};
\copyright American Astronomical Society. Reproduced with permission)}
\label{fig:LESAasymmetry}
\end{figure}

\subsubsection{Relevance of the symmetry energy}

The nuclear symmetry energy is another important parameter that determines the properties of 
the high-density EoS and that has a bearing on the evolution of PNSs in CCSNe \citep[the 
impact of the symmetry energy during the post-merger phase of coalescing binary NSs has 
recently been investigated, too, by][]{Most+2021}. It occurs in an 
expansion of the energy per nucleon of uniform matter composed of neutrons and protons
at zero-temperature in terms of the isospin asymmetry 
$\delta = (n_n-n_p)/n_B = 1-2Y_q$ \citep[e.g.,][]{Oertel+2017}: 
\begin{equation}
    E(n_B,\delta) = E_0(n_B) + E_\mathrm{sym}(n_B)\,\delta^2 + {\cal O}(\delta^4) \,,
    \label{eq:nucleonenergy}
\end{equation}
where $n_n$, $n_p$, 
$n_B$, $Y_q$ are the neutron number density, proton number density, baryon number
density $n_B = n_n+n_p$, and hadronic charge fraction (or proton fraction) 
$Y_q = Y_p = n_p/n_B$ (which is equal to $Y_e = Y_{e^-} - Y_{e^+}$ for
charge-neutral $n$-$p$-$e$ matter), respectively.
Close to nuclear saturation density $n_0$, both the energy per nucleon of 
symmetric matter (i.e., equal numbers of neutrons and protons), $E_0(n_B)$, 
and the symmetry energy (i.e., the difference between the energy of symmetric 
matter and pure neutron matter), $E_\mathrm{sym}(n_B)$, can be expanded in the
deviation $x = (n_B-n_0)/(3n_0)$ of the baryon density from $n_0$:
\begin{eqnarray}
E_0(n_B) &=& m_Bc^2 - B_0 + \frac{1}{2}Kx^2 + \frac{1}{6}Qx^3 + ... \,,
\label{eq:energysym} \\
E_\mathrm{sym}(n_B) &=& J + Lx + \frac{1}{2}K_\mathrm{sym}x^2 + ... \,.
\label{eq:symenergy} 
\end{eqnarray}
Here, $B_0$ is the binding energy of symmetric matter at saturation, $K$ the
incompressibility, $Q$ the skewness, $J$ the symmetry energy at saturation,
$L$ the slope parameter of the symmetry energy, and $K_\mathrm{sym}$ the 
incompressibility of the symmetry energy at saturation density. At nuclear
saturation density and for symmetric matter 
one has $\partial E/\partial n_B(n_B = n_0,\delta = 0) = 0$ by definition. 

The symmetry energy has immediate influence on the structure, thermodynamic 
state, and composition of NS matter, because
the pressure $P$, temperature $T$, and chemical potentials $\mu_i$
of the particles of species $i$ are thermodynamically given by derivatives of 
the energy per baryon,
\begin{equation}
    P = n_B^2\left(\frac{\partial E}{\partial n_B}\right)_{\!\! s,Y_i},\ \ 
    T = \left(\frac{\partial E}{\partial s}\right)_{\!\! n_B,Y_i},\ \ 
    \mu_i = \left(\frac{\partial E}{\partial Y_i}\right)_{\!\! n_B,s,Y_{j\neq i}}
    = \left(\frac{\partial E_i}{\partial Y_i}\right)_{\!\! n_B,s_i},
\label{eq:thermoderivatives}
\end{equation}
with $s$ being the entropy per baryon and $E_i$ and $s_i$ denoting the 
energy per baryon and entropy per baryon of species $i$.

A particularly interesting aspect was first discussed by \citet{Roberts+2012},
who investigated the effects of the symmetry energy on PNS convection during the 
SN explosion and the Kelvin-Helmholtz neutrino cooling phase of the hot compact 
remnant. The linear growth of Ledoux convection is driven by unstable 
gradients of composition (for $n$-$p$-$e$ matter including neutrinos expressed 
in terms of the net electron-type 
lepton fraction $Y_l = Y_e + Y_{\nu_e} - Y_{\bar\nu_e}$) and entropy. 
It is governed by the Ledoux criterion:
\begin{equation}
    C_\mathrm{L} = -\,\left(\frac{\partial P}{\partial n_B}\right)_{\!\! s,Y_l}^{\! -1}
    \cdot \left[\left(\frac{\partial P}{\partial s}\right)_{\!\! n_B,Y_l}
                      \frac{\mathrm{d}s}{\mathrm{d}r}
              + \left(\frac{\partial P}{\partial Y_l}\right)_{\!\! n_B,s}
                      \frac{\mathrm{d}Y_l}{\mathrm{d}r} \right] \,,
\label{eq:ledoux}
\end{equation}
where positive values of $C_\mathrm{L}$ indicate conditions that are unstable
against Ledoux convection. PNS convection takes place in the optically thick
regime for neutrinos, where neutrinos propagate via diffusion and are 
effectively in thermodynamic equilibrium with the stellar plasma. Therefore 
they provide a pressure component that contributes to establish pressure equilibrium
between rising and sinking buoyant fluid elements and their surroundings, for 
which reason neutrino effects should be taken into account in the Ledoux criterion.

For $T \approx 0$ and provided neutrino contributions are 
negligible, i.e., $Y_q = Y_l = Y_e$,  one gets
\begin{equation}
    \left(\frac{\partial P}{\partial Y_l}\right)_{\!\! n_B} \cong 
    \frac{\hbar c}{3}(3\pi^2)^{1/3} n_B^{4/3} Y_e^{1/3} - 
    4n_B^2 \,E_\mathrm{sym}'(n_B)\delta \,,
\label{eq:dpdy}
\end{equation}
where the first term on the rhs corresponds to the contribution by ultra-relativistic,
extremely degenerate electrons. The second term is derived from Eq.~(\ref{eq:nucleonenergy}), 
using the first relation in Eq.~(\ref{eq:thermoderivatives}) for the pressure, and is the 
contribution from the nuclear medium with $E_\mathrm{sym}'(n_B) = \partial E_\mathrm{sym}/\partial n_B = 
(\partial E_\mathrm{sym}/\partial x)(\mathrm{d}x/\mathrm{d}n_B) = (L+K_\mathrm{sym}x)/(3n_0)$.

The partial derivatives of the pressure with respect to density and entropy in
Equation~(\ref{eq:ledoux}) are always positive, which means that negative entropy
gradients always tend to destabilize against convection. In contrast, 
Equation~(\ref{eq:dpdy}) implies that $(\partial P/\partial Y_l)_{n_B,s}$ can
change its sign. It is positive for matter close to isospin symmetry (i.e.,
$\delta \sim 0$)
and thus high $Y_e$, whereas for more neutron-rich conditions, i.e., small $Y_e$,
it may become negative at high densities, depending on the slope of the nuclear 
symmetry energy $E_\mathrm{sym}(n_B)$ around and above nuclear saturation density. 
Thus the symmetry energy can have a direct influence
on the strength and duration of PNS convection. \citet{Roberts+2012} showed that for 
EoSs with a steeper increase of $E_\mathrm{sym}(n_B)$ with density, convection in the 
PNS mantle ceases at an earlier time than in cases where $E_\mathrm{sym}'(n_B)$ is 
smaller, because for steep $E_\mathrm{sym}(n_B)$ the second term in Equation~(\ref{eq:dpdy}) 
becomes increasingly dominant when the PNS mantle contracts and $Y_e$ drops. 
Eventually the term depending on $(\partial P/\partial Y_l)_{n_B,s}$ is
able to stabilize convection. This EoS dependent difference in the convective activity 
leads to characteristic differences in the temporal decline of the neutrino emission, for
example to distinctively different times for the break in the neutrino luminosities that 
signals the moment when convection ends within the PNS.

The impact of the nuclear symmetry energy on PNS convection has also a bearing on
a phenomenon called lepton-emission self-sustained asymmetry (LESA), which was first
witnessed in 3D CCSN simulations of the Garching group by \citet{Tamborra+2014a} and 
\citet{Tamborra+2014b}, and which was later confirmed by other works 
\citep{Janka+2016,OConnor+2018,Glas+2019,Powell+2019,Vartanyan+2019}. The LESA phenomenon
constitutes a global neutrino-emission asymmetry of the PNS, which manifests itself in 
a stable, long-lasting anisotropy of the electron lepton-number (electron neutrino
minus electron antineutrino) emission, for which the dipole term can become the dominant
multipole and even larger than the monopole component (Figure~\ref{fig:LESAasymmetry}, 
left panels). This implies that the individual $\nu_e$ and
$\bar\nu_e$ number fluxes exhibit hemispheric asymmetries on the level of some 10\% 
such that in one hemisphere the emission of $\nu_e$ strongly dominates the 
$\bar\nu_e$ emission, whereas their relative magnitudes are much more similar (or 
even reversed) in the opposite hemisphere. Also the energy luminosities of the
individual neutrino species and their summed luminosity show hemispheric differences
on the level of several percent. 

These emission asymmetries can be traced back to
PNS convection that is stronger in the hemisphere of the LESA dipole direction than 
on the opposite side (Figure~\ref{fig:LESAasymmetry}, right panels). The globally 
asymmetric convection builds up a hemispheric asymmetry of the $Y_e$ distribution 
within the PNS with higher $Y_e$ values in the convective layer and the surrounding
neutrinospheric region on the side of the LESA dipole maximum. This hemispheric difference
can be understood by the fact that strong convection replenishes the leptons lost 
through neutrino emission by convective transport of electrons out of the huge 
reservoir of the PNS core, whereas in the hemisphere with weaker convection this 
supply with fresh electron lepton number is far less efficient.

The large-amplitude, large-scale directional variations of the neutrino emission 
associated with the LESA phenomenon have consequences for the detection 
of CCSN neutrinos \citep{Tamborra+2014b}, for neutrino flavor oscillations
\citep{Glas+2020,Abbar+2021}, and for neutrino-induced PNS kicks \citep{Stockinger+2020}. 
Moreover, the different magnitudes of the electron neutrino flux
relative to the electron antineutrino flux also lead to a contrast in the
neutron-to-proton ratio (and thus $Y_e$) of the innermost, neutrino-heated SN 
ejecta on opposite sides of the PNS, because different amounts of $\nu_e$ and
$\bar\nu_e$ are absorbed according to the reactions of 
Equations~(\ref{eq:nuab1}) and (\ref{eq:nuab2}) 
\citep[see Figure~\ref{fig:z9.6ejectaasymmetry} and][]{Stockinger+2020}. 

It might be surmised that the nuclear symmetry energy through its impact 
on PNS convection could also have an influence on the LESA. If $Y_e$ decreases
faster in the hemisphere opposite to the LESA dipole direction, a steeper
slope of $E_\mathrm{sym}(n_B)$ will tend to stabilize convection in this
hemisphere through a change in the sign of the pressure derivative in 
Equation~(\ref{eq:dpdy}), whereas Ledoux convection on the other side of the 
PNS, where $Y_e$ is still higher, can continue without such a damping effect.
This contrast in the Ledoux conditions in both hemispheres can therefore sustain 
the LESA dipole asymmetry. Such a possibility was already pointed out by 
\citet{Janka+2016}, and indeed 3D CCSN simulations with different nuclear EoSs 
(those of Figure~\ref{fig:s19-shockradii})
reveal that high LESA dipole amplitudes exist for longer post-bounce evolution
periods if the symmetry energy of the EoS increases more steeply with density.

The fingerprints of LESA neutrino-emission asymmetries on the innermost
ejecta must also be expected to have implications for CCSN nucleosynthesis.
A detailed investigation of this possibility, based on 3D explosion models,
has still to be performed; for preliminary results making use of 2D SN
simulations, see \citet{Wanajo+2018}.

\subsubsection{Relevance of the nucleon mean field potentials}

The nuclear symmetry energy also plays a role for the charged-current reactions
of neutrinos with unbound nucleons, because these processes are modified by
the mean field potentials that describe the interactions of nucleons in dense media 
\citep[for comprehensive discussions, see, e.g.,][]{Fischer+2014,Hempel2015,Oertel+2017}. 
Within mean-field models the single-particle energies of nucleons are written as 
the sum of a kinetic part that depends on the effective nucleon mass $m_i^\ast$ 
($i = n,\,p$) and a mean-field interaction potential. In the nonrelativistic limit 
they are approximated by
\begin{equation}
    E_i \cong m_ic^2 + \frac{p_i^2}{2m_i^\ast} + U_i \,,
    \label{eq:singlenucenergy}
\end{equation}
and the nonrelativistic potential difference $\Delta U$ between neutrons and 
protons is given by the interaction part of the symmetry energy,
$E_\mathrm{sym}^{\mathrm{int}}$, and corrections depending on higher orders in
the asymmetry parameter $\delta = 1-2Y_p$:
\begin{equation}
    \Delta U = U_n - U_p = 
    4(1-2Y_p)E_\mathrm{sym}^\mathrm{int}(n_B) + {\cal O}((\Delta Y_p)^2) 
    \cong 4(1-2Y_p)E_\mathrm{sym}^{\mathrm{int},0}(n_B) \,,
    \label{eq:deltaU}
\end{equation}
where $\Delta Y_p = Y_p - 0.5$, the quantity $E_\mathrm{sym}^{\mathrm{int},0}$ 
is the interaction part of the symmetry energy at zero temperature, and 
higher-order terms in $Y_p$ are ignored in the last expression \citep{Hempel2015}.
Similarly, for the chemical potential difference one obtains in the same 
approximation \citep{Hempel2015,Most+2021}  
\begin{equation}
    \hat\mu = \mu_n-\mu_p \cong 
    (m_n - m_p)c^2 + 4(1-2Y_p)E_\mathrm{sym}^\mathrm{int}(n_B) \,.
    \label{eq:muhat}
\end{equation}
Both of the quantities in Equations~(\ref{eq:deltaU}) and (\ref{eq:muhat}) 
enter the calculation of the charged-current reactions of Equations~(\ref{eq:nuab1})--(\ref{eq:nuem2}). In this way the nuclear symmetry 
energy has a direct bearing on the neutrino emission of the PNS and thus,
indirectly, on the nucleosynthesis conditions in the CCSN ejecta.

In the neutron-rich conditions of the nascent NS, the mean-field potential 
difference between neutrons and protons is positive and can reach values up 
to several 10\,MeV. Due to this difference of the neutron and proton energies,
the energies of $\nu_e$ produced by electron captures are decreased and the 
energies of $\bar\nu_e$ created by positron captures are increased. Moreover, 
the mean-free-path for $\nu_e$ absorption
is reduced. Overall, the spectral differences between the electron neutrinos 
and antineutrinos radiated by the PNS are amplified. \citet{Martinez-Pinedo+2012}
and \citet{Martinez-Pinedo+2014}
as well as \citet{Roberts+2012b} and \citet{Roberts2012} found that this leads to 
relatively enhanced absorption of $\bar\nu_e$ in the neutrino-driven wind and
thus to slightly neutron-rich conditions at least during a transient period
of the wind evolution. This effect is, however, reduced or even erased by PNS
convection, because the convective flows carry electron lepton number from the
non-convective central core of the PNS to the PNS mantle layers and thus 
enhance the emission of $\nu_e$ relative to $\bar\nu_e$ 
\citep{Huedepohl2013,Groote2014,Mirizzi+2016,Pascal+2022}. Robust r-process
conditions in neutrino-driven winds of newly formed NSs are therefore not 
facilitated by the mean-field effects in the charged-current weak reactions.

\subsubsection{Closing remarks}

The EoS of hot PNS matter thus influences CCSN explosions and their ejecta 
properties in multiple ways, directly by determining the formation of either a NS
or a BH, and indirectly by governing the time when the explosion sets in, the
contraction of the new-born NS, and the characteristics of the neutrino emission 
of the compact remnant. In this section we reported on several examples how 
physical effects connected to the high-temperature nuclear EoS can affect the 
neutron-to-proton ratio in the innermost, neutrino-heated
SN ejecta. But also the explosion energy and large-scale explosion asymmetries
(e.g., connected to the LESA phenomenon) may depend on the nuclear EoS.

These questions are barely explored to date, mainly because the 
self-consistent, ab-initio 3D modelling of stellar collapse and explosion 
with elaborate neutrino transport methods has only recently become possible 
and such 3D simulations are still computationally demanding. Also the implications
of 3D explosions with detailed neutrino physics for CCSN nucleosynthesis have
still to be investigated. It is well possible that characteristic fingerprints
in the chemical abundances of special types of SNe could provide clues on
the high-temperature EoS in nascent NSs, for example the presence
of weak r-process elements in ECSN-like explosions (Section~\ref{sec:ECSNrprocess})
or the existence of third-peak r-process nuclei in CCSNe triggered by a hadron-quark 
phase transition (Section~\ref{sec:QCDSNe}). Such special signatures might yield
information on the finite-temperature EoS complementary
to the messages carried by the GW and neutrino signals that will be measured
for a future Galactic stellar collapse event and that will be deduced from
the post-merger ring-down GW and kilonova emission of nearby binary NSs.


\section{Compact-object mergers}\label{sec:com}

This part of the article addresses the merging of two compact objects and, in particular, covers the mass ejection and nucleosynthesis of these events. The different phases and the general dynamics of a merger are described in Section~\ref{sec:phases}. The various mass ejection processes are addressed in Section~\ref{sec:massejection}. Section~\ref{sec:rp} provides an overview on the rapid neutron-capture process (r-process). Messengers from COM and current observations are discussed in Sections~\ref{sec:mess} and~\ref{sec:obs}. The EoS impact on the ejecta and nucleosynthesis is detailed in Section~\ref{sec:eosimpact}. 

The presentation strives for a pedagogical introduction rather than a detailed review of the latest advances in the field or a description of the historical developments. The references in this handbook article are intentionally limited to a relatively small number to make it easier to navigate through and to present a selection as a start for more detailed reading. The reader is encouraged to follow up and to consult studies which could not fit here especially on more advanced or related topics which could not be addressed. Also, references are avoided for more general descriptions, e.g. in Sections~\ref{sec:phases} and~\ref{sec:rp}, where a number of review articles are available. More literature can also be found in these reviews, e.g.~\cite{Arnould+2007,Sneden2008,Thielemann2011,Fernandez+2016,Baiotti2017,Metzger2019,Horowitz2019,Siegel2019,Bauswein2019,Shibata2019,Radice2020,Ciolfi2020,Nakar2020,Perego2021,Cowan+2021,Kyutoku2021,Rosswog2022}.

\subsection{Merger Dynamics and Phases} \label{sec:phases}
\begin{figure}
\begin{center}
\includegraphics[width=0.8\columnwidth]{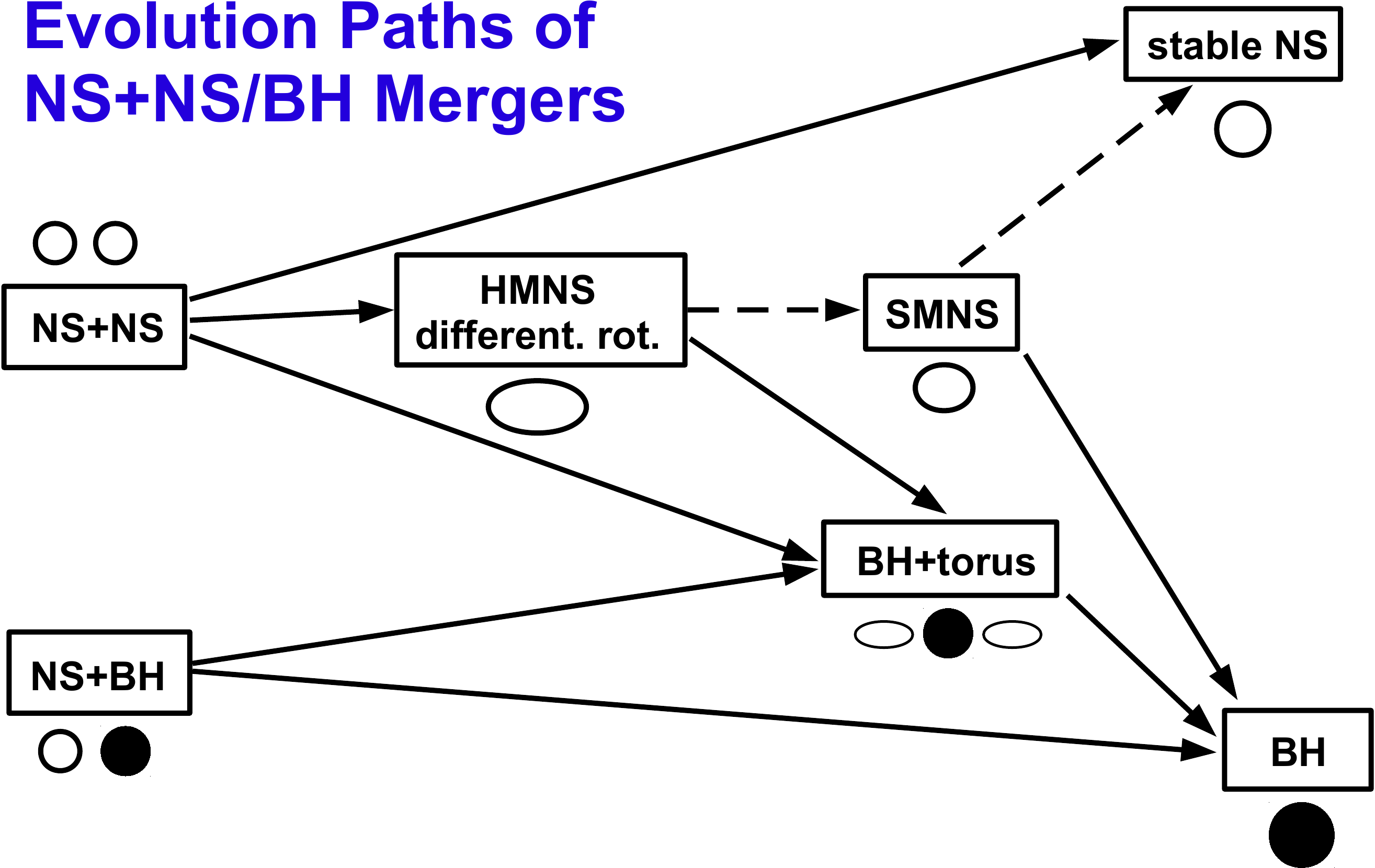}
\end{center}
\caption{Evolutionary paths of COMs. The fate of a compact object binary depends predominantly on its mass and the high-density EoS. The evolution involves many different time scales. SMNS refers to supermassive NSs, i.e. objects with a mass that exceeds the maximum mass of non-rotating NSs and which are stabilized against the gravitational collapse by rigid rotation. Differential rotation can provide even more stability, and differentially rotating neutrons stars whose mass exceeds the maximum mass of SMNSs are call hypermassive (HMNS). See main text for explanations. Figure from~\cite{Just+2015}; \copyright Oxford University Press on behalf of the Royal Astronomical Society (RAS). Reproduced with permission}
\label{fig:phases}
\end{figure}

\begin{figure}
\begin{center}
\includegraphics[width=0.8\columnwidth]{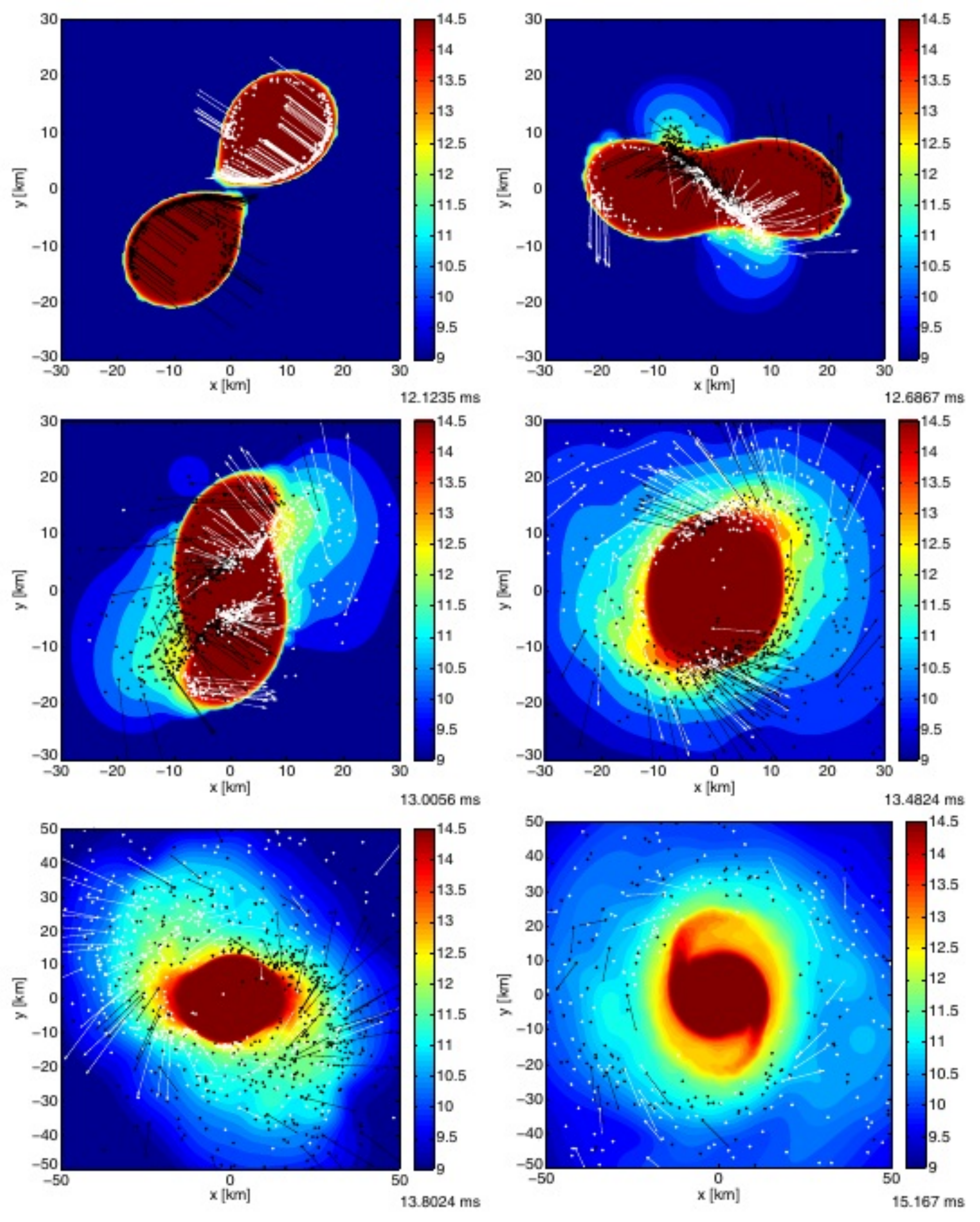}
\end{center}
\caption{Rest-mass density evolution (color-coded is the $log_{10}\rho$ with $\rho$ in $\mathrm{g/cm^3}$) in the equatorial plane of a 1.35-1.35~$M_\odot$ NS merger with the DD2 EoS~\citep{Hempel+2010,Typel+2010}. Dots trace fluid elements which become gravitationally unbound until the end of the simulation. Arrows indicate the current velocity of the fluid elements. Figure from~\cite{Bauswein+2013}; \copyright American Astronomical Society. Reproduced with permission}
\label{fig:snap}
\end{figure}

The merging of two NSs proceeds through different phases and can produce different final objects. Various physical mechanisms govern the evolution in these stages of the merger and shape the observable signals. Figure~\ref{fig:phases} provides an overview, and Fig.~\ref{fig:snap} illustrates the evolution by snapshots from hydrodynamical simulations of the early dynamical merger phase.

{\bf Binary evolution:} The merger of NSs in a binary system is the result of continuous GW emission, reducing the angular momentum and energy of the system and leading to an ``inspiral'' of the binary components, i.e. a continuous and accelerated decline of the orbital separation. The inspiral is driven by point-particle dynamics, where post-Newtonian effects and general relativistic effects, respectively, become increasingly important as the binary components come closer and thus speed up. In the final phase of the inspiral, the orbital period decreases to a few milliseconds and the stellar components experience tidal deformations during the last revolutions. In general, the evolution is dominantly determined by the binary masses of the system and to some extent by the spins and the finite size of the stellar components. 

{\bf Merger dynamics:} The stars finally coalesce with a relatively large impact parameter since the system features a significant amount of angular momentum. For typical NS masses of 1.3 to 1.4~$M_\odot$~\citep[see e.g.][]{Lattimer2012} the total mass of the system will likely exceed the maximum mass $M_
\mathrm{max}$ of non-rotating NSs. The remnant may however avoid the gravitational collapse because of the rapid rotation. During the coalescence temperatures rise and can reach several 10~MeV providing additional thermal pressure support. Only if the total mass exceeds a threshold binary mass $M_\mathrm{thres}$, the merger remnant collapses into a BH within less than a millisecond. In the case of such a ``prompt collapse'' most of the matter is swallowed by the forming event horizon but material with sufficient angular momentum forms a torus around the central BH. The exact threshold mass depends on the EoS and is not precisely known. It is also influenced by the binary mass ratio and the intrinsic spins of the binary components if sufficiently large. For currently viable EoS models $M_\mathrm{thres}$ ranges from about 2.8~$M_\odot$ to 3.5~$M_\odot$~\citep[][]{Bauswein2021}. Considering this range the likely outcome of a merger with typical binary masses of 2.7~$M_\odot$ is the formation of a rotating NS remnant.

{\bf Remnant evolution:} The merger remnant is initially strongly deformed and oscillating. The cores of the binary components form a rotating double-core structure with a shearing layer in between, which is created during the first contact of the stars (see Fig.~\ref{fig:snap}). In this contact layer the Kelvin-Helmholtz instability develops and highest temperatures are found there. The collision excites significant quasi-radial and quadrupolar oscillations of this remnant structure. During the merging process tidal arms form at the outer edges of the stars. For asymmetric binaries symmetry is broken and one primary tail forms from the tidally elongated lighter companion star, which is wrapped around the more massive component. 

The tidal arms settle into a more axisymmetric inflated torus around the central region. The initially very complex velocity field of the double-core structure evolves into an axisymmetric differentially rotating object, while the oscillations are damped on time scales of several 10 milliseconds. After the initial dynamical merger stage a quasi-equilibirum is reached and the remnant enters a phase of secular evolution. Possibly, an $m=1$ one-armed spiral instability develops, which breaks axisymmetry and which can persist for many dynamical time scales~\citep{Paschalidis2015}. Literally, the remnant develops a ``bump'' which is wobbling around. 

The evolution of the system is driven by angular momentum redistribution mostly by magnetic fields (generating turbulent motion) and losses of energy and angular momentum through GWs and neutrinos. These processes will drive the central remnant towards a state of uniform rotation, while centrifugal and thermal support is drained from the inner core. Redistribution and losses of angular momentum and energy destabilize the remnant, which can trigger a delayed gravitational collapse. This delayed BH formation can take place soon after merging, i.e. in the dynamical phase of the remnant evolution, or on longer secular time scales. The life time of the remnant depends sensitively on the total mass, i.e. how close it is to the threshold mass for prompt BH formation. Obviously, the binary mass ratio also affects the life time as it determines the total amount of angular momentum in the remnant and its initial distribution throughout the central object. If a delayed collapse occurs, matter of the outer parts of the remnant with sufficient centrifugal support will form a torus around the central BH. 

{\bf Torus evolution:} As in the case of a prompt collapse, the BH torus system created after a delayed collapse will undergo further evolution. Again, this stage is driven by angular momentum transport by viscous and magnetic effects in combination with neutrino radiation cooling of the torus. (Note that the literature often uses the term disk and torus interchangable albeit the geometry is closer to an inflated torus rather than a thin disk). As a result of this evolution the largest fraction of the initial torus mass is accreted by the BH. However, as angular momentum is transported outwards and various processes heat the torus, a substantial amount of matter becomes gravitationally unbound (see discussion below). Typical time scales of the secular evolution before or after a delayed collapse can be coarsely estimated to be of the order of one second (times scale of angular momentum redistribution in the remnant NS, accretion time scale of a viscous torus and the cooling time scale). 

{\bf Remnant classification:} For not too large total binary mass the system may avoid the gravitational collapse until it reaches a phase of uniform rotation. Ultimately, centrifugally supported torus material could partly fall back onto the central NS remnant such that an isolated, uniformly rotating, cold NS is left on longer time scales. The literature has established the term ``supermassive'' NS (SMNS) for a system with the mass exceeding $M_\mathrm{max}$ but stabilized by {\it uniform} rotation (see~\cite{Paschalidis2017} for a review on rotating NSs). Uniform rotation can stabilize stars up to roughly 1.2~$M_\mathrm{max}$, where fast rotation (and thus more stabilization) is limited by mass shedding from the surface. Differential rotation allows for faster rotation in the center and can thus support more mass. Stars with a mass beyond the threshold of uniformly rotating NSs of about 1.2~$M_\mathrm{max}$ are referred to as ``hypermassive''~\citep{Baumgarte2000}. The terms ``supermassive'' and ``hypermassive'' are often used in the context of merger remnants with hypermassive remnants being systems which cannot be stabilized by uniform rotation and which thus undergo a delayed collapse in contrast to supermassive remnants where uniform rotation is sufficient to prevent BH formation.

However, it is conceivable that a supermassive remnant, i.e. a uniformily rotating NS, will be magnetized, leading to magnetic dipole emission, and thus its rotation slows down (similar to the known pulsar spin-down). Hence, if the mass of the object exceeds $M_\mathrm{max}$, it may also collapse into a BH (Fig.~\ref{fig:phases}). Only for very light binary systems the final object may be a stable NS, i.e. if the remnant's mass after all mass loss epsiodes during the secular evolution (to be discussed in more detail below) is smaller than the maximum mass of non-rotating NSs. $M_\mathrm{max}$ is probably not significantly above 2.0~$M_\odot$ implying that at most the lightest binaries will result in a stable NS, e.g.~\citep{Lattimer2012,Oertel+2017,Margalit+2017,Rezzolla+2018,Ruiz+2018,Shibata2019a}.

{\bf NS-BH mergers:} This article focuses on NS mergers, but large parts of the discussion apply to NS-BH mergers as well. For these systems, it is reasonable to expect that the BH will be significantly more massive than the NS, and thus the binary mass ratios will be more extreme as compared to NS mergers. Also, for these systems the spin of the BH can have a significant impact on the dynamics, because it can be larger than that of NSs, which is usually assumed to be small~\citep{Tauris2017}.

There can be two fundamentally different outcomes of a NS-BH merger depending on the EoS and the exact binary parameters including the BH spin~\citep[e.g.][]{Janka1999,Janka2002,Rosswog2005,Shibata2008,Duez2010,Foucart2011,Kyutoku2011,Etienne2012,Foucart2013,Foucart2014,Bauswein2014,Kyutoku2015,Kyutoku2021}. If the binary mass ratio is not too extreme, the NS is tidally elongated and disrupted. The tidal arm forms a torus around the BH and some matter from the tip of the tidal tail may become gravitationally unbound. For more extreme mass ratios, however, the NS may be swallowed by the BH as a whole and no matter will remain outside the BH.

For a fixed mass ratio, systems with a faster spinning BH lead to more matter outside the BH. Generally, tidal effects are more pronounced in NS-BH mergers as compared to NS mergers because of the more extreme binary mass ratio, which leads to a more pronounced tidal disruption of the NS. For favorable binary parameters (high BH spin, less extreme mass ratio) a significant amount of matter may become gravitationally unbound during the disruption, i.e. on a dynamical time scale. As in the case of NS mergers, a substantial amount of matter may be ejected on secular time scales from the torus.

\subsection{Mass ejection}\label{sec:massejection}
As described above NS mergers can take different evolutionary paths and result in different final objects, depending on the high-density EoS, the total binary mass, the binary mass ratio and the spins. In particular the duration and detailed evolution of the different stages are determined by these dependencies~\citep{Fernandez+2016,Baiotti2017,Metzger2019,Siegel2019,Bauswein2019,Radice2020,Shibata2019,Bernuzzi2020a,Nakar2020,Perego2021,Rosswog2022}. For a significant range of binary masses an initially very dynamical, violent merger evolves towards a more secular phase, which at some point undergoes another dramatic change when the NS remnant collapses to a BH. After most of the remnant has been swallowed by the BH on very short time scales ($<\mathrm{ms}$), the BH torus system again enters a long-term evolution until all mass is either expelled or accreted by the remaining BH. During all these stages matter is ejected from the system, i.e. it becomes gravitationally unbound~\citep{Ruffert1997,Rosswog1999,Rosswog2000,Oechslin2007,Metzger2009,Metzger2008,Dessart2009,Goriely2011,Korobkin2012,Fernandez2013,Bauswein+2013,Hotokezaka2013,Perego2014,Siegel2014,Metzger2014,Just+2015}. 

This scenario can serve as a prototype for the discussion of mass ejection bearing in mind that for higher total binary masses the remnant may collapse promptly or on relatively short time scales into a BH. In this case the dynamical and secular mass ejection from the NS remnant is either missing or limited by its life time. In contrast, a mass ejection episode from a BH torus configuration may be absent for very light binaries which avoid the gravitational collapse completely or sufficiently long. These basic considerations already exemplify that the characteristics and relative contributions of the different mass ejection episodes strongly depend on the parameters of the merger like binary masses and EoS, which determine its evolution and outcome.

Also, the mass ejection from NSBH mergers follows qualitatively the aforementioned stages with a strong tidal component producing dynamical ejecta during merging and the formation of a torus from which secular ejecta originate~\citep{Janka1999,Janka2002,Rosswog2005,Foucart2013,Bauswein2014,Foucart2014,Kyutoku2015,Kyutoku2021}. Obviously, a phase of a quasi-stable NS remnant is missing in these systems. 

\begin{figure}
\begin{center}
\includegraphics[width=0.8\columnwidth]{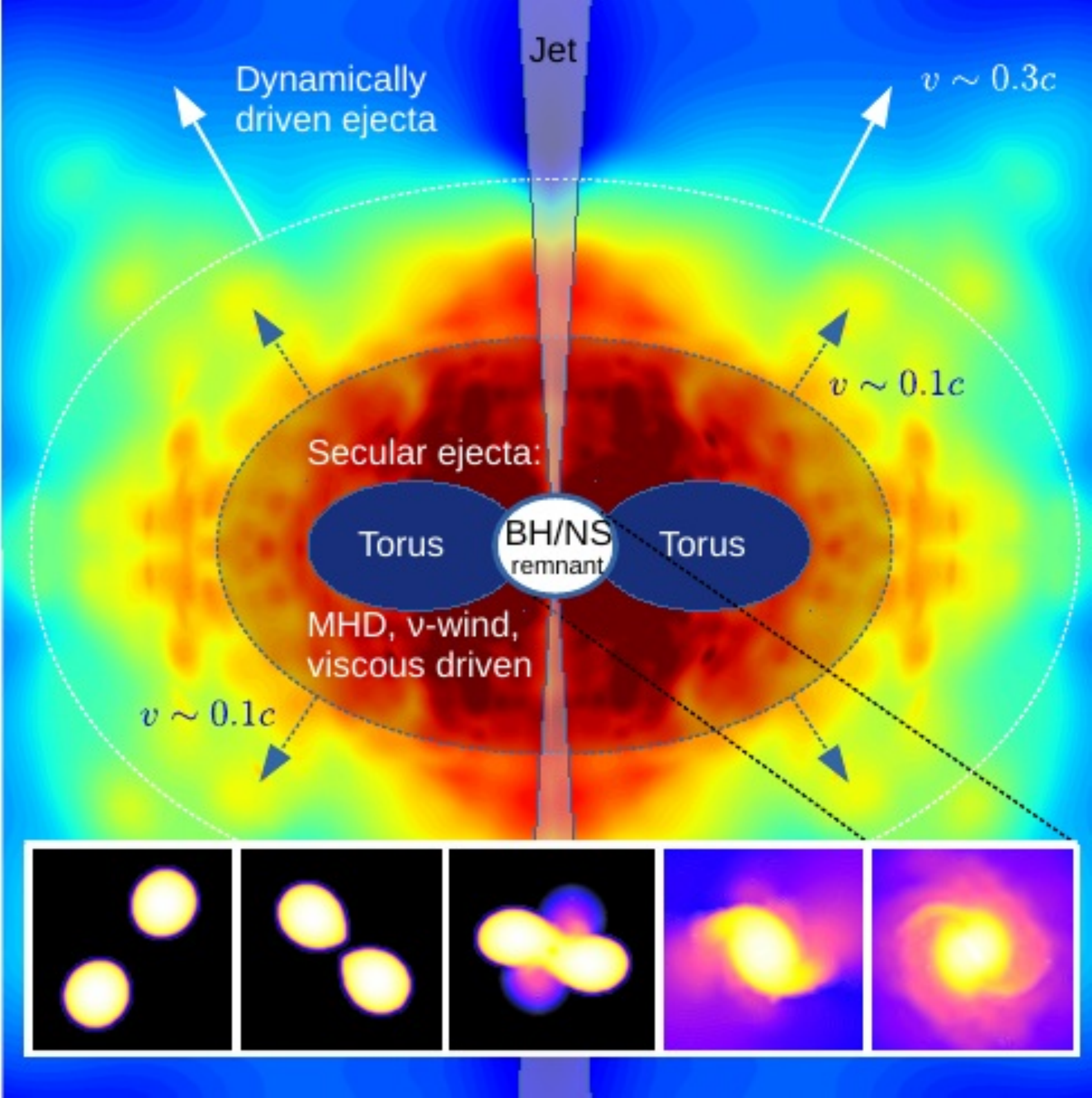}
\end{center}
\caption{Overview of different ejecta components and ejection mechanisms. Dynamical ejecta are launched first in a quasi-spherical manner with relatively high velocities of several ten per cent of the speed of light. Hence the dynamical component envelopes the secular ejecta, which typically feature somewhat lower velocities. If a jet emerges from the central object, it drills through the previously ejected material}
\label{fig:sketch}
\end{figure}

The literature often distinguishes dynamical ejecta and secular, long-term ejecta, i.e. material becoming unbound on longer time scales and at a lower rate. The latter ejecta component is sometimes referred to as postmerger ejecta or disk/torus ejecta if it originates from the torus. The classification of dynamical and secular ejecta is not very strict and mostly refers to the different stages of the merger. In particular, it does not specify a specific ejection mechanism in detail yet. In fact, several mechanisms contribute to unbind matter during the different stages of the merger (Fig.~\ref{fig:sketch}).

{\bf Dynamical mass ejection:} Figure~\ref{fig:snap} displays the phases of the early merger evolution by snapshots of the density distribution in the equatorial plane from a relativistic hydrodynamical simulation~\citep[][which includes a description of the basic ejection mechanisms]{Bauswein+2013}. The calculation follows the evolution of two stars with a mass of 1.35~$M_\odot$ and adopts the DD2 EoS~\citep{Hempel+2010,Typel+2010}. The dots trace fluid elements which eventually become gravitationally unbound during the dynamical evolution of the first 10~ms after merging. A majority of the ejecta originates from the contact interface between the two stars (upper right panel). This matter is shocked in the collisional layer and squeezed out in all directions, slipping over the respective other star (middle left panel). Right after the collision, the NS remnant bounces back and while expanding and rotating strongly, it pushes into the surrounding matter previously squeezed out and thereby, it unbinds a first massive wave of ejecta material (middle right panel). The rotating remnant undergoes several more cycles of contraction and expansion with decreasing amplitude, and thus the ejection of matter continues. Also, spiral arms form at the outer edges of the remnant and run into the surrounding low-density material (lower right panel). For rather asymmetric binary systems a sizable fraction of the ejecta stems from the tip of the massive tidal tail, which developed from the disruption of the lighter companion star.

These mechanisms can unbind several $10^{-3}~M_\odot$ up to a few 0.01~$M_\odot$ within only a few milliseconds, hence the term ``dynamical'' ejecta~\citep[e.g.][]{Rosswog1999,Ruffert2001,Oechslin2007,Goriely2011,Korobkin2012,Hotokezaka2013,Bauswein2013,Palenzuela2015,Sekiguchi2015,Sekiguchi2016,Kastaun2016,Foucart2016,Lehner2016,Radice2016a,Ciolfi2017,Dietrich2017a,Dietrich2017b,Bovard2017,Radice+2018,ArdevolPulpillo2019,Vincent2020,Combi2022}. Since the remnant has not yet reached an equilibrium state, it continues to inject energy and angular momentum into its envelope, which can contribute to the mass ejection at a lower rate. For instance, the quadrupolar oscillation of the remnant persists for several 10 milliseconds and possibly a long-lived $m=1$ deformation develops~\citep{Paschalidis2015}. This may increase the ejecta mass by several  $10^{-3}~M_\odot$ on time scales of several 10~milliseconds if the instability and deformation persist sufficiently long and with sufficient strength~\citep{Nedora2019,Nedora2021}.

Obviously these mechanisms can only operate as long as the remnant NS does not collapse and they are absent if the merger leads to direct BH formation. If a BH promptly forms, dynamical mass ejection is greatly reduced as the support from the NS remnant is missing~\citep{Bauswein+2013,Hotokezaka2013,Kiuchi2019}. Only small amounts of ejecta 
directly emerge from the collision interface. For direct-collapse cases of asymmetric mergers the tidal ejection may lead to unbound matter of several $10^{-3}~M_\odot$ depending on the binary mass ratio. 

The tori forming after a prompt gravitational collapse are less massive in comparison to systems that undergo a delayed collapse~\citep[e.g., ][]{Hotokezaka2011,Just+2015,Radice+2018,Kiuchi2019}. Again, unequal-mass mergers lead to more torus material as more material is shed from the tidal arm of the lighter companion. Generally, the reduced torus mass in prompt collapse events implies that there is less secular mass ejection from these systems (see discussion below).

\begin{figure}
\begin{center}
\includegraphics[width=0.8\columnwidth]{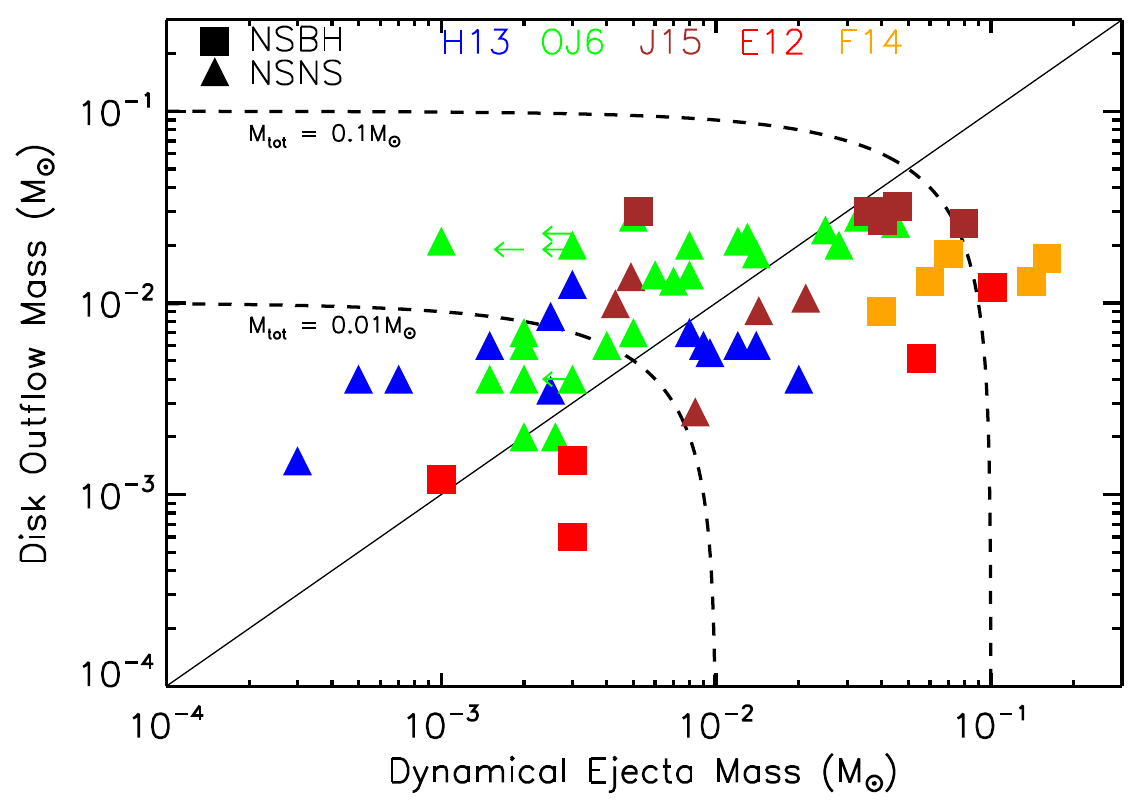}
\end{center}
\caption{Compilation of simulation results for the dynamical ejecta and unbound matter from a torus. Figure from~\cite{Wu+2016}; \copyright Oxford University Press on behalf of the Royal Astronomical Society (RAS). Reproduced with permission}
\label{fig:dynsec}
\end{figure}

{\bf Secular ejecta:} 
In systems that avoid the gravitational collapse, additional effects contribute to the unbound mass on longer time scales of hundreds of milliseconds to several seconds. The NS remnant featuring temperatures of several 10~MeV emits copious amounts of neutrinos of all flavors with luminosities of the order of $10^{53}~\mathrm{erg/sec}$~\citep{Ruffert1997,Rosswog1999,Sekiguchi2011,Sekiguchi2015,Palenzuela2015}. A fraction of these neutrinos is reabsorbed in the outer layers of the remnant depositing sufficient energy to drive a wind~\citep{Duncan+1986,Dessart2009,Perego2014,Martin2015,Fujibayashi2017}. This wind ejection will be more pronounced closer to the rotational axis because for geometric reasons the neutrino emission is most intense above and below the oblate-shaped remnant, i.e. towards the polar directions. The neutrino driven ejecta can amount to several $10^{-3}~M_\odot$ or more \citep[see, e.g., ][]{Perego2014,Fujibayashi2020,Nedora2021} while the exact quantity will strongly depend on the life time of the remnant. This ejecta component is less neutron rich because the rate of electron-neutrino absorptions is close to the rate of electron-antineutrino absorptions, resulting typically in a comparable number of neutrons and protons. It should be emphasized that it is not straightforward to unambiguously identify this ejecta component.

Energy deposition by neutrinos can also play a role in the subsequent BH-torus phase after the gravitational collapse of the remnant~\citep{Ruffert1999,Fernandez2013,Just+2015}. Naturally, neutrino-driven wind ejecta are significantly more massive during the life time of the NS remnant compared to the BH-torus phase, simply because the NS is much more massive and hotter, and therefore a  stronger source of neutrinos than the torus.

Long-term simulations of the magnetized NS remnant indicate the possibility of winds driven by magnetic forces, although various mechanisms are discussed and details of the ejection depend sensitively on the structure and strength of the magnetic field, which is an active field of research~\citep{Siegel2014,2018ApJ...856..101M,Kiuchi2018,Ciolfi2020a,Moesta2020}. For instance, in the case that a strong, large-scale helical field develops in the NS remnant, it could drive a polar outflow that can become as massive as $\sim0.01~M_\odot$~\citep{Ciolfi2020a}.
Magnetohydrodynamic simulations suggest that the role of magnetic fields in driving outflows depends to some extent on the geometry and strength of the initial magnetic fields carried by the merging NSs, which is not exactly known and may even vary in different binary systems. 

At any rate magnetic fields are important because they govern the secular evolution of the torus before and after BH formation~\citep[e.g.][]{Siegel2018,Christie2019,Fernandez2019,Miller2019,MurguiaBerthier2021,Hayashi2022,Fahlman2022,Just2022a}. A major source of ejecta stems from the torus either around the central NS or the BH if the former collapsed~\citep[e.g.][]{Surman2008,Metzger2009,Lee2009,Fernandez2013,Perego2014,Metzger2014,Just+2015}. 
Since matter in the torus is in nearly Keplerian motion and magnetized, the flow is expected to be unstable to the so-called magneto-rotational instability (MRI)~\citep{Balbus1998}. The MRI instigates turbulent matter motion, which drives angular momentum transport and generates heat through dissipation of kinetic into thermal energy~\citep[e.g.][]{Siegel2018,Fernandez2019,Just2022a,Hayashi2022}. Additionally, the recombination of nucleons into helium and heavier nuclei releases energy, which heats the material~\citep{Lee2005,Metzger2009}.

These effects will unbind a significant fraction of the torus of about 10 to 40 per cent over the life time of the torus of seconds. Considering a typical torus mass of 0.1~$M_\odot$~\citep[e.g.][]{Ruffert1999,Oechslin2006,Shibata2006,Hotokezaka2013}, the torus or disk ejecta can be massive and in fact dominate other mass ejection channels. Figure~\ref{fig:dynsec} compiles data for the torus ejecta and the dynamical ejecta from different simulations providing an overview of typical numbers~\citep{Wu+2016}.

Since MRI-induced turbulence behaves like a viscosity when averaged over turbulent fluctuation length scales, the ejecta driven by this ``turbulent viscosity'' are often called viscously driven ejecta. The approximate similarity between a laminar and an MRI-turbulent flow is the reason why in practice many studies adopt the method of viscous hydrodynamics instead of a more complex magnetohydrodynamic description to model accretion disks in numerical simulations \citep[e.g.][]{Fernandez2013,Just+2015,Fujibayashi2020a}. The literature has established different terms referring to this type of mass ejection, and uses synonymously disk ejecta, torus ejecta, viscous ejecta, torus wind or more generally just post-merger ejecta, long-term ejecta, or secular ejecta. As indicated previously, hot accretion tori can also launch neutrino-driven winds, though with significant masses only for very massive tori with $M\gtrsim 0.1~M_\odot$. It should also be mentioned that viscous hydrodynamics are employed to study the evolution of massive NS remnants and their ejecta \citep[e.g.][]{Radice2018a,Fujibayashi2018}.

It is needless to say that the properties of the torus ejecta such as their mass do not only depend on the secular evolution of the torus but on the dynamical formation of the torus in the first place setting its initial conditions. This implies that the dynamical merger phase and the secular mass ejection processes are tightly interconnected and are not independent from each other, albeit many studies focus on individual ejecta components because of the tremendous complexity and computational challenge to model the different phases.

A {\bf word of caution} is in order with regard to the ejecta masses and other ejecta or torus properties quoted here and in the literature in general. Most of the understanding of the mass ejection processes of NS mergers and their remnants relies on numerical simulations. These calculations model the relevant physics with different degrees of sophistication and employ different numerical schemes, details of which are beyond the scope of this text~\citep[see e.g.][]{Baumgarte2010,Rezzolla2013,Marti2015,Rosswog2015,Shibata2016,Duez2019}. Even by the aid of supercomputers it is challenging to achieve high numerical resolution to resolve for instance the steep density gradients of NSs, turbulent matter motion or the detailed structure and evolution of mass outflows, which typically expand away from the central object, i.e. into regions of coarser numerical resolution. Also, numerical simulations typically follow the evolution of a merger and its remnant only for a finite time, and hence ejecta properties can only be monitored during the simulated time but cannot be followed until the outflow is completely released from the system and reaches its final (so-called homologous) structure. This bears potentially two issues: matter becoming unbound after the end of the simulation is not taken into account and the analysis of ejecta relies on criteria to estimate which fluid elements would in fact become unbound. Typical choices are the assumption of ballistic motion or criteria which consider that internal energy may be converted to kinetic energy during the outflow (Bernoulli criterion)~\citep[e.g.][]{Foucart2021}. However, these criteria do not reliably model ejecta physics beyond the simulation time, e.g., the collision of a faster ejecta cloud catching up with slower moving material, or energy injection on longer time scales by nuclear heating or other processes~\citep[e.g.][]{Grossman2014,Rosswog2014,2018ApJ...856..101M,Nativi2021,Klion2022,Neuweiler2022,Haddadi2022}. Unless BH formation is directly simulated, the analysis of torus properties is affected by similar limitations, e.g. different criteria to approximately identify torus material before BH formation actually took place~\citep{Oechslin2006,Shibata2006}. 

In essence, all these considerations indicate that numbers quoted in the literature characterizing ejecta properties should be taken with a grain of salt. Uncertainties of, for instance, ejecta masses may coarsely be estimated to be at least some tens of per cent or possibly even more than a factor 2. This may be inferred by comparing calculations of the same system with different numerical resolutions or the results by different groups using different codes~\citep[e.g.][]{Bauswein+2013,Hotokezaka2013,Dietrich2017a,Radice+2018}. Similarly, other ejecta properties may be affected by significant uncertainties including velocities, temperatures and the composition. For instance, the distribution of the electron fraction $Y_e$ as the crucial parameter determining the neutron-richness and thus the details of the r-process (see below), is influenced by the approximate treatment of neutrino radiation, incomplete or approximate sets of weak interaction rates, and the not fully reliable determination of the fluid temperature evolution, since weak reaction rates typically scale with high powers of the temperature~\citep[e.g.][]{Foucart2018,Just2022a}.

{\bf Further ejecta properties:} It is common to all ejecta components that the matter expands very rapidly with up to several ten per cent of the speed of light~\citep[e.g.][]{Hotokezaka2013,Bauswein2013,Fernandez2013,Perego2014,Siegel2014,Metzger2014,Just+2015,Fujibayashi2017,Radice+2018}. This scale may not be surprising considering that the material needs to overcome the enormous gravitational attraction of the compact central object (the escape velocity from an isolated NS is several ten per cent of the speed of light). The different ejecta components usually feature a broad velocity distribution with dynamical ejecta typically being the fastest component. Being ejected first and quasi-spherically, i.e. in all directions, the outermost layers of the outflow are likely to be comprised of dynamical ejecta. 

The majority of ejecta originates from deeper crust layers of the NSs, which are very neutron rich. Note that the less neutron-rich outer crust of a NS is not very massive ($<10^{-4}~M_\odot$) and can thus not contribute significantly to the unbound matter. The dynamical ejecta for instance mostly stem from the inner crust of the NSs with densities of about $10^{14}~\mathrm{g/cm^3}$, which has a proton fraction of less than five per cent, i.e. an electron fraction of $Y_e<0.05$ adopting charge neutrality. The initial electron fraction is set by the condition of neutrino-less beta-equilibrium in the progenitor stars, i.e. by the balance of weak interactions. The initial composition, however, can change significantly by weak interactions. Even if matter expands fast like the dynamical ejecta, weak reactions can be sufficiently fast to significantly alter the initial electron fraction~\citep[e.g.][]{Fernandez2013,Perego2014,Metzger2014,Wanajo2014,Sekiguchi2015,Just+2015,Goriely2015,Palenzuela2015,Wu+2016,Foucart2016,Radice2016a,Bovard2017,Fujibayashi2017,Papenfort2018,Radice+2018,Martin2018,ArdevolPulpillo2019,Miller2019,Fujibayashi2020,Fujibayashi2020a,Just2022a,Combi2022}.

The most relevant weak interactions for setting the $Y_e$ in outflows from COMs are given by Eqs.~(\ref{eq:nuab1}) to~(\ref{eq:nuem2}). During the violent collision in a NS merger, temperatures reach up to several 10 MeV, i.e. well above the rest mass of electrons and positrons. Under these conditions numerous electron-positron pairs are created, which can be captured by the protons and neutrons. Right after the sudden increase of the temperatures when both NSs coalesce, the newly created positrons are captured by neutrons. As there are more neutrons available, positron captures are favored and the initially low electron fraction increases. In addition, the matter is exposed to intense neutrino radiation. As a consequence, this material reabsorbs a large number of electron neutrinos and electron anti-neutrinos, which typically drives the electron fraction towards high values.

Usually the $Y_e$ of most of the ejecta remains below 0.5 for all ejecta components, while the exact impact of weak interactions obviously depends on the (thermodynamic) history of the ejected fluid elements and thus on the mass ejection channel and its details.

For instance, the electron fraction in the dynamical ejecta shows a broad distribution ranging from about 0.05 to 0.45~\citep[e.g.][]{Wanajo2014,Sekiguchi2015,Goriely2015,Palenzuela2015,Foucart2016,Radice2016a,Bovard2017,Radice+2018,Papenfort2018,Martin2018,ArdevolPulpillo2019,Combi2022}. During the dynamical mass ejection material expanding along the poles is exposed to stronger neutrino irradiation as compared to the equatorial outflow (which is shielded by the massive torus). Consequently, polar ejecta are less neutron rich with $Y_e$ being higher by roughly 0.1 compared to the equatorial ejecta. This can have a significant impact on the nucleosynthesis in the outflows. 

Predominantly neutrino-driven ejecta along the poles are usually less neutron rich because the reabsorption of neutrinos raises the electron fraction in the wind to values above $\sim 0.3 $ up to even 0.5~\citep[e.g.,][]{Perego2014,Just+2015,Fujibayashi2017,Fujibayashi2020a}.

Weak interactions also play a very important role in hot accretion tori and in particular to regulate the $Y_e$ of the corresponding ejecta~\citep{Fernandez2013,Just+2015,Siegel2018,Miller2019,Fujibayashi2020a,Fujibayashi2020,MurguiaBerthier2021,Just2022a}. The conditions in a typical accretion torus are such that the electron fraction of the torus matter is roughly determined by a beta-equilibrium of the rates given in Eqs.~\ref{eq:nuab1} and~\ref{eq:nuem2}. Since electrons are mildly degenerate in such a torus, positrons are suppressed, resulting in a beta-equilibrium characterized by moderately neutron rich conditions with $Y_e~\sim 0.1-0.3$. Hot accretion tori emit numerous neutrinos, which is essential in reprocessing the outflowing material and setting its composition. During the expansion from the torus, material is subject to neutrino emission and absorption reactions, which both tend to increase $Y_e$. As a result, the values of $Y_e$ in the ejecta are higher than in the torus, and typically distributed within $0.25 \lesssim Y_e \lesssim 0.5$. 
See~\cite{Just2022a} for a detailed discussion of the evolution of accretion tori and the conditions in their secular ejecta.

In~\cite{Metzger2019,Nakar2020} tables can be found which summarize various properties of the different ejecta components and provide an overview.

\subsection{Nucleosynthesis} \label{sec:rp}
All ejecta components provide favorable conditions for the r-process creating heavy elements through successive captures of neutrons onto seed nuclei~\citep{Burbidge1957,Cameron1959}. See~\cite{Lattimer1976,Lattimer1974,Symbalisty1982,Eichler1989} for the first suggestions of the r-process taking place in the ejecta of COMs,  and~\cite{Freiburghaus1999,Metzger2010,Roberts2011,Goriely2011,Korobkin2012,Bauswein+2013,Perego2014,Just+2015,Wu+2016} for some first r-process calculations based on simulation data. 
During the outflow basically the entire ejecta are converted to heavy nuclei of different nuclear mass numbers between $A=60$ and $A=240$ including elements like silver, gold, platinum and uranium. See, e.g.,~\cite{Arnould+2007,Thielemann2011,Horowitz2019,Arnould2020,Cowan+2021,Siegel+2022} for reviews on the r-process. The r-process is responsible for the formation of about half of all heavy elements beyond the iron group. This explains the significant scientific efforts in understanding this nucleosynthesis process and especially in elucidating the astrophysical production sites, which has been a major puzzle for decades with CCSNe being the most popular candidate for many years but less favored in the last decade (see Section~\ref{sec:ccsn}).

Ejecta outflows from mergers expand very rapidly with up to a few ten per cent of the speed of light resulting in a rapid drop of the density by several orders of magnitude in the first few milliseconds, which is accompanied by a decrease of the temperature. Once the temperature falls below $\sim 1$~MeV (10 Giga-Kelvin), protons recombine with some of the neutrons to form seed nuclei. Because of the neutron-rich conditions there are hundred or more neutrons per seed left. The neutron density is so high that neutron captures are the dominant reaction, i.e., they are faster than $\beta$-decays and any seed nucleus becomes heavier and increasingly neutron-rich. The crucial role of this reaction explains the term {\it rapid} neutron-capture process. Notably, neutron captures are critical to form heavy nuclei because they do not experience the strong Coulomb repulsion of heavy nuclei, which prevents positively charged particles to be captured.

\begin{figure}
\begin{center}
\includegraphics[width=0.8\columnwidth]{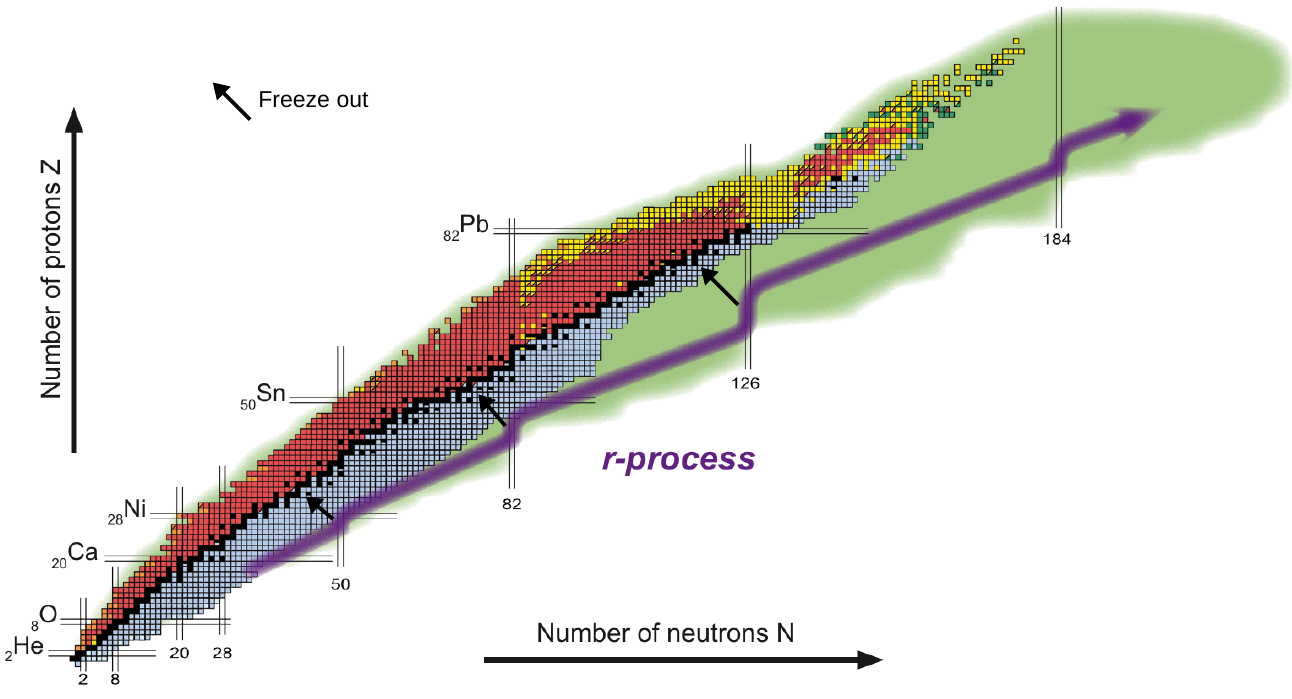}
\end{center}
\caption{Chart of the nuclei with the blue arrow indicating the path of the r-process crossing closed neutron shells at $N=50$, $N=82$ and $N=126$. The black arrows visualize the nuclear flow during the ``freeze out'' towards the valley of stability by $\beta^-$ decays. Colors of individual nuclei display dominant decay channels: blue $\beta^-$ decay, red $\beta^+$ decay, yellow $\alpha$ decay, black stable nuclei. The green area roughly indicates unstable nuclei which have not yet been measured. In the vicinity of the path of the r-process these nuclei decay predominantly by $\beta^-$ decays or fission for high mass numbers $A=N+Z$. Figure adapted from EMMI, GSI / Different Arts}
\label{fig:chart}
\end{figure}

Nuclei cannot become arbitrarily neutron-rich because for too large neutron excess it is energetically unfavorable to bind additional neutrons. In the chart of the nuclei (see Fig.~\ref{fig:chart}), seed nuclei ``move'' by successive neutron captures far away from the valley of stability to the very right, i.e. close to the so-called neutron drip line, beyond which no nuclei exist and no additional neutron captures can take place. Nuclei in the vicinity of the neutron drip line are highly unstable and quickly undergo a $\beta^{-}$-decay, i.e. a neutron inside a nucleus is converted to a proton. In the chart of the nuclei, a $\beta^{-}$-decay moves the nucleus one step diagonally to the upper left, i.e. a little away from the neutron drip line. Being away from the neutron drip line, neutron captures become possible again and will immediately take place as neutron captures are the dominant reaction in the neutron-rich environment of NS merger ejecta. By a sequence of these two consecutive reactions, i.e. neutron capture and $\beta^{-}$-decay, the seed nuclei will move along the neutron drip line to higher and higher mass number as long as enough neutrons are available for frequent neutron captures. 

Once neutrons are exhausted and densities drop, the neutron capture rate decreases. At this point $\beta^{-}$-decays become the dominant reactions. Hence, the ensemble of nuclei that has been built up along the neutron drip line, moves diagonally to the upper left towards the valley of stability. The $\beta^{-}$-decay rates are faster further away from the valley of stability and the ``freeze out'' slows down as the nuclei approach the valley of stability. Also other reactions like $\alpha$-decays may play a role during the freeze out. Neutrons are used up after about a second, which marks the end of the r-process, and nuclei move initially quickly towards the valley of stability. Closer to the valley of stability the nuclear half lives times increase and thus the freeze out takes place on many different time scales to reach the final abundance distribution only as very long-lived nuclei have decayed (up to Gyrs) \citep[see e.g.][for the time evolution of the abundance]{MendozaTemis2015}.

\begin{figure}
\begin{center}
\includegraphics[width=0.8\columnwidth]{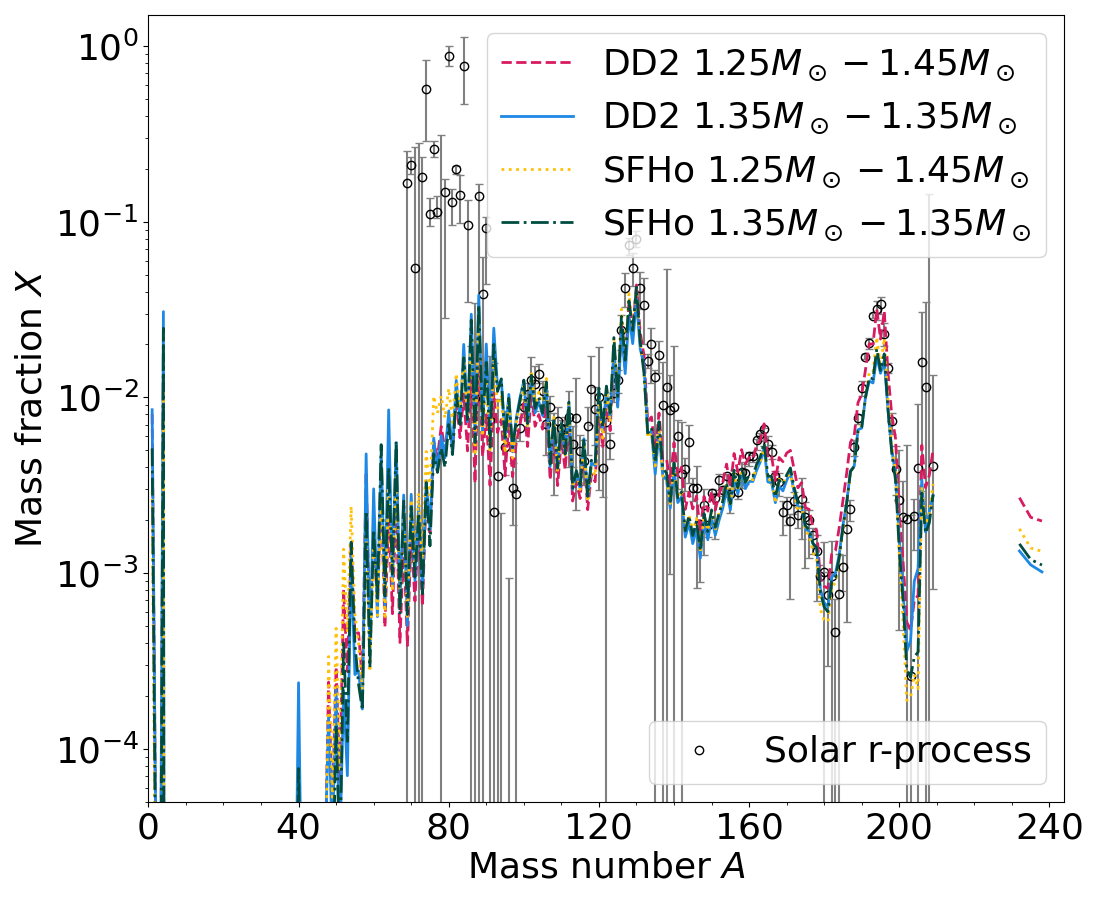}
\end{center}
\caption{Abundance distributions of the dynamical ejecta as functions of mass number $A$ for simulations with different binary masses and varied EoS. The black dots display the observed solar r-process abundance pattern. Figure from~\cite{Kullmann2022a}; \copyright Oxford University Press on behalf of the Royal Astronomical Society (RAS). Reproduced with permission}
\label{fig:abueos}
\end{figure}

Following the path of the r-process along the neutron drip line, it is evident that the seeds pass nuclei with magic neutron numbers, i.e. closed neutron shells with $N=50$, $N=82$ and $N=126$. For these nuclei being in comparison more stable, reaction rates are slower. Considering the ``flow'' of the immense number of nuclei moving along the r-process path, the nuclei will move slower through the region with magic neutron numbers. Consequently, during the flow nuclei with closed neutron shells will be in comparison more abundant. These overabundances will be dragged along during the freeze out. Hence, the final abundances will carry an imprint of the closed neutron shells once the freeze out reaches the valley of stability. The r-process path crosses the closed neutron shells close to the neutron drip line, where the magic nuclei have mass numbers $A$ of about 80, 130 and 195, respectively. Beta decays do not change the mass number but convert roughly 15 to 25 protons to neutrons during the freeze out. Hence, the overabundances occur approximately at $A=80$, $A=130$ and $A=195$ in the valley of stability with corresponding atomic numbers of $Z\approx 35$, $Z\approx 55$ and $Z\approx 80$. The respective peaks in the abundance distribution are referred to as the first, second and third r-process peak (see Fig.~\ref{fig:abueos}).

{\bf Observed abundance pattern:} This characteristic abundance pattern is universally found in nature and was for a long time in fact the most convincing argument for the r-process to take place although the astrophysical production site was not identified yet; see e.g. the reviews by~\cite{Arnould+2007,Sneden2008,Thielemann2011,Horowitz2019,Arnould2020,Cowan+2021,Siegel+2022} for a more exhaustive discussion of the historical developments and the observations. The elemental distribution with distinct r-process peaks is for instance observed in the solar system (see Fig.~\ref{fig:abueos}) but also in so-called metal-poor stars, i.e. old stars that formed earlier in the Galactic evolution and were enriched by only a few or possibly a single nucleosynthesis event. The fact that the relative abundance distribution is so similar in these different observations (although with respectively very different absolute abundances) points to a certain robustness and universality of the r-process. It should be mentioned that the interpretation of abundance measurements often relies on subtracting the contributions by other nucleosynthesis processes such as, for instance, the slow neutron-capture process, which takes place in other sites, and is assumed to be sufficiently well known (Fig.~\ref{fig:chart}).

{\bf R-process in mergers:} The basic description of the r-process above illustrates why it is so important to reliably determine the ejecta composition in terms of $Y_e$. It is the crucial parameter to set the initial neutron-to-seed ratio and thus the evolution of the r-process as it dictates the number of available neutrons to be captured. For smaller $Y_e$ nuclei with larger mass numbers are reached by the nuclear flow through the chart of the nuclei. In order to create third-peak elements, the electron fraction should not exceed $\approx 0.25$~\citep[see e.g.][for a comprehensive parameter study]{Lippuner2015}. This is considered to be an important threshold because it similarly determines whether significant amounts of lanthanides are produced. Because of their atomic structure these elements feature a particularly high opacity, which strongly effects the electromagnetic emission of the ejecta~\citep{Barnes2013,Kasen2013,Tanaka2013}. 

The ejecta produce electromagnetic emission in the optical, infrared and ultraviolet (\cite{Li1998,Kulkarni2005,Metzger2010}, see Section~\ref{sec:mess}). This astronomical transient is called kilonova refering to its luminosity being roughly thousand times higher than that of novae. The term ``kilonova'' was proposed in~\citet{Metzger2010}. Sometimes also ``macronova'' is used~\citep{Kulkarni2005}. Generally, the kilonova will be redder and dimmer if lanthanides are present, while bluish emission components would require the absence of these elements~\citep{Barnes2013,Kasen2013,Tanaka2013,Metzger2014}.

For the production of low-mass number elements, e.g. around the first peak, the electron fraction should not be too low because otherwise all seed nuclei are driven to higher mass numbers. Apart from $Y_e$, the r-process path is additionally affected by the entropy and the expansion time scale of the outflow, which influence the neutron-to-seed ratio and the reaction rates. See e.g.~\cite{Lippuner2015} for a parameter study with analytic trajectories.

In addition to neutron captures and $\beta^{-}$-decays other nuclear reactions may become important under certain conditions. For instance, for high-entropy environments photo-dissociation may play a role and establish an equilibrium between neutron captures and dissociations by energetic photons (dubbed n-$\gamma$ $\gamma$-n equilibrium). In this case the nuclear flow does not reach too close to the neutron drip line but proceeds to higher mass numbers somewhat closer to the valley of stability.

Under very neutron-rich conditions ($Y_e$ below $\sim 0.2$ or even as low as just a few per cent) the neutron-to-seed ratio will significantly exceed 200, which would be the number of neutrons required to reach beyond the third r-process peak. In such a neutron-rich environment, nuclei will attain even higher mass numbers and enter a region in the chart of the nuclei, where fission processes (neutron-induced fission, spontaneous fission, $\beta$-delayed fission, photo-fission) become likely at about $A\sim 250$~\citep[see, e.g., ][for dedicated studies of fission in the context of merger ejecta]{Goriely2013,Eichler2015,Mumpower2018,Vassh2019,Giuliani2020,Lemaitre2021}. When a nucleus fissions, it splits in two fission fragments with mass number of roughly $A/2$ and a few free neutrons. The fission fragments follow some distribution and do not need to be equally heavy (asymmetric fission). In any case the fragments are located at mass numbers of roughly $A=120$ and can thus start again capturing neutrons and proceed along the r-process path. Possibly the fragments reach the fission region again and a few cycles can occur. The process is called fission recycling and yields generally very robust and solar-like r-process abundance patterns above $A\approx 130$. 

It is important to emphasize that the nuclear reaction rates determining the r-process are not very well known~\citep{Arnould+2007,Horowitz2019,Arnould2020,Cowan+2021}. The r-process involves very neutron-rich, heavy, particularly short-lived nuclei, many of which can currently not be measured by laboratory experiments. Hence, describing reaction rates as necessary input for nuclear network calculations to model the r-process and to calculate the final abundance pattern, largely relies on theoretical models, which extrapolate properties of known nuclei. One of the most important ingredients is the nuclear mass model, which is why mass measurements of very neutron-rich nuclei are highly desirable to benchmark such theoretical calculations. In particular, the construction of the FAIR accelerator facility may allow to measure the properties of nuclei which are currently inaccessible \citep[see, e.g., ][]{Cowan+2021}. Also, fission rates and the fission fragment distribution exhibit large uncertainties.

All nuclear models reproduce the gross structure of the abundance pattern with distinct peaks. The uncertainties of the abundances of individual elements or nuclei with a given mass number, however, can amount to factors of several for a given set of astrophysical conditions~\citep[see, e.g.,][for sensitivity studies involving different nuclear physics input]{Goriely2013,Eichler2015,MendozaTemis2015,Mumpower2015,Mumpower2016,Martin2016,Mumpower2018,Holmbeck2019,Vassh2019,Nikas2020,Giuliani2020,Lemaitre2021,Barnes2021,Kullmann2022}. The impact of the nuclear reaction models on the final properties of the merger ejecta (abundance pattern and heating rate) is thus sizable. It may therefore not be straightforward to disentangle these nuclear uncertainties from variations in the outflow properties resulting from the EoS or the binary masses. In addition, in theoretical studies the numerical error from the hydrodynamical simulations can be significant and one cannot expect to be able to determine the conditions in the ejecta ab initio for a given EoS and binary masses.

A significant number of studies employs hydrodynamical simulation data of individual or several ejecta components to compute the respective abundances in the outflow by nuclear network calculations~\citep[some examples include][]{Freiburghaus1999,Goriely2011,Korobkin2012,Bauswein+2013,Perego2014,Wanajo2014,Just+2015,Martin2015,Wu+2016,Lippuner2017,Radice+2018,Curtis2021,Just2022a,Kullmann2022a,Perego2022,Kullmann2022,Fujibayashi2022}. In practice, a number of tracers particles in the ejecta monitor the thermodynamical evolution and composition of unbound fluid elements. These trajectories are then post-processed with a nuclear network code. Knowing how much mass a tracer represents one can compute the final elemental distribution by weighting the trajectory accordingly.

Generally, the different types of merger ejecta result in robust abundance patterns that closely follow the solar distribution in some range of mass numbers $A$. It is, however, also apparent in these calculations that the abundances of individual trajectories extracted from ejecta simulations can in fact show significant differences from a solar-like distribution, and only the appropriate average resembles the solar abundance pattern. This observation cautions that calculations considering individual trajectories or too few tracers may not be representative of a particular astrophysical outflow, which can only be characterized by a distribution of trajectrories with different properties ($Y_e$, entropy and expansion time scale)~\citep[e.g.][]{Kullmann2022}. Note that typical dynamical merger simulations cover only a few 10 milliseconds and usually do not follow the ejecta outflow to very low densities. Hence, for network calculations capturing the r-process on time scales of seconds, trajectories need to be extrapolated which may introduce additional uncertainties.  Also, it is important to include the heat generated by the r-process consistently in network calculations, i.e. considering the generation of thermal energy by the decay products, which changes the reaction rates~\citep[see e.g.][]{MendozaTemis2015}.

{\bf Heating:} The r-process is not only important for the formation of heavy elements but also because it provides the energy source for electromagnetic emission of the ejecta~\citep{Li1998,Kulkarni2005,Metzger2010,Roberts2011,Goriely2011,Korobkin2012,Metzger2019,Nakar2020}. As described, the r-process involves sequences of many radioactive decays during the freeze-out process. These decays deposit heat in the ejecta as charged decay products thermalize in the surrounding medium (neutrinos escape unhindered in this stage; gamma rays deposit only a fraction of their energy). The energy release amounts to a few MeV per nucleon, where the major contributions come from beta decays, while alpha decays and fission contribute for very neutron-rich conditions. The decay chains include nuclei with a large spread in their half life times. This implies that the heating occurs on different time scales; and adding up the contributions of many exponential decays of different half life times effectively results in a power law decay of the heating rate. The heating rate, i.e. energy release per time per mass, follows approximately
\begin{equation}
    \dot{Q}(t)=2\times 10^{10}\mathrm{erg/g/s} \left( \frac{t}{ 1\mathrm{d}} \right) ^{-1.3}
\end{equation}
\citep{Metzger2010,Korobkin2012}.
This expression should be multiplied by a thermalization efficiency \citep[see, e.g.,][]{Metzger2010,Barnes2016,Waxman2018,Hotokezaka2020}, which is of the order of about 0.5, to determine the actual energy deposition in the ejecta and account for the fact that not all energy is actually deposited in the ejecta. Because of the $t^{-1.3}$ behavior, the energy deposition drops by several orders of magnitude from minutes to weeks after merger. The composition of the ejecta as well as the nuclear model of the r-process have some moderate influence on the heating rate~\citep[e.g.][]{Hotokezaka2020}. Network calculations show that the universal behavior is a result of an ensemble of trajectories, while the heating rate of individual fluid element histories can deviate significantly. This may suggest that the local energy deposition in a certain region of the ejecta may not necessarily follow a universal behavior but rather a local heating rate, see, e.g.,~\citet{Just2022}.

It has been pointed out that already a minor fraction of free neutrons that are not captured during the r-process may provide an early intense source of heating~\citep{Metzger2015}. As little as $10^{-5}~M_\odot$ of neutrons could significantly alter the electromagnetic emission during the first hours after merging (recall the half life of free neutrons of about 10~mins). Considering the challenges to resolve ejecta properties in simulations, it is not yet fully settled whether or not a part of the dynamical ejecta can have conditions such that neutrons are left after the r-process. See, e.g.,~\cite{Dean2021} and references therein.

At later times of weeks after the merger, the heating can be dominated by individual nuclei with half lives of this order, which leads to bumps in the heating rate and thus the bolometric light curve that are potentially observable with the James Webb Space Telescope, a space-borne infrared telescope~\citep{Zhu2018,Wu2019a}.

\subsection{Messengers}\label{sec:mess}
One of the most fascinating aspects of NS mergers is the fact that they can be observed through very different messengers, which carry complementary information about the physical system and in particular about the EoS. Combining several messengers is thus especially informative.

{\bf Gravitational waves:} The current network of ground-based GW detectors is sensitive to GWs with a frequency between a few 10~Hz and a few kHz~\citep{Aso2013,Acernese2015,Collaboration2015}. During the inspiral the GW signal of a binary is largely dominated by the (point-particle) orbital motion, and the GW frequency equals twice the orbital frequency. This means that although the binary evolution is driven by GW emission over millions of years, only during the last minute before merging a binary will actually be detectable by a GW instrument. Only in this last phase the system reaches frequencies in the range that is accessible by ground-based instruments. Note that the projected spaceborne LISA GW detector is sensitive to lower frequencies and can thus detect binaries long before merging~\citep{AmaroSeoane2022}.

The signal of a binary, in particular its frequency evolution, is predominantly governed by a single parameter with the dimension of a mass \citep[see for instance][]{Maggiore2008,Sathyaprakash2009,Creighton2011,Blanchet2014}. This chirp mass,
\begin{equation}
\mathcal{M}=\frac{(M_1 M_2)^{3/5}}{(M_1+M_2)^{1/5}}\,,
\label{eq:chirp}
\end{equation}
is given by the masses of the two binary components, and the term chirp mass refers to the increase of the frequency and amplitude with time. Being the dominating parameter describing the signal, $\mathcal{M}$ is measured with very good precision. For not too large mass asymmetries $M_1\neq M_2$, the chirp mass mostly scales with the total mass and is less sensitive to the mass ratio $q$. Hence, a measurement of  $\mathcal{M}$ informs approximately about the total mass of a NS binary.

At higher post-Newtonian order additional parameters like the binary mass ratio, spins, and finite-size effects influence the binary evolution and the GW signal. These effects only appear when the velocity reaches a fraction of the speed of light. Hence, they affect the signal only during the late inspiral, when orbital frequencies and thus velocities increase. In this phase the binary evolves very fast, and detectors are less sensitive in this frequency range, which is why the corresponding parameters are more difficult to measure. 

Combining $\mathcal{M}$ and the mass ratio $q=M_1/M_2$ one can compute the individual masses of the binary components and thus the total mass. Obtaining the mass ratio and the total mass is important since both clearly affect the dynamics during and after merging and in particular the mass ejection. 

The spins of the NSs enter through an effective spin parameter, which contains the contributions from both NSs. It is generally difficult to measure the effective spin parameter and in particular the individual spins through GWs~\citep{Farr2016}. In addition, the effective spin and the mass ratio are degenerate.  It is often assumed, based on stellar and binary evolution arguments, that spins of NSs in binary systems are rather small, i.e. compared to the orbital motion~\citep{Tauris2017}. In fact, the rotation periods of pulsars in known binary NSs are long compared to the orbital period \citep[the fastest known pulsar in a binary has a spin period of 17~ms][]{Stovall2018}. Also note that tidal locking during the late inspiral phase is estimated not to play a role~\citep[][]{Bildsten1992,Kochanek1992,Lai1994}. If this generally holds true for all binaries, spin effects are less relevant for the dynamics of a NS merger.

{\bf EoS constraints from the GW inspiral:} Finite-size effects on the GW signal are described by the tidal deformability $\Lambda$, which quantifies the quadrupolar deformation of a star by a given tidal field~\citep{Flanagan2008,Hinderer2008,Hinderer2010,Damour2010,Damour2012}. In the case of a binary the tidal field is created by the respective companion star. The tidal deformability is defined as property of an isolated static NS and is fully determined by the EoS and mass. For this reason the measurement of the tidal deformability is of immense interest to constrain the high-density EoS. The tidal deformabilty (also called tidal polarizability) of a NS is given by 
\begin{equation}
\Lambda=\frac{2}{3}k_2\left( \frac{c^2 R}{G M} \right)^{\! 5}
\label{eq:tidaldeform}
\end{equation}
with the tidal Love number $k_2$ and the stellar mass $M$ and radius $R$. The definition of $\Lambda$ illustrates that tidal effects strongly depend on the mass and the radius or the inverse compactness, respectively. The influence of $k_2$ is relatively minor, and for a fixed mass the tidal deformability provides a good proxy of the radius. Hence, a measurement of $\Lambda$ effectively constrains the NS radius.

A larger tidal deformability leads to an accelerated inspiral~\citep[see, e.g.,][for reviews]{Chatziioannou2020,Dietrich2021}. In a binary system the effect is described by a mass-weighted linear combination of the tidal deformabilities of the individual binary components. This combined tidal deformability is given by 
\begin{equation}
\tilde{\Lambda}=\frac{16}{13}\,\frac{(M_1+12 M_2)M_1^4\Lambda_1+(M_2+12 M_1)M_2^4\Lambda_2}{(M_1+M_2)^5}
\label{eq:combtidaldeform}
\end{equation}
and is the parameter which is actually inferred from a measurement. The binary mass ratio does not affect $\tilde{\Lambda}$ too strongly, in the sense that binaries of the same chirp mass but with different $q$ have very similar $\tilde{\Lambda}$~\citep[e.g.][]{De2018}. Hence, for EoS constraints it is not too essential to measure the mass ratio accurately.

{\bf Postmerger GW emission:} Gravitational waves are not only emitted during the binary inspiral, but also afterwards if a rapidly rotating NS remnant forms. The violent oscillations of the remnant generate GWs mostly in the range of a few kHz~\citep{Xing1994,Shibata2006,Oechslin+2007}. If the remnant does not collapse, the signal can persist for several 10~ms over which the oscillations are slowly damped. The dominant oscillation mode of the remnant occurs as a pronounced peak in the GW spectrum and is potentially measurable, although the sensitivity of current detectors declines in this frequency range. Importantly, the frequency $f_\mathrm{peak}$ of the peak tightly scales with the NS radius and other NS parameters~\citep{Bauswein+2012PhRvD,Bauswein+2012PRL,Takami+2015,Bernuzzi2015,Bauswein2019}. Hence, measuring  $f_\mathrm{peak}$ can provide important constraints on the EoS.

The postmerger GW spectrum contains additional subdominant features~\citep[e.g.][]{Stergioulas+2011,Bauswein2015,Takami+2015,Bauswein2019}. Those can potentially also inform about the EoS although they may be harder to measure. Detecting such features can be rewarding because they may reveal the detailed dynamics of the remnant, which may be important to connect to the dynamical mass ejection. For instance, one of the subdominant peaks emerges from a coupling of the dominant quadrupolar oscillation and the quasi-radial mode~\citep{Stergioulas+2011}. Hence, this feature can in principle provide information on the radial oscillation of the remnant, which is relevant for the dynamical mass ejection (see above).

Finally, it should be emphasized that the presence or absence of postmerger GW emission informs about the outcome of the merger. Obviously, this is highly relevant for EoS constraints since combined with the mass measurement from the inspiral phase it allows to constrain the threshold binary mass for prompt collapse~\citep{Hotokezaka2011,Bauswein2013}. As described above, the merger outcome is also important for mass ejection and thus the nucleosynthesis and the resulting kilonova signal, which exemplifies the importance to detect postmerger GWs. For a sufficiently loud GW signal it may in principle be possible to measure the life time of the remnant. The ring down of the BH forming during the gravitational collapse of the remnant occurs at several kHz and is thus unlikely to be measured~\citep{Shibata2006}. 

{\bf Extrinsic parameters:} Detecting GWs is vital in the context of nucleosynthesis because the signal carries important information about additional properties of the binary which are not intrinsic to the source~\citep{Sathyaprakash2009}. The GW amplitude scales with the luminosity distance of the source, which is largely degenerate with the inclination angle of the binary. Obviously both parameters are invaluable to interpret kilonovae and to link the electromagnetic signal to the mass ejection. At typical distances of several 10~Mpc to some 100~Mpc cosmological redshift plays a role~\citep{Maggiore2008,Sathyaprakash2009,Creighton2011}. For instance, the chirp mass inferred from the frequency evolution, which is measured by the detector, has to be corrected by adopting a value for the Hubble constant or using, if available, the redshift of the host galaxy~\citep{Abbott+2019PhRvX...9a1001A}.

The GW signal also provides a rough localization of the event on the sky (mostly through the different arrival times in the network of detectors), which can be used to trigger and guide electromagnetic follow-up observations to search for the kilonova~\citep{Abbott2020}. More generally, GW detections will yield an estimate of the merger rate in the local universe and the properties of the binary population such as the distribution of masses. The former is obtained by knowing the sensitivity of the GW instruments and thus the volume which is monitored by the detectors, for a given duration of the observing run~\citep{Abbott+2017PhRvL.119p1101A}. These data on the rate and masses of the binaries are essential to quantify the overall production of heavy elements by mergers in the local universe. This may help to assess whether mergers are the dominant contributors of heavy elements or if there are other sites producing significant amounts of r-process elements.

{\bf Kilonovae:} The most important messenger for accessing the nucleosynthesis of COMs are kilonovae~\citep{Li1998,Kulkarni2005,Metzger2010,Roberts2011,Goriely2011,Metzger2019,Nakar2020}. Based on the properties inferred from simulations one can expect that the majority if not the entire ejecta undergo the r-process. As described above this includes many radioactive decays like beta-decays, gamma-decays, alpha-decays and fission, which deposit heat in the expanding ejecta cloud. This leads to quasi-thermal emission peaking at optical and infrared wavelength. As a result of the many contributions from decays with different half lives, the total heating rate follows roughly a power-law decline and most of the energy is generated in the first seconds.

Very soon after the merger ($<$~1 second) the ejecta expand homologeusly, i.e. fluid elements move with a constant velocity except for small changes by working against the gravitational potential and gaining energy through the radioactive decays~\citep{Rosswog2014,Grossman2014,Kawaguchi2021,Klion2022,Neuweiler2022,Haddadi2022}. Possibly, there could also be additional heating contributions or large scale reconfigurations of the ejecta by energy injection from the remnant \citep[e.g.][]{2018ApJ...856..101M,Nativi2021}. Homologous expansion implies that the distribution of mass as function of velocity $m(v)$ remains constant (self-similar), which means that the ejecta have arranged themselves such that faster moving material is ahead of the more slowly expanding shells. A typical distribution $m(v)$ shows the bulk of the ejecta with a velocity of about 0.1~c with a steep decline at higher velocities~\citep[e.g.][]{Metzger2015,Radice+2018}.

At early times, i.e., in the first hours after the merger, the densities in the ejecta are so high that a major fraction of the expanding outflow is optically thick. Radiation can only escape from the surface layers and most of the radioactive energy is used for expansion work in this phase of the explosion. The photosphere as the area from where photons can stream out, is located in the surface layers, i.e. at very high velocities. As the ejecta expand the density drops and so does the optical depth. Hence, the photosphere moves further inside the ejecta to lower velocities and a larger fraction of the outflow becomes transparent. Here the term ``moving inside'' does not refer to a radial location in a fixed reference frame but to a frame comoving with the expanding ejecta. 

{\bf Basic dependencies:} A first quantitative assessment is made by comparing the expansion time scale and an estimate of the photon diffusion time \citep[see][]{Arnett1980,Metzger2010,Metzger2019}. An average radius of the ejecta is given by $R=v t$ with $t$ being the time since merging and with a constant mean velocity $v$ adopting homologeous expansion. The mean density of the ejecta is given by $\rho=\frac{3 M}{4\pi R^3}$ with $M$ being the mass of the ejecta. Here spherical symmetry is assumed. The photon diffusion time scale can be estimated by $t_\mathrm{diff}\sim \frac{R}{c}\tau$ with the optical depth $\tau$. The latter is given by $\tau=\rho \kappa R$ with the opacity $\kappa$. The opacity is a material constant and depends on the composition and temperature of the ejecta. 

Equating the expansion time scale $t_\mathrm{ex}=R/v$ and the diffusion time scale yields $t=\frac{3 M \kappa}{4\pi c v t}$, which describes the situation when most of the radiation can escape from the outflow. This phase is reached after
\begin{equation}
t_\mathrm{peak}=\sqrt{\frac{3}{4\pi c}}\,\sqrt{M}\,\sqrt{\kappa}\,\sqrt{\frac{1}{v}}
\label{eq:tpeak}
\end{equation}
and determines a time scale for the emission peak of the kilonova as the time when most of the outflow reaches transparency, while the heating continuously declines. For typical values of the ejecta properties this simple estimate yields typical time scales of hours and days for the evolution of the kilonova.

The luminosity at $t_\mathrm{peak}$ is given by the radioactive heating at that time, i.e., only the contributions from those decay products which efficiently thermalize in the ejecta. Hence, $L_\mathrm{peak}\sim M \dot{Q}(t_\mathrm{peak})$, recalling that $\dot{Q}(t)$ roughly follows a power law with index $-1.3$ (with $\dot{Q}$ giving the energy output per time per mass). Inserting $t_\mathrm{peak}$ yields a typical peak luminosity of $10^{41}$ to  $10^{42}~\mathrm{erg/s}$ as order of magnitude estimate.

The temperature of the emission can be obtained from the Stefan-Boltzmann law as
\begin{equation}
T_\mathrm{peak}=\left(\frac{L_\mathrm{peak}}{4 \pi \sigma R^2(t_\mathrm{peak})} \right)^{\! 1/4}
\label{eq:temppeak}
\end{equation}
with the Stefan-Boltzmann constant $\sigma$. Hence, the radiation temperature can also be estimated by $M$, $v$ and $\kappa$ with typical values of several 1000~K, i.e., an emission peaking in the wavebands between the UV and the infrared. These considerations show that kilonovae are a phenomenon which is accessible by optical telescopes.

These estimates provide the order of magnitude of the kilonova properties. A more refined calculation takes into account the structure of the ejecta, i.e., a distribution $m(v)$. This yields an estimate of
\begin{equation}
    t_\mathrm{peak}\approx 1.6\,\mathrm{d} \,\left(\frac{M}{10^{-2}~M_\odot} \right)^{\! 1/2}  \left(\frac{v}{0.1 c} \right)^{\! -1/2}  \left( \frac{\kappa}{1~\mathrm{cm^2/g}} \right)^{\! 1/2} 
\end{equation}
and
\begin{equation}
    L_\mathrm{peak}\approx  10^{41} \,\mathrm{\frac{erg}{s}} \, \left( \frac{\epsilon_\mathrm{th}}{0.5}\right) \left( \frac{M}{10^{-2}~M_\odot}\right)^{0.35} \left(\frac{v}{0.1 c} \right)^{0.65} \left( \frac{\kappa}{1~\mathrm{cm^2/g}} \right)^{\! -0.65}
\end{equation}
(see~\citep{Metzger2019}). The luminosity expression contains an additional factor $\epsilon_\mathrm{th}$ accounting for the thermalization efficiency of the radioactive decay products to deposit their energy in the surrounding plasma. Although assumed to be constant here, it in fact follows some time evolution and depends on the exact composition \citep[see][]{Barnes2016,Hotokezaka2020}.

Overall, these expressions illustrate that the observable kilonova features strongly depend on the properties of the ejecta and r-process. This implies, in turn, that observations of kilonovae can provide information on the ejecta and by this on the EoS, which strongly affects the mass ejection.

The basic dependencies of the above formulae are straightforward to understand. The opacity determines how easily light can escape from the ejecta and thus affects the time the outflow needs to reach transparency. More massive ejecta imply a higher density and thus the emission will peak later since the outflow needs to expand for a longer time to reach a sufficiently low density. Similarly, a higher velocity yields a short peak time scale as the dilution proceeds faster and the ejecta reaches a stage of transparency earlier. Reaching transparency earlier implies a higher luminosity because radioactive heating is stronger at earlier times. Obviously, a more massive ejecta cloud produces a higher luminosity since the total energy output is higher.

{\bf Kilonova modeling and opacities:} The equations above clearly demonstrate the significant impact of the opacity of the ejecta. It is important to emphasize that the opacity sensitively depends on the composition of the ejecta~\citep{Barnes2013,Kasen2013,Tanaka2013}. In particular, lanthanides and actinides feature very high opacities, which are two orders of magnitude above typical opacities of iron group elements ($0.1~\mathrm{cm^2/g}$). Therefore, the initial electron fraction of the ejecta has a critical influence on the kilonova since it determines the production of lanthanides. For $Y_e<0.2$ significant amounts of these elements are produced and a typical opacity is of the order of $30~\mathrm{cm^2/g}$, whereas for $Y_e>0.3$ the lanthanide fraction is very small and a typical value of the opacity is about $3~\mathrm{cm^2/g}$ \citep[see][]{Tanaka2020}. This exemplifies that kilonova properties in principle provide access to the composition of the ejecta.

Various further refinements and more advanced models exist to predict the bulk properties of kilonovae, which cannot be described in detail here~\citep[e.g.][and references therein]{Grossman2014,Villar2017,Perego+2017,Waxman2018,Waxman2019,Metzger2019,Nakar2020,Hotokezaka2020}. These models can for instance include an angle-dependence, a structure of the density and composition in the ejecta,  several ejecta components with different velocities and composition (opacity or lanthanide fraction) and a time dependence of the thermalization efficiency. Also, these calculations can predict the time dependence of the light curve or the emission in certain filter bands, i.e. the luminosity in a certain range of wavelength. Such filters are used by telescopes and provide information on the color evolution of an astronomical transient.

The most sophisticated modeling of kilonovae is based on radiative transfer calculations, which follow the propagation of photons through the ejecta cloud partially taking into account detailed atomic interactions~\citep[e.g.][]{Kasen2013,Tanaka2013,Kasen2015,Kasen+2017,Tanaka2018,Wollaeger2018,Kawaguchi2018,Bulla2019,Watson2019,Kawaguchi2021,Domoto2021,Perego2022,Just2022,Gillanders2022,Collins2022,Neuweiler2022,Vieira2022}. This allows a detailed modeling of the spectra of kilonovae beyond black body emission and provides means to interpret features, which can be connected to specific elements. In a further step the presence or absence of certain elements may be related to details of the mass ejection and the underlying EoS.

In this regard it is important to emphasize that for a large fraction of the heavy r-process elements none or only incomplete atomic data are available. The complexity of the atomic structure of these elements represents a challenge, which is important to be addressed for a full comprehension of the spectral properties of kilonovae and the determination of elemental abundances in the outflow~\citep{Kasen2013,Tanaka2013,Tanaka2020,Fontes2020}.

Kilonovae contain a wealth of information about the merger ejecta and thus the underlying physics and especially the EoS, which will be further discussed in Section~\ref{sec:eosimpact}. But it should not remain unmentioned that kilonova observations and their interpretation are important in a broader scientific context. The localization of a merger event by GWs is relatively coarse (typically many square degrees), and only by detecting the electromagnetic counterpart the exact position can be determined~\citep{Abbott+2017PhRvL.119p1101A,Abbott2020}. This allows to identify the host galaxy and its cosmological redshift. This information can be used to constrain the Hubble constant and to be incorporated in the parameter estimation of the GW signal~\citep{Abbott2017,Abbott2020}. Also, it can be instructive to connect to the properties of the host galaxy and to interpret the spatial offset between merger and host galaxy. These relate to questions of star formation, stellar evolution and binary evolution like the delay time between NS binary formation and merger~\citep{Hotokezaka2018,Tauris2017,Beniamini2019}. The emission of the kilonova features a certain angle dependence, which is why a detailed interpretation can reveal the orientation of the merger, which is, for instance, helpful for GW data analysis. Finally, finding kilonovae in so-called blind surveys, i.e. without GW trigger but by regular wide-field observations, may provide better estimates of the rates of mergers, which is important to address the overall contribution of COM for r-process nucleosynthesis.

{\bf Other electromagnetic counterparts:} In addition to kilonovae, NS mergers also lead to other phenomena that produce electromagnetic emission, which will be briefly mentioned for completeness~\citep{Metzger2012,Nakar2020}. Driven by magnetic fields and possibly aided by neutrino annihilation, powerful relativistic outflows may be launched from a BH torus system that forms after a NS merger. These outflows are concentrated in a small solid angle around the rotational axis and are referred to as jets. Some models consider jet formation from magnetized NS remnants. In any case, these relativistic jets are launched after mass ejection has started, and thus the jet drills through the ejecta. Farther out internal shocks in the jet produce intense forward-beamed gamma-ray emission, which is observable from cosmological distances~\citep[see e.g.][]{Eichler1989,Nakar2007,Berger2014,Nakar2020}. These so-called short GRBs are already observed for decades by gamma-ray satellites. The attribute ``short'' refers to the length of the gamma-ray emission (below two seconds) and distinguishes the short bursts from the long GRBs, which originate from the collapse of massive stars (see Section~\ref{sec:magex}).

The relativistic jets run into the interstellar medium and produce the so-called afterglow by synchrotron emission, which is observable from radio to X-rays. Very similar processes are expected to occur when the mildly relativistic ejecta from a merger interact with the interstellar medium resulting in radio remnants, which evolve on time scales of weeks to many years~\citep{Nakar2011,Piran2013,Nakar2020}. These phenomenona are called radio flares or radio remnants and are in principle observable from all directions, whereas short GRBs can only be detected if the jet points towards the earth. In the context of r-process nucleosynthesis, short GRBs and potentially radio remnants are interesting to estimate the rates of mergers and to study the environments in which mergers occur. Also, short GRBs have been used as trigger to identify kilonova emission with GRB130603B being a famous example~\citep{Tanvir2013,Berger2013}.

{\bf Neutrinos:} Neutron star mergers are very strong emitters of neutrinos since the forming remnant features temperatures of a few to several 10 MeV. Typical neutrino luminosities are of the order of a few $10^{52}$ to $10^{53}~\mathrm{erg/s}$~\citep{Ruffert1997,Rosswog1999,Sekiguchi2011,Sekiguchi2015,Palenzuela2015,Foucart2016,Lehner2016,Radice2016a,Bovard2017,Radice+2018,ArdevolPulpillo2019,Vincent2020,Combi2022}. The emission of electron anti-neutrinos dominates over electron neutrinos and heavy-flavor neutrinos. Mean energies of the neutrinos are roughly 15~MeV. Although these numbers are roughly comparable to those of CCSNe (see Section~\ref{sec:ccsn}), it is unlikely to directly detect a neutrino signal from a NS merger. The reason is that COMs are much less frequent than SNe and will thus on average take place at larger distances. While neutrino signals from CCSNe can be detected for events from the neighborhood of the Milky Way, it is statistically very unlikely that a COM will occur at a similar distance. Hence, the neutrino emission of COMs should not be considered as direct messenger but it does have a very strong observational imprint on the kilonova by determining the electron fraction of the ejecta and thus their opacity.

There is also the possibility that high-energy neutrinos (in the range of $10^{13}-10^{17}$ eV) are produced in the relativistic jet of a short gamma-ray burst~\citep{Waxman1997,Tamborra2016,Pitik2021}, but their production depends sensitively on the jet properties and has not been confirmed by IceCube Neutrino Observatory detections yet.

\subsection{Observations: GW170817 and more}\label{sec:obs}
There is a general agreement in the astrophysical community that the very first measurement of a NS merger with GWs marks a breakthrough. Not only it was the first unambiguous detection of a nearby NS merger but also it was a multi-messenger observation with electromagnetic emission discovered throughout the whole electromagnetic spectrum from gamma-rays to radio. The interpretation of these observations settled many open questions in the context of COMs and in particular gave new insights into the properties of high-density matter, r-process nucleosynthesis, and the connection between mergers and high-energy phenomena such as short GRBs. 

{\bf System parameters and EoS constraints:} In August 2017, the network of GW instruments consisting of Advanced LIGO and Advanced Virgo detected a NS merger at a distance of about 40~Mpc~\citep{Abbott+2017PhRvL.119p1101A,Abbott+2019PhRvX...9a1001A}. The event was dubbed GW170817 with the number referring to the detection date. The system had a chirp mass of 1.186~$M_\odot$ and a binary mass ratio of $0.7<q=M_1/M_2<1$~\citep{Abbott+2019PhRvX...9a1001A}. These parameters correspond to the mass ranges of 1.17 -- 1.36~$M_\odot$ and 1.36 -- 1.59~$M_\odot$ for the individual component masses of the binary. These masses are in line with those typically found in NS binary systems tat are observed by radio telescopes. The spins of the individual NSs are consistent with zero but generally not very well constrained for the analysis adopting a high-spin prior. 

The combined tidal deformability $\tilde{\Lambda}$ was determined to be roughly in the range of $100<\tilde{\Lambda}<700$~\citep{Abbott+2019PhRvX...9a1001A}. The exact value depends to some extent on the waveform models and the chosen priors which are employed in the parameter estimation. This is a relatively broad range of values for $\tilde{\Lambda}$, but in particular the upper limit represents a valuable constraint on the properties of high-density matter. The tidal deformabilty scales tightly with the NS radius, and the observation can thus be translated to an upper limit on NS radii~\citep{Abbott+2018PhRvL.121p1101A,De2018}. Note that this conversion is not significantly affected by the fact that the binary mass ratio was not well constrained, because for the given mass range NS radii are expected not to vary strongly. In essence, the inspiral GW signal constrains NS radii to be smaller than about 13.5~km and thus excludes very stiff nuclear matter. The measurement can be directly converted to a bound on the EoS, namely the pressure at twice nuclear saturation density should not exceed a value of about $6.2\times 10^{34}~\mathrm{dyn/cm^2}$. The exact numbers of these limits depend on details and assumptions of the analysis~\citep{Abbott+2018PhRvL.121p1101A,Chatziioannou2020}.

No postmerger GW signal was detected for GW170817~\citep{Abbott+2019PhRvX...9a1001A}. With the instrumental sensitivity at that time this is expected even if a postmerger GW signal was present. Hence, the non-detection does not permit any conclusions on the outcome of the merger, where the measurement of postmerger GWs represents a smoking-gun signature for the formation of a NS remnant as opposed to a direct gravitational collapse. With further improvements of the sensitivity in the kHz range in the future, the detection of postmerger GWs will come within reach and promises additional EoS constraints and insights into the merger outcome~\citep[e.g.][]{TorresRivas2019}.

{\bf Kilonova detection:} Important information on the EoS but also on r-process nucleosynthesis was inferred from the identification of the electromagnetic counterpart in the optical~\citep{SoaresSantos2017,Cowperthwaite+2017,Nicholl2017,Chornock2017,Arcavi+2017,Smartt+2017,Shappee2017,Kilpatrick2017,Pian+2017,Kasliwal2017,Tanvir+2017,Villar2017,Waxman2018,Waxman2019}. The GW signal provided a coarse localization of the event on the sky within an area of a few 10 square degrees and triggered an intensive follow-up campaign with optical telescopes. The electromagnetic counterpart was found about 11~hours after the GW detection and pinpointed the merger to a position near the galaxy NGC 4993, which has a distance consistent with the luminosity distance inferred from the GW amplitude and was thus identified as host galaxy of GW170817. In the following days and weeks essentially all major astronomical facilities including X-ray and radio telescopes observed the electromagnetic counterpart of GW170817~\citep{Abbott+2017ApJ...848L..12A}. The astronomical transient is referred to as AT~2017gfo. Independent of the GW detection a GRB was observed by the Fermi and INTEGRAL satellites about 1.7 seconds after the coalescence~\citep{Goldstein2017,Savchenko2017,Abbott+2017ApJ...848L..13A}, which on its own would probably not have enabled the discovery of the kilonova because of the less well constrained position on the sky. 

The bolometric luminosity of the kilonova, i.e. the total energy emission in the optical, ultraviolet and infrared, faded from $10^{42}~\mathrm{erg/s}$ to $10^{40}~\mathrm{erg/s}$ within about 10 days. The transient was observed in many different filter bands and several high-quality spectra, e.g. with the Very Large Telescope (VLT) X-Shooter instrument, have been recorded at different epoches~\citep{Pian+2017,Smartt+2017}. The radiation resembled a black-body emission as expected for a kilonova with the peak of the emission shifting from initially bluish colors to red and infrared. See also~\cite{Villar2017,Waxman2018} for a compilation of the different observations.

The color, the temporal evolution, and the luminosity of the transient are in excellent agreement with predicted properties of a kilonova, i.e. r-process heated, rapidly expanding outflows composed of heavy elements. Qualitatively the properties of the electromagnetic counterpart were clearly distinguished from other known astronomical transient phenomena such as, in particular, SNe. Overall, the observations provide very compelling evidence that r-process nucleosynthesis took place in the outflow of GW170817. NS mergers are therefore the first and so far only confirmed astrophysical site of r-process element formation.

As already indicated in the previous section, numerous models exist to describe the kilonova emission with different degrees of sophistication. Some of these models have been employed to extract the physical parameters of the ejecta, e.g., their mass, outflow velocity, and composition or opacity. Clearly, the interpretation and the inferred ejecta parameters depend on the details of these analyses and their assumptions.

\begin{figure}
\begin{center}
\includegraphics[width=0.8\columnwidth]{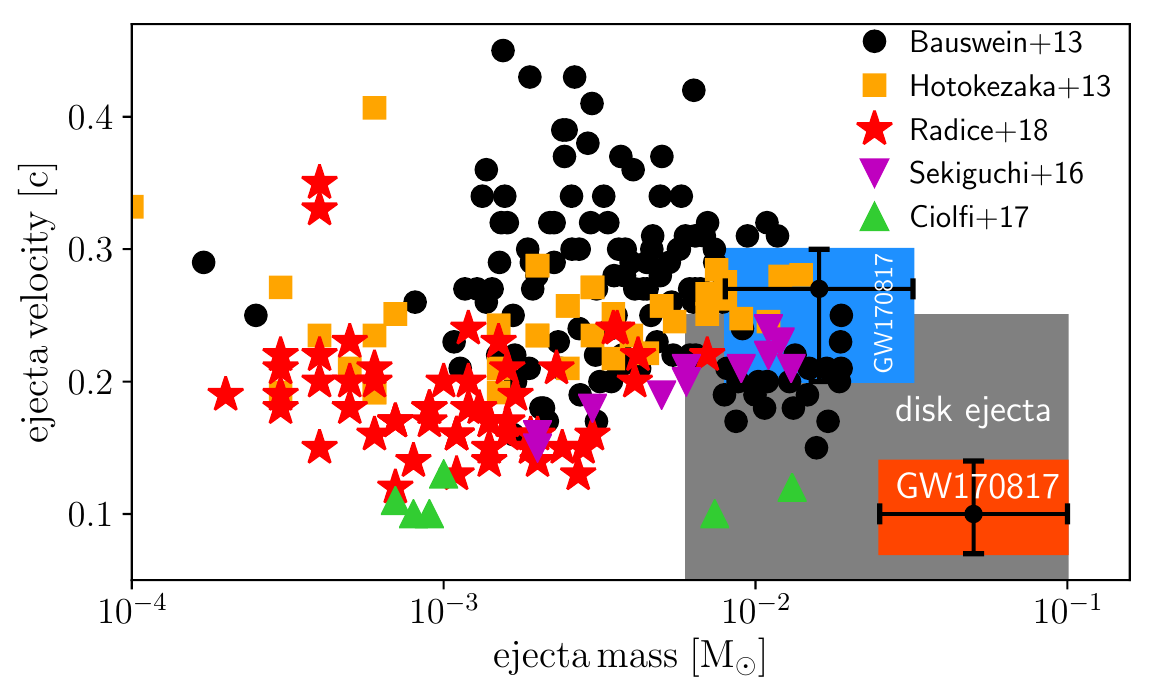}
\end{center}
\caption{Compilation of simulation results for the dynamical ejecta mass and outflow velocity in comparison to inferred values from the electromagnetic counterpart of GW170817. Figure from~\cite{Siegel2019}; with kind permission of The European Physical Journal (EPJ)}
\label{fig:mejv}
\end{figure}

Many models interpret the data as to show a faster moving blue component and a red low-velocity ejecta component~\citep[e.g.][]{Cowperthwaite+2017,Chornock2017,Evans2017,Kasen+2017,Kasliwal2017,Kilpatrick2017,Perego+2017,Tanaka2017,Villar2017,Smartt+2017,Tanvir+2017,Troja+2017,Rosswog2018,Bulla2019}. The colors ``blue'' and ``red'' refer to the opacity of the components and thus to their respective lanthanide fraction, which sensitively affects the opacity and is mostly determined by the electron fraction of the ejecta. These models obtain a velocity of 0.2$-$0.3 $c$ and mass in the range of about 0.02~$M_\odot$ for the blue component and $v\sim 0.15~c$ and a mass of roughly 0.03~$M_\odot$ for the red component. However, also models with only a single component involving a range of velocities in combination with the time evolution of the opacity as a result of the temperature and density evolution do fit the data well~\citep{Smartt+2017,Waxman2018}. Essentially all models do infer a total ejecta mass of about 0.03 to 0.06~$M_\odot$. See, e.g.,~\cite{Cote2018} for a compilation of literature results on the ejecta mass. Notably these numbers are too high to be explained by the dynamical ejecta alone if one considers literature values from simulations. Figure~\ref{fig:mejv} compares estimated ejecta properties of GW170817 with data from simulations for dynamical mass ejection~\citep{Siegel2019}.  

\begin{figure}
\begin{center}
\includegraphics[width=0.8\columnwidth]{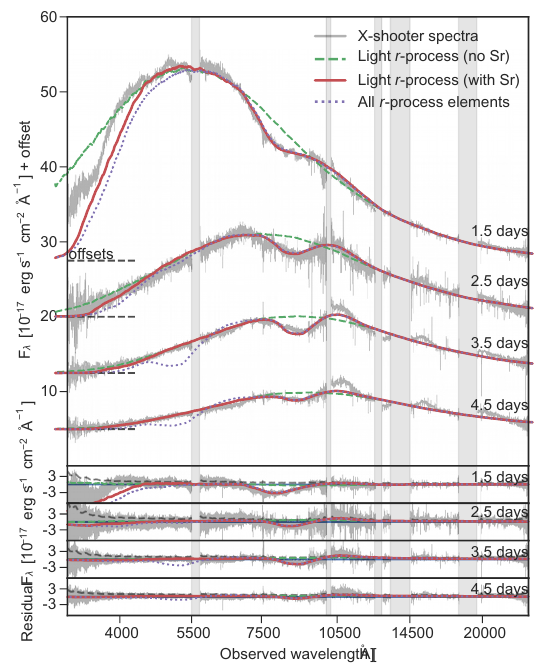}
\end{center}
\caption{Spectra of the electromagnetic counterpart of GW170817 at different epochs. The figure shows data from VLT/X-Shooter, which roughly reveal black-body emission with additional features imprinted. The emission peak shifts to redder colors. Credit: D. Watson \& J. Selsing from data taken at the ESO telescopes. See also~\cite{Watson2019}}
\label{fig:spectrum}
\end{figure}

The interpretation of the color evolution of the kilonova as a result of the opacity and the lanthanide content hints to the underlying nucleosynthesis and electron fraction of the ejecta. Lanthanide-poor conditions imply that the r-process did not proceed to very high mass numbers, i.e. a blue component would not have a typical solar r-process abundance pattern including a significant fraction of lanthanides. A red, high-opacity ejecta component may be associated with the formation of  heavy r-process elements in a low-$Y_e$ environment. This provides some clues on the composition of the outflow (Fig.~\ref{fig:spectrum}). 

A direct probe of the elemental distribution may be obtained from interpreting spectral signatures of the kilonova~\citep{Pian+2017,Smartt+2017}. While the emission roughly resembles a black body, the spectrum of AT2017gfo clearly contains additional spectral features. For instance, the spectrum at 1.5 days shows a prominent absorption feature at 810~nm (Fig.~\ref{fig:spectrum}). This has been interpreted by~\cite{Watson2019} to be blue-shifted strontium lines moving with a velocity of about 0.25 to 0.3~$c$, see also~\cite{Domoto2021,Gillanders2022,Domoto2022,Vieira2022}.
At later epochs the dip in the spectrum moves to larger wavelengths indicating a slower expansion velocity consistent with the expectation that deeper ejecta layers become visible at later times. Strontium is produced by the slow neutron-capture process as well as the r-process. With a mass number of about $A\approx 88$ strontium is a rather light r-process element indicating the presence of an ejecta component with relatively high $Y_e$, which is required to create such lighter elements.

{\bf More merger detections:} GW170817 was clearly an outstanding event, the interpretation of which is still ongoing. But there were about a handful more GW observations of COMs likely involving NSs apart from the numerous BH-BH binary detections~\citep{Abbott2021,GWTC-3:2021arXiv211103606T}. For none of the detections an electromagnetic counterpart was found, which, considering the estimated distances, did not come as a surprise. From a nucleosynthesis point of view these events may not be very informative.

However, considering the small number of events, the inferred binary masses are somewhat surprising as they do not follow the commonly expected distribution of NS masses in compact binaries. For instance, GW190425 had an unusually high total mass of about 3.4~$M_\odot$~\citep{Abbott+2020ApJ...892L...3A}. Here, ``unusual'' refers to the mass typically found in double NS systems containing a pulsar, which have system masses in the range of about 2.7 to 2.8~$M_\odot$~\citep[e.g.][]{Lattimer2012}. Also GW190814 was unusual with one binary component having a mass of 23.2~$M_\odot$ (obviously a BH) and the companion having a mass of 2.6~$M_\odot$~\citep{Abbott2020a}. For the latter it is not obvious whether it was a BH or a NS. If it was a NS, it either implies a rather high maximum NS mass $M_\mathrm{max}$ or it was rapidly spinning, which may challenge typical evolution scenarios and is surprising since among the few thousand known NSs none is rotating that fast. If the companion was a BH, the detection is challenging the notion that there may be a ``mass gap'' between NSs and BHs. There are also some candidate events with relatively low signal-to-noise ratio, which were not classified as confirmed detections.

In any case the next observing runs of the GW detector network are very likely to discover several more COMs with the prospect to find electromagnetic counterparts as the instrumental sensitivity increases and thus provides more accurate localizations. Also the capabilities of optical and infrared observatories are increasing with, for instance, the James Webb Space Telescope or the European Extremely Large Telescope, which promise better observational data of more distant events.

While GW detections are the primary tool to identify nearby COMs, it should be mentioned that also short GRBs can in principle be used to infer kilonova data from on average more distant events. For example, GRB130603B features excess emission in the afterglow light curve of the GRB, which is consistent with kilonova emission from a few 0.01~$M_\odot$ of ejecta~\citep{Tanvir2013,Berger2013}.

\subsection{Equation-of-state impact}\label{sec:eosimpact}
{\bf Motivation:} As already mentioned, the EoS is the entity which most sensitively affects the merger dynamics and thus the mass ejection apart from the binary masses. Hence, the nucleosynthesis and the kilonova carry a strong imprint of the EoS, while the GW inspiral signal constrains the binary masses with good accuracy for near-by mergers. Addressing the impact of the EoS on mass ejection and nucleosynthesis serves two purposes, which are connected but follow different directions.

(1) While GWs do provide information on the EoS (Sect.~\ref{sec:mess}), there is still need for additional and complementary constraints from interpreting kilonovae, which is only possible through the modeling and interpretation of these transients. A growing number of studies in fact combines constraints from different messengers including GWs and kilonovae~\citep[e.g.][]{Margalit+2017,Bauswein+2017,Shibata+2017,Radice2018c,Coughlin2018,Ruiz+2018,Most2018,Rezzolla+2018,Capano2020,Dietrich2020,Bauswein2021,Raaijmakers+2021,Huth2022}.

(2) In turn, adopting current knowledge about the EoS from merger observations, from other astronomical measurements, from laboratory experiments or from theoretical developments, the comprehension of the EoS influence is critical to predict the outcome of COMs. This is relevant to assess the total production of r-process elements throughout the cosmic history and to interpret for instance the abundances of metal-poor, i.e., old stars in the Milky Way and its neighborhood. The nucleosynthetic yields per merger event are a specific input for modeling the chemical evolution of the Galaxy and the Universe \citep[see, e.g.,][]{Cowan+2021} for a review and specific references on chemical evolution aspects). Clearly, these far-reaching goals require more information beyond the EoS influence such as the mass distribution of the binary populations, their merger rates and their delay times until coalescence. These ingredients may be obtained from a large sample of future merger observations and their interpretation. Also, a theoretical understanding of the EoS dependence is necessary to predict and bracket the possible range of kilonova properties. This may be useful for example in blind astronomical surveys, i.e., observing campaigns which are not triggered by specific events, to identify and classify COMs.

{\bf Simulations:} The influence of the EoS on the mass ejection is only accessible through numerical simulations. Considering the complexity of the mass ejection through different channels, the modeling of the ejecta and their EoS dependence represents a major challenge and is still subject to uncertainties as described in Section~\ref{sec:massejection}. To date mass ejection in COMs is qualitatively very well understood and the EoS dependence of several properties is quantitatively described, albeit usually within some uncertainty range, which on its own may not yet be fully quantified. At the same time it should be appreciated that there exists only a small subset of simulations which cover the whole mass ejection process from the dynamical merger phase to the late secular postmerger evolution in a fully consistent calculation or connect the different phases of mass ejection by some sort of mapping between different codes~\citep[e.g.,][]{Perego2014,Just+2015,Martin2015,Fernandez2017,HosseinNouri2018,Fujibayashi2020,Most2021a,Hayashi2022,Fujibayashi2022}. But generally, there is some decoupling between the modeling of  different merger stages, in particular in large-scale surveys based on many calculations with a systematic variation of the input parameters like binary masses and EoS. For this reason the discussion below follows the current status in the literature and addresses specific aspects and questions of the EoS and binary-mass dependence referring to Section~\ref{sec:massejection} for the detailed  description of the different mass ejection channels. The discussion largely focuses on the ejecta mass as one of the most important properties and the quantity that shows the largest variations with respect to the EoS influence.

Describing the EoS impact quantitatively requires not only to numerically quantify ejecta parameters but also to define characteristics of the EoS to express functional dependencies. EoSs used in merger simulations are derived within different models and there is no commonly used overarching framework to define universal model parameters. However, it is possible to extract nuclear parameters from the EoS such as the symmetry energy and its slope or the incompressibility. These parameters are useful to describe the EoS but are only characteristic of a certain density range (at saturation density). 
Hence, it is obligatory to employ the stellar properties of cold, non-rotating NSs to characterize a given EoS \citep[but see][for an exploration of the impact of the slope of the symmetry energy]{Most+2021}. Stellar parameters such as the radius, compactness, tidal deformability or maximum mass are uniquely linked to the EoS through the Tolman-Oppenheimer-Volkoff equations of stellar structure. As integrated bulk properties these quantities characterize well the behavior of the EoS across a larger density range and, in particular, they are useful to describe the dynamics of a COM.

Stellar parameters of cold, non-rotating NSs are determined by the cold EoS in neutrinoless beta-equilibrium, whereas COMs reach finite temperatures and finite neutrino chemical potentials during merging. Therefore, in simulations the full temperature and composition dependence of the EoS has to be taken into account together with neutrino transport. Still, the EoS slice at $T=0$ and beta-equilibrium describes the basic properties of the model, where finite temperature and composition changes add moderate deviations.

{\bf Merger outcome, threshold mass for prompt collapse:}
The EoS affects mass ejection and r-process nucleosynthesis in many, partially subtle ways. The most basic influence is on the immediate outcome of a NS merger, i.e., the EoS dictates if for a given binary mass configuration a NS remnant or a BH-torus system forms~\citep{Shibata2006,Baiotti2008}. This obviously has significant impact on the ejecta properties, since it influences the very early dynamical mass ejection and it determines if there is a phase of mass ejection from a NS remnant at all. This very basic characteristic of a NS merger is quantitatively described by the threshold binary mass $M_\mathrm{thres}$ for prompt collapse~\citep{Hotokezaka2011,Bauswein2013}. Systems with a total binary mass $M_\mathrm{tot}$ exceeding $M_\mathrm{thres}$ directly form a BH, whereas binaries with $M_\mathrm{tot}<M_\mathrm{thres}$ lead to the formation of an at least transiently stable rotating NS remnant.

Using numericcal simulations $M_\mathrm{thres}$ is found to depend in a systematic way on the EoS, which can be expressed through stellar parameters of non-rotating NSs. With good accuracy $M_\mathrm{thres}$ follows
\begin{equation}\label{eq:mthr}
    M_\mathrm{thres}(R_{1.6},M_\mathrm{max},q)=c_1 M_\mathrm{max}+c_2 R_{1.6}+c_3 +c4 (1-q)^3 M_\mathrm{max}+c5 (1-q)^3 R_{1.6}
\end{equation}
with $c_1=0.578$, $c_2=0.161$, $c_3=-0.218$, $c_4=8.987$ and $c_5=-1.767$~\citep{Bauswein2021}. In Equation~(\ref{eq:mthr}) the EoS dependence is expressed through the maximum mass of non-rotating NSs, $M_\mathrm{max}$, and the radius of a non-rotating NS with a mass of 1.6~$M_\odot$. The latter may be replaced by other quantities like, for instance, the tidal deformability at a given mass, which similar to the radius is a measure of the NS compactness. The fact that $M_\mathrm{thres}$ correlates with $M_\mathrm{max}$ may not be surprising since $M_\mathrm{max}$ determines the threshold for BH formation of non-rotating NSs. Equation~(\ref{eq:mthr}) also contains an explicit dependence on the binary mass ratio $q$. For small deviations from $q=1$ the threshold mass does not change strongly compared to the equal-mass case. For systems with significant mass asymmetry, i.e. $q\lesssim 0.8$, the threshold mass is significantly reduced reflecting the fact that the merger remnant in this case features less total angular momentum and an angular momentum distribution with less rotation in the center. Based on these insights one may expect that generally NS merger remnants of asymmetric binaries are less stable as compared to the equal-mass case of the same total mass.

\begin{figure}
\begin{center}
\includegraphics[width=0.8\columnwidth]{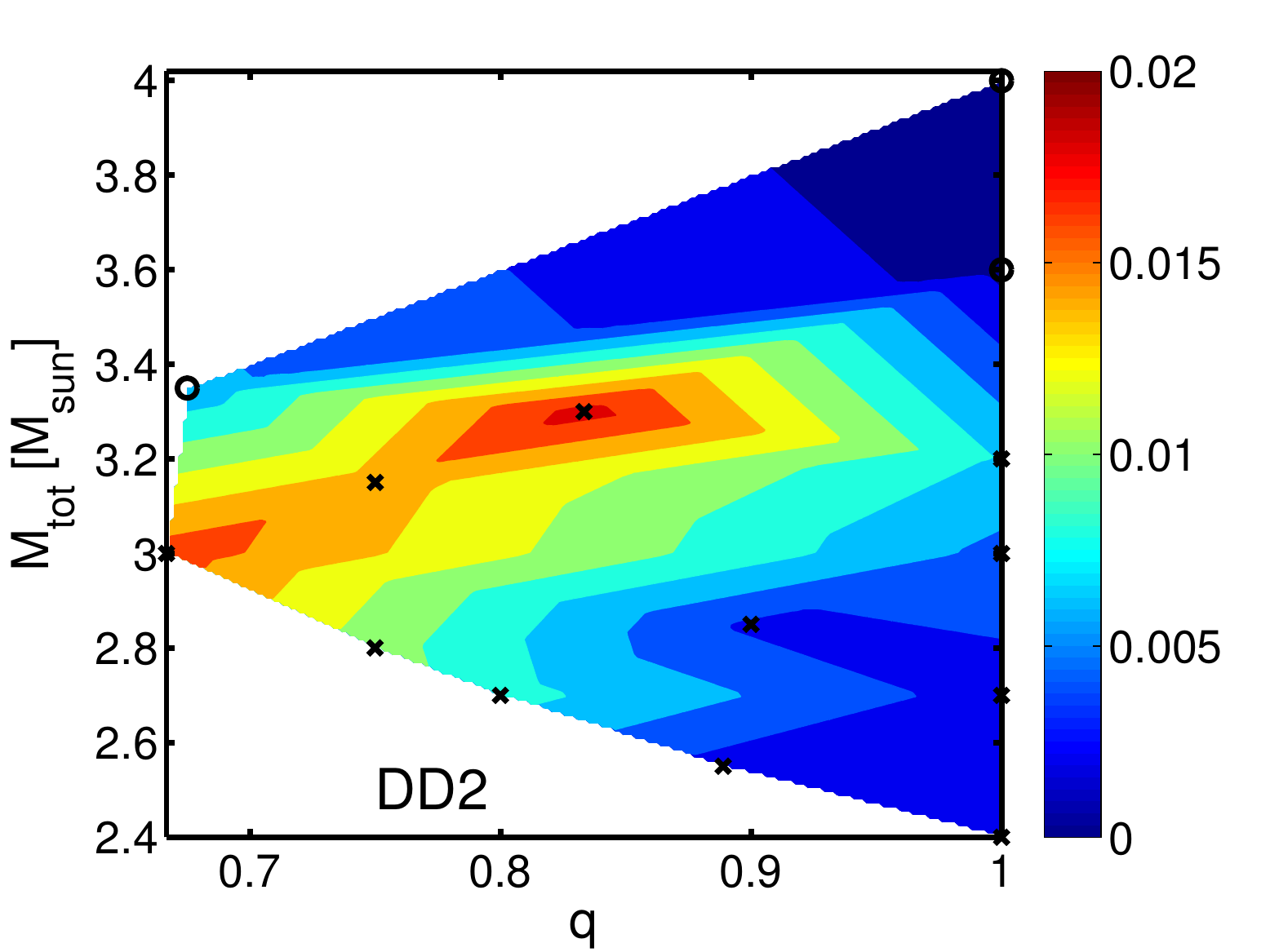}
\end{center}
\caption{Dynamical ejecta mass (color-coded in $M_\odot$) as function of the binary mass ratio $q$ and the total binary mass $M_\mathrm{tot}$. Figure from~\cite{Bauswein+2013}; \copyright American Astronomical Society. Reproduced with permission}
\label{fig:mejbinmass}
\end{figure}

{\bf Dynamical ejecta dependencies:} Already early studies of the impact of the EoS and binary masses on the dynamical ejecta identified basic dependencies by evaluating a large sample of simulations with different high-density models~\citep{Hotokezaka2013,Bauswein+2013,Dietrich2017a}. (i) For a fixed EoS the ejecta mass increases with the binary mass asymmetry. (ii) Typically, for a fixed EoS the ejecta mass mildly increases with the total binary mass. (iii) If a prompt collapse occurs ($M_\mathrm{tot}>M_\mathrm{thres}$), the dynamical ejecta mass is strongly reduced. These dependencies are visualized for an exemplary EoS model in Fig.~\ref{fig:mejbinmass}. As described in more detail in Section~\ref{sec:massejection}, simulation results do show some uncertainties, which should be kept in mind when one quantitatively interprets such ejecta properties and relations. (iv) For a fixed binary mass configuration, soft EoSs typically lead to more ejecta, if the binary mass asymmetry is not too large and the system does not directly form a BH. Mergers with softer EoSs resulting in smaller NS radii collide with a larger impact velocity and feature more violent postmerger oscillations \citep[see, e.g.,][]{Bauswein+2013}. This leads to enhanced mass ejection.

While all of these dependencies are physically understandable it should be emphasized that especially the EoS dependence is not an exact relation but features some scatter of at least several ten per cent. 

Some studies have summarized these simulation results and dependencies by developing fit formulae, which provide the ejecta mass or other ejecta properties as functions of the binary masses and parameters describing the EoS, i.e., NS parameters~\citep[e.g.][]{Dietrich2017,Coughlin2018,Krueger2020,Nedora2022}. Some of those models include results from several different sets of simulations, which on the one hand might trigger the question of consistency (different resolution, numerical treatment, physical effects, ways of defining and extracting ejecta) but on the other hand provides a better and more representative overview. \cite{Henkel2022} collect and review several of such fit formulae.

\begin{figure}
\begin{center}
\includegraphics[width=0.8\columnwidth]{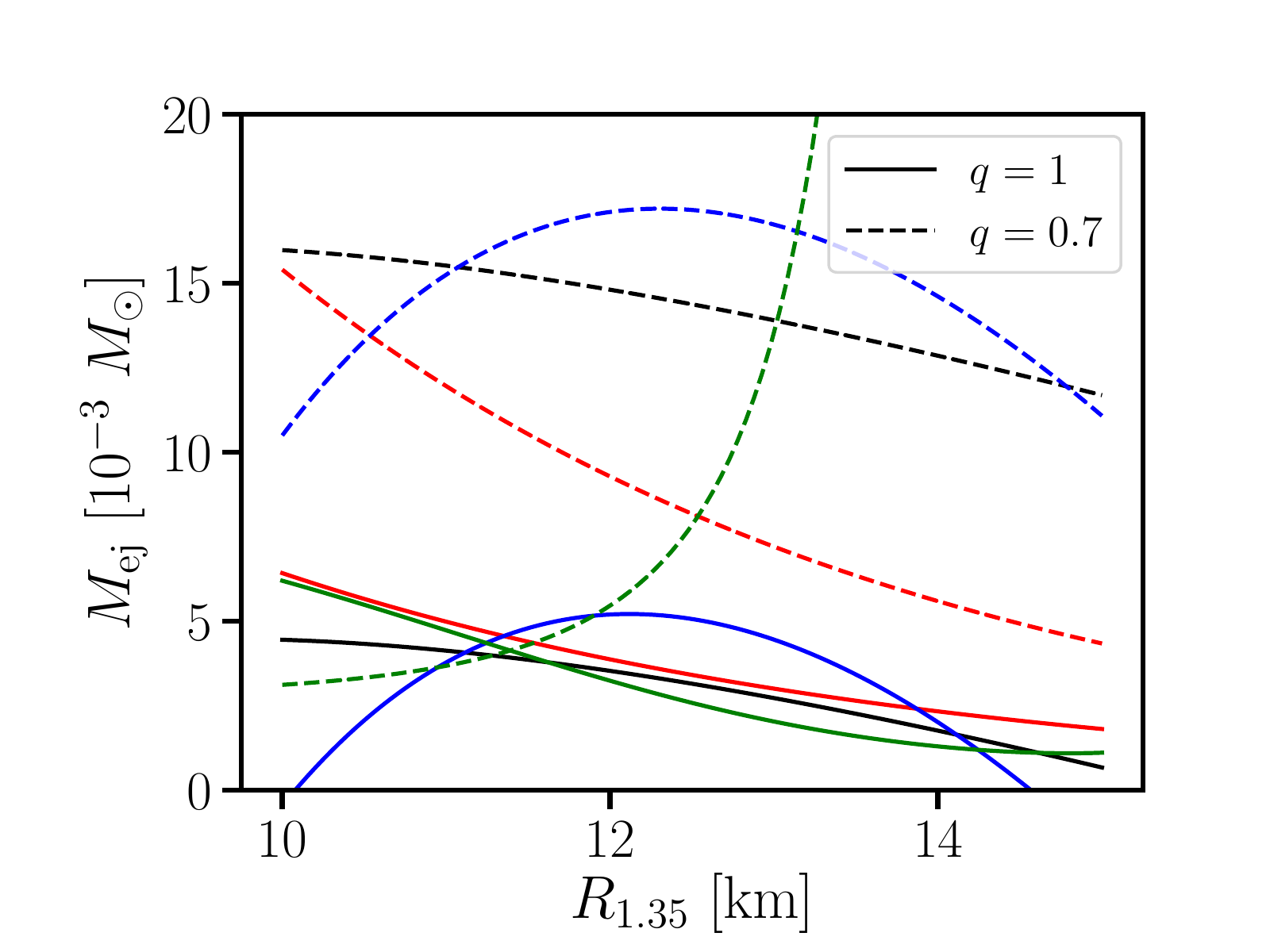}
\end{center}
\caption{Dynamical ejecta mass as function of the radius $R_{1.35}$ of a 1.35~$M_\odot$ NS (representing the EoS dependence) as predicted by different fit formulae (black:~\citealt{Dietrich2017}; red:~\citealt{Coughlin2018}; blue:~\citealt{Krueger2020}; green:~\citealt{Nedora2022}; additional assumptions are adopted to present all fit formulae as functions of $R_{1.35}$). A total mass of 2.73~$M_\odot$ is assumed, where the solid curves refer to a mass ratio of $q=1$ and the dashed curves to $q=0.7$. This roughly spans the possible range of binary mass ratios of GW170817}
\label{fig:mejR}
\end{figure}

Following this study to exemplify the EoS dependence, Fig.~\ref{fig:mejR} shows the ejecta mass as function of the NS radius of a 1.35~$M_\odot$ NS. See e.g.~\cite{Henkel2022} for more details. For this plot binary masses are chosen such that they represent the ones of GW170817 with the solid curves assuming an equal-mass binary and the dashed lines refering to a mass ratio of $q=0.7$. The different colors refer to different fit formulae developed in the literature. It should be noted that additional assumptions and relations~\citep{Timmes1996,De2018,Bauswein2021} were adopted in order to compose the various fit formulae in a single plot like Fig.~\ref{fig:mejR}. For instance, some relations are not given as function of the NS radius or compactness but of the tidal deformability, which can be converted to a radius estimate using the relations of \citep[e.g.][]{De2018}.

For equal-mass binaries the relations broadly agree with each other with the main difference being that the fit formula of~\cite{Krueger2020} displays a reduction of ejecta mass for soft EoSs, i.e., models with small NS radii. Systems with a binary mass asymmetry show a significantly larger scatter, which results from the chosen simulation models entering the fit formula. Note that for the sake of clarity, Fig.~\ref{fig:mejR} does not display error bars, which for some of the fit formulae have been estimated and are usually relatively large (several ten per cent). The importance of such fit formulae lies in the fact that they can be seen as a form of meta-study summarizing simulation results and that they can be employed to constrain EoS parameters, when ejecta properties such as the ejecta mass are estimated by an interpretation of the kilonova (see above). Numerous studies have followed this route, often incorporating estimates of the ejecta velocity and including contributions from the secular ejecta~\citep[e.g.,][]{Coughlin2018,Coughlin2019,Radice2019,Capano2020,Dietrich2020,Breschi2021,Raaijmakers+2021,Huth2022}.

The fit formulae for the averaged ejecta velocities show that one typically expects an increase of the velocity for soft EoSs, i.e., small NS radii~\citep[e.g.,][]{Nedora2022}. This may be expected because as argued before, those systems lead to a more violent merging and strongly excited remnant oscillations, which enhances the outflow velocity. There also seems to be the tendency that asymmetric binaries lead to somewhat smaller ejecta velocities for the same total mass and the same EoS model, which may again be explained by the remnant dynamics being less violent in these cases. Typical average outflow velocities are between roughly 0.15~$c$ and 0.35~$c$.

The electron fraction of the dynamical ejecta shows typically a broad distribution between about 0.05 and 0.45 with a mildly pronounced maximum in the range between 0.2 and 0.3~\citep[e.g.][]{Palenzuela2015,Sekiguchi2015,Foucart2016,Lehner2016,Radice2016a,Bovard2017,Radice+2018,ArdevolPulpillo2019,Combi2022,Vincent2020}. The distributions of $Y_e$ exhibit a mild dependence on the EoS, but it is not straightforward to identify a clear systematic dependence. Tentatively the outflow becomes less neutron rich with larger binary mass asymmetry and with the EoS stiffness parameterized e.g. by the NS radius or its tidal deformability. Overall, there is a significant scatter in these tentative relations \citep[see, e.g., the compilation in][and the fit formulae therein]{Nedora2022}, and one should keep in mind that only simulations which include the reabsorption of neutrinos in some way can yield a reliable distribution of the electron fraction in the dynamical ejecta. 
Moreover, the variations in the distribution are so modest that the impact on the final r-process abundance patterns is usually small~\citep{Kullmann2022a,Radice+2018}. This can be seen in Fig.~\ref{fig:abueos}, which shows the outcome of nuclear network calculations for merger simulations with different EoSs and binary mass ratios. A rather robust feature in most simulations is that the electron fraction is higher in directions towards to poles where neutrino reabsorption is more effective~\citep[e.g.][]{Palenzuela2015,Sekiguchi2015,Foucart2016,Lehner2016,Bovard2017,Radice+2018,ArdevolPulpillo2019,Vincent2020,Combi2022,Kullmann2022a}.

{\bf Torus mass and secular ejecta:} Another important and possibly even dominant component of the ejecta stems from the secular evolution of a torus around the central object (either a BH or a rotating NS). Identifying torus material and thus defining torus properties can be ambiguous before the gravitational collapse of the remnant, because one attempts to estimate the amount of matter escaping the forming BH, which may only form after some further dynamical evolution. Nonetheless, one can observe some clear dependencies of the torus mass~\citep[e.g.][]{Shibata2006,Oechslin2006,Oechslin2007,Hotokezaka2013,Dietrich2017a,Radice+2018}. For the same EoS and the same total binary mass, asymmetric systems lead to more massive tori, which is understandable from the fact that these mergers lead to a tidal disruption of the lighter binary component with its larger fraction of the system's oribtal angular momentum. For fixed binary masses, mergers with stiffer EoSs corresponding to larger NS radii typically result in higher torus masses. These systems are less strongly bound and they are associated with longer remnant life times and hence there is more time to redistribute angular momentum from the central regions of the remnant to the outer layers. Typical torus masses range from several $10^{-3}~M_\odot$ to a few 0.1~$M_\odot$ if no direct collapse occurs.

The torus mass is strongly reduced for systems where prompt BH formation takes place~\citep[e.g.,][]{Shibata2006,Just+2015,Radice+2018}. In these cases the torus mass may not exceed $10^{-3}$~$M_\odot$ unless the binary system is strongly asymmetric such that the tidal disruption of the lighter component directly produces a massive torus in the merging phase. If the remnant survives for only a short period of time, the torus mass can easily reach 0.1~$M_\odot$ especially for asymmetric binaries \citep[see also][]{Kiuchi2019} for a detailed exploration of the transition region). In combination with the considerations about the dynamical ejecta, this suggests a pronounced difference in the kilonova properties of prompt-collapse and delayed-collapse systems and emphasizes the importance to understand the exact threshold mass for direct BH formation.

\begin{figure}
\begin{center}
\includegraphics[width=0.8\columnwidth]{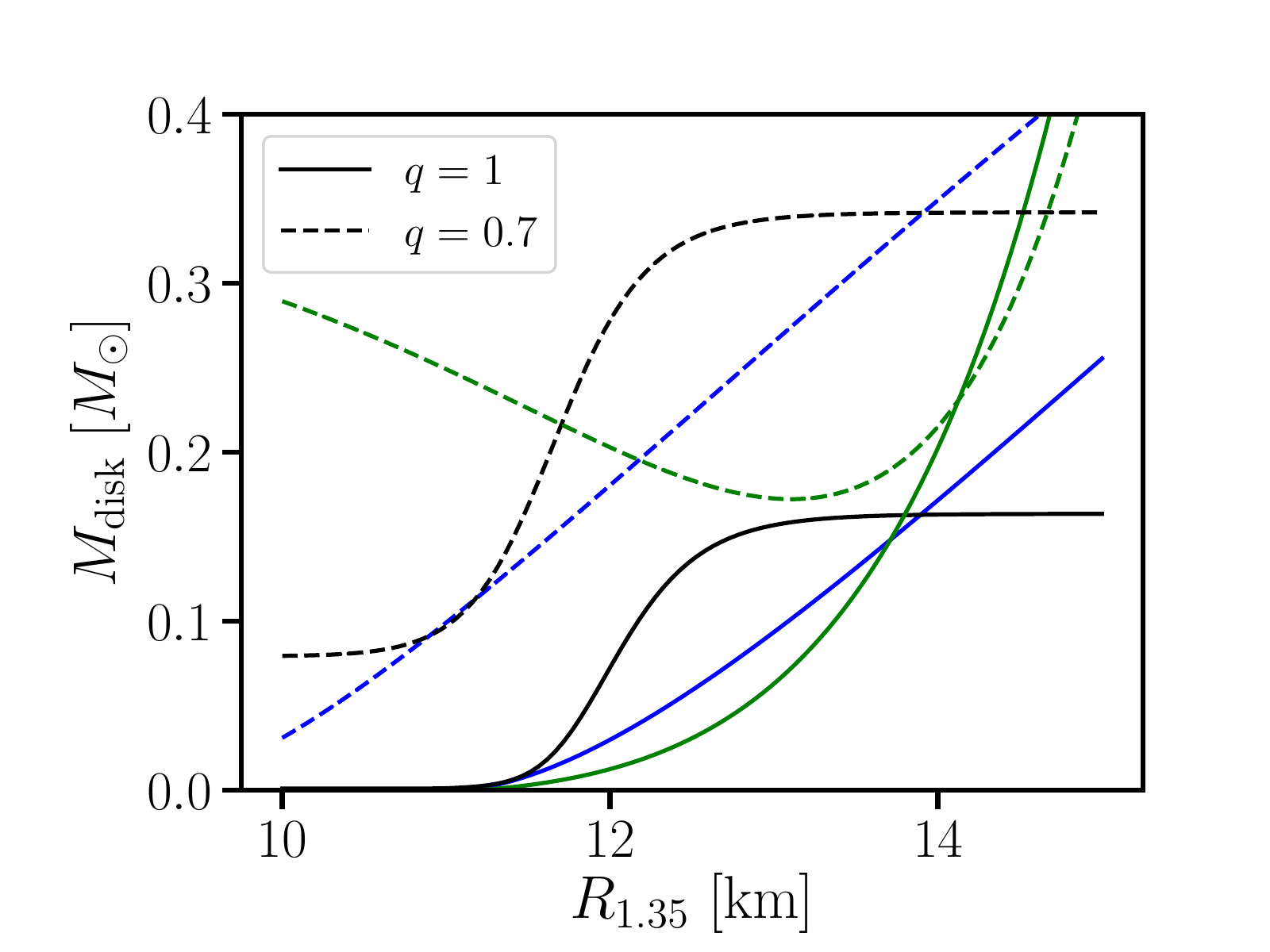}
\end{center}
\caption{Torus mass as function of the radius $R_{1.35}$ of a 1.35~$M_\odot$ NS (representing the EoS dependence) as predicted by different fit formulae (black:~\citealt{Dietrich2020}; blue:~\citealt{Krueger2020}, green:~\citealt{Nedora2022}; additional assumptions are adopted to present all fit formulae as functions of $R_{1.35}$). A total mass of 2.73~$M_\odot$ is assumed, where the solid curves refer to a mass ratio of $q=1$ and the dashed curves to $q=0.7$. This roughly spans the possible range of binary mass ratios of GW170817}
\label{fig:mdiskR}
\end{figure}

Fit formulae have also been developed for the torus mass expressing the binary mass and EoS dependence~\citep[e.g.][]{Radice+2018,Dietrich2020,Krueger2020,Nedora2022}. The differences between different relations that are available in the literature can be significant (factor two). See, e.g.,~\cite{Henkel2022} for a more detailed discussion and references to various formulae. It is clear that the different relations are not only affected by which function is chosen for the fit, but in particular also by the set of models included. Figure~\ref{fig:mdiskR} displays torus masses obtained from different fit formulae. Again, there is a rough agreement between the different relations for $q=1$, while there is a larger scatter for asymmetric binaries.

Most of the torus material will be accreted onto the BH on a longer time scale. Only a fraction of roughly 20 per cent becomes gravitationally unbound during this evolution~\citep{Just+2015,Fernandez2015a,Siegel2018,Fernandez2020,Fujibayashi2020a,Just2022a}. For typical torus masses of $\sim 0.1~M_\odot$ one can thus expect a few $10^{-2}~M_\odot$ of torus ejecta. Hence, the amounts of torus ejecta may exceed the dynamical ejecta mass (see also Fig.~\ref{fig:dynsec}).

It is important to note that it is challenging to compute the exact fraction of unbound torus material since it depends on a detailed modeling of magneto-hydrodynamic processes and of neutrino transport effects. In addition, the fraction of the torus which is ejected, does depend on physical parameters of the system and through this on the initial binary masses and the EoS. \cite{Just2022a} for instance provide a systematic study of the impact of the torus properties on the ejecta as well as an assessment of various modeling ingredients. For instance, more massive or faster spinning BHs increase the fraction of the torus material becoming unbound. Similarly, the average electron fraction of the ejected torus matter is dependent on the properties of the system. Typically, the electron fraction of the torus ejecta is slightly above 0.3 and thus the r-process creates predominantly nuclei of the first and second r-process abundance peaks, whereas species with higher mass number are less abundant in these outflows. 

{\bf Phase transition to deconfined quark matter:} A particular EoS aspect which is hardly explored yet, is the impact of a hadron-quark phase transition on the mass ejection. If deconfined quark matter occurs in a NS, it can alter the stellar structure significantly and thus, in principle, also affect the merger dynamics and mass ejection. Current studies of the dynamical phase suggest that there may not be an unambiguous imprint in the sense that the occurrence of a phase transition would lead to a characteristic increase or decrease of the mass ejection or significantly alter the elemental abundance pattern~\citep{Bauswein2019b,Prakash2021}. But more work is required on this aspect.

{\bf EOS constraints:} 
The EoS dependence as discussed is the basis for various constraints on the properties of NSs and high-density matter, which are derived from the multi-messenger observation of GW170817. These approaches can be roughly classified according to how and which information from the kilonovae is made use of. The methodology and the detailed assumptions within these classes partially differ, which if why the quantitative results can show some differences. Also, the association of a specific work to one of these classes below is not always unambiguous.

(i) Some works conclude that because of the brightness of the kilonova no direct BH formation took place in GW170817, and, hence, the threshold binary mass for prompt BH formation is larger than the measured total binary mass of GW170817. This can be translated into a lower bound on NS radii or the tidal deformability since more compact NSs yielding a prompt collapse would lead to a dimmer electromagnetic emission~\citep{Bauswein+2017,Radice2018c,Most2018,Bauswein2019b,Koeppel2019,Capano2020,Bauswein2021}. In summary, this argument leads to the conclusion that nuclear matter cannot be very soft and NS radii should be roughly larger than about 11~km.

(ii) A number of studies rely on the conclusion that GW170817 did form a NS remnant and employ additional information or assumptions, e.g., from the detailed properties of the ejecta or the observation of a short GRB about 2 seconds after merging. The latter suggest that a BH formed at some point during the remnant evolution~\citep{Margalit+2017,Shibata+2017,Rezzolla+2018,Ruiz+2018,Margalit2019,Capano2020}. This can be converted to an upper limit on the maximum mass $M_\mathrm{max}$ of non-rotating NSs, which should be smaller than about 2.2~$M_\odot$.

(iii) Another class of interpretations of the observational data directly connects the properties of the kilonova light curve and inferred ejecta properties to NS parameters employing empirical relations, which are obtained from simulations~\citep[e.g.][]{Coughlin2018,Coughlin2019,Radice2019,Dietrich2020,Breschi2021,Raaijmakers+2021,Huth2022}. These studies are based on fit formulae as described above. Clearly, this approach is in comparison more ambitious as the fit formulae include significant uncertainties and, moreover, the inference of ejecta properties from the observations introduces additional uncertainties. The bounds on NS radii are roughly in the range between 11 and 13~km. Note that most of the studies above actually combine EoS constraints from different sources, including, for instance, observations by NICER, the GW inspiral signal, theoretical calculations of nuclear matter properties, heavy-ion collisions, pulsar observations or the limits derived by the studies listed in class (ii).

\subsection{Outlook - Mergers}\label{sec:outmergers}

These considerations of the different ejecta components show that the total yields, i.e., the combination of dynamical ejecta and secular ejecta from a massive NS evolution and the disk wind, depend sensitively on the EoS and the binary masses. The discussion also indicates that it is not easy to exactly determine the specific contributions of the individual mass ejection channels and thus the total yields of a given event. To date only a subset of models is striving for an inclusion or combination of all ejecta components, and a fully consistent, systematic broad survey is still missing~\citep[e.g.,][]{Perego2014,Just+2015,Martin2015,Fernandez2017,HosseinNouri2018,Fujibayashi2020,Most2021a,Fujibayashi2022}. Some of these studies~\citep[e.g.,][]{Fujibayashi2022} indicate that not every merger event may produce nucleosynthetic yields that follow the solar distribution, although some models result in a solar abundance pattern over a wide range of mass number. For instance, the calculations of~\cite{Fujibayashi2022} suggest that the remnant life time may affect the abundance distribution and hence the nucleosynthetic output depends on the binary masses. It is clear that future work should further improve the modeling of individual ejecta components but in particular address how those can be consistently combined. Also, there is a growing number of studies that consider the impact of different types of possible neutrino flavor conversions on the nucleosynthesis in a merger environment. This aspect could not be covered here, and the reader is referred to the literature, e.g.,~\cite{Zhu2016,Malkus2016,Wu2017,Richers2021,Li2021,Just2022b} (and references therein).

Future GW detections of COMs and associated observations of electromagnetic counterparts will provide valuable information on the nucleosynthesis in these events. These measurements will improve the current estimates of the merger rate in the local Universe and the distribution of masses within the binary population. Also the identification of host galaxies can help to better understand the environments of mergers. This may possibly also provide clues on delay times until merging and kick velocities. The GW signal alone or in combination with the electromagnetic counterparts will further constrain the EoS and NS properties. More kilonova observations in combination with more advanced models will allow to determine bulk ejecta properties like their masses and potentially the elemental composition of the outflow. Combining this information will provide a more comprehensive picture of the nucleosynthetic yields of COMs clarifying whether mergers produce r-process elements with a solar distribution and whether they are the dominant source of r-process elements.

\section{Summary and perspectives}

The phenomena and multi-messenger signals connected to CCSNe and COMs described in this
chapter render these violent astrophysical events superb and unique cosmic laboratories for
nuclear physics and astro-particle physics, especially for testing and exploiting the
processes in high-density matter at temperatures up to about 100\,MeV. Neutrinos, GWs, and
the products of explosive nucleosynthesis carry information of the physics inside and in
the close vicinity of the hot NSs that are formed or collide in such environments.
Their computational modeling in three dimensions, even including general relativity and magnetic
fields, has made enormous progress over the past decade, and basic dependencies of these 
messengers on the properties of the nuclear EoS begin to emerge from systematic studies.
However, further improvements of the models are needed for high-fidelity predictions of
neutrinos and GWs, for example with respect to the numerical resolution especially in 
magnetohydrodynamic simulations or concerning the treatment of the neutrino transport, 
which still lags behind state-of-the-art CCSN models in the majority of elaborate COM models
\citep[e.g.,][and references therein]{Foucart2022}. Further theoretical exploration is also 
needed to solve the nagging problem of potential consequences of neutrino-flavor oscillations 
in the dense media of newly born and colliding NSs. 
Successively improved computational modeling and a deeper fundamental understanding of the 
neutrino physics will ultimately also permit to reliably connect the electromagnetically 
observable emission with the properties of the nuclear and super-nuclear EoS of hot NS matter, 
which governs the dynamical evolution, ejecta mass, neutrino production, and thus also 
the nucleosynthetic outputs as well as light curves, spectra, X-ray, and $\gamma$-ray 
signals of the explosive events.

Not being enshrouded by solar masses of stellar material, COMs may offer the cleaner 
astrophysical sites for addressing these questions, and in the era of GW astronomy 
they are certainly the more frequent observable multi-messenger sources compared to CCSNe.
One COM per decade seen in GWs and electromagnetic emission seems likely and even a few 
events in this time span may be possible, whereas neutrinos and GWs from CCSNe will be 
measurable only for a stellar death in the Milky Way and its close cosmic neighborhood. 
With the considerable recent progress in the numerical model building, 
however, even the historical neutrino detection connected to SN~1987A has
gained new interest and relevance for deriving tighter constraints on nuclear and particle
processes in hot NS environments \citep[e.g.,][]{Bollig+2020,Caputo+2022,Fiorillo+2022},
and the possible detection of a NS in SN~1987A \citep{Cigan+2019,Page+2020} will further 
propel these potentials.
Moreover, CCSNe are enormously diverse and multi-faceted producers of chemical elements
and the far more abundant cosmic phenomena (presently 1--2\,events per century in Milky Way 
type galaxies compared to maybe a few events per 100,000 years for COMs). Both circumstances 
define the outstanding role of SNe for the chemical evolution of galaxies 
and the creation of stars, planets, and life in the Universe. A next Galactic CCSN or
stellar BH formation will happen, and we should get prepared for maximizing the harvest 
from the flood of its data.

\begin{acknowledgement}
HTJ is grateful for support by the Deutsche Forschungsgemeinschaft (DFG, German
Research Foundation) through Sonderforschungsbereich (Collaborative Research
Center) SFB~1258 ``Neutrinos and Dark Matter in Astro- and Particle Physics (NDM)''
and under Germany's Excellence Strategy through Cluster of Excellence ORIGINS (EXC-2094)-390783311. AB acknowledges support by the European Research Council (ERC) under the European Union's Horizon 2020 research and innovation program under grant agreement No.~759253, support by DFG --Project-ID 279384907-- SFB~1245 and DFG --Project-ID 138713538-- SFB~881 (``The Milky Way System'', subproject A10) and support by the State of Hesse within the Cluster Project ELEMENTS.
\end{acknowledgement}


\end{document}